\documentclass[twocolumn,twocolappendix]{aastex63}
\usepackage{xspace}
\usepackage{hyperref}
\usepackage{multirow}
\usepackage{graphicx}
\usepackage{enumitem}
\usepackage{amssymb}
\usepackage{amsmath}
\usepackage{float}
\usepackage{hyperref}

\newcommand{\chandra}{\textit{Chandra}\xspace}
\newcommand{\xmm}{XMM-\textit{Newton}\xspace}
\defcitealias{lamassa16}{L16}
\defcitealias{ueda14}{U14}
\defcitealias{aird15}{A15}
\defcitealias{buchner15}{B15}
\defcitealias{ananna19}{A19}
\defcitealias{yuan22}{Y22}

\shorttitle{S82X: spectral analysis \& evolution of AGN}
\shortauthors{Peca et al.}
\graphicspath{{./}{}}

\begin{document}

\title{On the cosmic evolution of AGN obscuration and the X-ray luminosity function: XMM-Newton and Chandra spectral analysis of the 31.3 deg$^2$ Stripe 82X}

\author[0000-0003-2196-3298]{Alessandro Peca$^{\star}$}
\affiliation{Department of Physics, University of Miami, 
Coral Gables, FL 33124, USA; $^{\star}$alessandro.peca@miami.edu}

\author[0000-0002-1697-186X]{Nico Cappelluti}
\affiliation{Department of Physics, University of Miami, 
Coral Gables, FL 33124, USA; $^{\star}$alessandro.peca@miami.edu}

\author[0000-0002-0745-9792]{C. Megan Urry}
\affiliation{Yale Center for Astronomy \& Astrophysics, 52 Hillhouse Avenue, New Haven, CT 06511, USA}
\affiliation{Department of Physics, Yale University, P.O. Box 208120, New Haven, CT 06520, USA}

\author[0000-0002-5907-3330]{Stephanie LaMassa}
\affiliation{Space Telescope Science Institute, 3700 San Martin Drive, Baltimore, MD 21210, USA}

\author[0000-0001-5544-0749]{Stefano Marchesi}
\affiliation{INAF – Osservatorio di Astrofisica e Scienza dello Spazio di Bologna, Via Piero Gobetti, 93/3, 40129 Bologna, Italy}
\affiliation{Department of Physics and Astronomy, Clemson University, Kinard Lab of Physics, Clemson, SC 29634, USA}

\author[0000-0001-8211-3807]{Tonima Tasnim Ananna}
\affiliation{Department of Physics and Astronomy, Dartmouth College, 6127 Wilder Laboratory, Hanover, NH 03755, USA}

\author[0000-0003-0476-6647]{Mislav Balokovi\'{c}}
\affiliation{Yale Center for Astronomy \& Astrophysics, 52 Hillhouse Avenue, New Haven, CT 06511, USA}
\affiliation{Department of Physics, Yale University, P.O. Box 208120, New Haven, CT 06520, USA}

\author[0000-0002-1233-9998]{David Sanders}
\affiliation{Institute for Astronomy, University of Hawaii, 2680 Woodlawn Drive, Honolulu, HI 96822, USA}

\author[0000-0002-5504-8752]{Connor Auge}
\affiliation{Institute for Astronomy, University of Hawaii, 2680 Woodlawn Drive, Honolulu, HI 96822, USA}

\author[0000-0001-7568-6412]{Ezequiel Treister}
\affiliation{Instituto de Astrofísica, Facultad de Física, Pontificia Universidad Católica de Chile, Casilla 306, Santiago 22, Chile}

\author[0000-0003-2284-8603]{Meredith Powell}
\affiliation{Kavli Institute of Particle Astrophysics and Cosmology, Stanford University, 452 Lomita Mall, Stanford, CA 94305, USA}

\author[0000-0003-2971-1722]{Tracey Jane Turner}
\affiliation{Eureka Scientific, Inc., 2452 Delmer Street, Suite 100, Oakland, CA 94602-3017, USA}

\author[0000-0002-1306-1545]{Allison Kirkpatrick}
\affiliation{Department of Physics \& Astronomy, University of Kansas, Lawrence, KS 66045, USA}

\author[0000-0003-4056-7071]{Chuan Tian}
\affiliation{Yale Center for Astronomy \& Astrophysics, 52 Hillhouse Avenue, New Haven, CT 06511, USA}
\affiliation{Department of Physics, Yale University, P.O. Box 208120, New Haven, CT 06520, USA}

\begin{abstract}

We present X-ray spectral analysis of \xmm and \chandra observations in the 31.3 deg$^2$ Stripe-82X (S82X) field. Of the 6181 unique X-ray sources in this field, we analyze a sample of 2937 candidate active galactic nuclei (AGN) with solid redshifts and sufficient counts determined by simulations. Our results show an observed population with median values of spectral index $\Gamma=1.94_{-0.39}^{+0.31}$, column density log$\,N_{\mathrm{H}}/\mathrm{cm}^{-2}=20.7_{-0.5}^{+1.2}$ and intrinsic, de-absorbed, 2-10 keV luminosity log$\,L_{\mathrm{X}}/\mathrm{erg\,\,s}^{-1}=44.0_{-1.0}^{+0.7}$, in the redshift range 0-4. We derive the intrinsic, model-independent, fraction of AGN that are obscured ($22\leq\mathrm{log}\,N_{\mathrm{H}}/\mathrm{cm}^{-2}<24$), finding a significant increase in the obscured AGN fraction with redshift and a decline with increasing luminosity. The average obscured AGN fraction is $57\pm4\%$ for log$\,L_{\mathrm{X}}/\mathrm{erg\,\,s}^{-1}>43$. This work constrains the AGN obscuration and spectral shape of the still uncertain high-luminosity and high-redshift regimes (log$\,L_{\mathrm{X}}/\mathrm{erg\,\,s}^{-1}>45.5$, $z>3$), where the obscured AGN fraction rises to $64\pm12\%$. We report a luminosity and density evolution of the X-ray luminosity function, with obscured AGN dominating at all luminosities at $z>2$ and unobscured sources prevailing at log$\,L_{\mathrm{X}}/\mathrm{erg\,\,s}^{-1}>45$ at lower redshifts. Our results agree with the evolutionary models in which the bulk of AGN activity is triggered by gas-rich environments and in a downsizing scenario. Moreover, the black hole accretion density (BHAD) is found to evolve similarly to the star formation rate density, confirming the co-evolution between AGN and host-galaxy, but suggesting different time scales in their growing history. The derived BHAD evolution shows that Compton-thick AGN contribute to the accretion history of AGN as much as all other AGN populations combined.

\end{abstract}


\section{Introduction} \label{sec:intro}
Active galactic nuclei (AGN) are the manifestation of gas accretion onto supermassive black holes (SMBHs). During this growth phase, AGN produce powerful radiation detectable across the entire electromagnetic spectrum (\citealt{antonucci93,urry95}). 
The evolution of SMBH growth and its relation to the surrounding environment are still a matter of active study due to the need for unbiased multiwavelength samples with high spectroscopic completeness and large enough volume to reach high luminosities and redshifts.
AGN selection at optical-UV wavelengths is biased against obscured sources
(i.e., with column densities $\log N_{\rm H}/\mathrm {cm}^{-2} > 22$)
because the primary emission from accretion is scattered or absorbed by dust and gas along our line of sight, either on small circumnuclear scales or in dense clouds in the host galaxy.
In the infrared and optical, AGN can also be diluted by stellar emission from the host galaxy.

For these reasons, X-ray selection is a leading approach for defining nearly unbiased samples (e.g., \citealt{brandt15}). 
In particular, hard X-rays ($E\gtrsim 2$ keV) penetrate even large obscuring column densities, leading to relatively complete AGN samples (e.g., \citealt{hickox18}). While in the local Universe these studies are possible with telescopes sensitive at energies above 10 keV, like \textit{Swift-BAT} and \textit{NuSTAR} (e.g., \citealt{marchesi19,lamassa19,koss22}), at higher redshifts the relevant penetrating emission enters the \chandra and \xmm $\sim$0.5-10 keV energy band. This allows us to leverage the extensive amount of data available in both the \chandra and \xmm archives.
At the same time, the volume density of AGN drops dramatically at the highest luminosities ($\log L_{\rm X}/\mathrm{erg\,\, s}^{-1}>45$) and redshifts ($z>3$) (e.g., \citealt{hasinger08}, \citealt{ueda14}, hereafter \citetalias{ueda14}, \citealt{aird15}, hereafter \citetalias{aird15}, \citealt{buchner15}, hereafter \citetalias{buchner15}, \citealt{miyaji15}, and \citealt{ananna19}, hereafter \citetalias{ananna19}), leading deep pencil-beam X-ray surveys to miss the most powerful accretors. According to popular models (e.g., \citealt{hopkins09, treister12}), these sources represent evolutionary key phases where the bulk of the mass is accreted onto the central SMBH. Due to their low space density, large area surveys become crucial to collect them in samples large enough to perform population studies.
This is why at present, the census of high-luminosity, high-redshift, and highly obscured sources, as well as their evolution, is still poorly known.

Here we address this gap using the large-volume Stripe 82X survey (S82X, \citealt{lamassa13a,lamassa13b,lamassa16}), which covers an area of $\sim$31.3 deg$^2$ with \xmm and \chandra observations. 
While S82X is not the only large area survey observed in hard X-rays (e.g., XMM-XXL, \citealt{pierre16}; CDWFS, \citealt{masini20}) which covers the bright portion of the flux-area plane (see e.g., figure 1 in \citealt{nanni20}), 
it benefits from unprecedented multiwavelength coverage. S82X was observed from the UV to the radio wavebands, including from facilities no longer available like \textit{Spitzer} and \textit{Herschel} (see \citealt{lamassa16, ananna17} for details on the datasets).
While X-ray photons are a powerful tool to uncover and study the black hole activity, supporting information about the AGN and their host galaxies is needed for a reliable analysis. In particular, the rich S82X multiband coverage allows the estimate of both spectroscopic and high-quality photometric redshifts, which are fundamental for X-ray spectral fitting (e.g., \citealt{salvato09,ananna17,peca21}).

In this work we 
perform a detailed 
X-ray spectral analysis of the S82X sources, using both \xmm and \chandra data.
We build the selection function with respect to parameters like flux and $N_{\rm H}$, and derive the intrinsic $N_{\rm H}$ and $L_{\rm X}$ distributions for the S82X sample.
Finally, we determine the X-ray luminosity function (XLF) at different redshifts, in order to measure its evolution and inform population synthesis models in the high luminosity and high redshift regimes.

The paper is organized as follows. In \S \ref{sec:data} we introduce the S82X datasets used in this work. We present the X-ray spectral analysis procedures in \S \ref{sec:analysis} and results in \S \ref{sec:results}. In \S \ref{sec:evol} we discuss the evolution of the fraction of AGN that are obscured and derive the intrinsic distributions.
The total, obscured, and unobscured XLFs, and the black hole accretion density are also shown and discussed in this section. In \S \ref{sec:future_perspectives} we discuss the implications of this work. \S \ref{sec:summary} summarizes the work.
Throughout this paper, we assumed a $\Lambda$CDM cosmology with the fiducial parameters $H_0=70$ km s$^{-1}$ Mpc$^{-1}$, $\Omega_m=0.3$, and $\Omega_{\Lambda}=0.7$.
Errors are reported at the 90\% confidence level if not specified otherwise.

\section{Data} \label{sec:data}
\begin{figure}[!tp]
    \center 
    \includegraphics[scale=0.4]{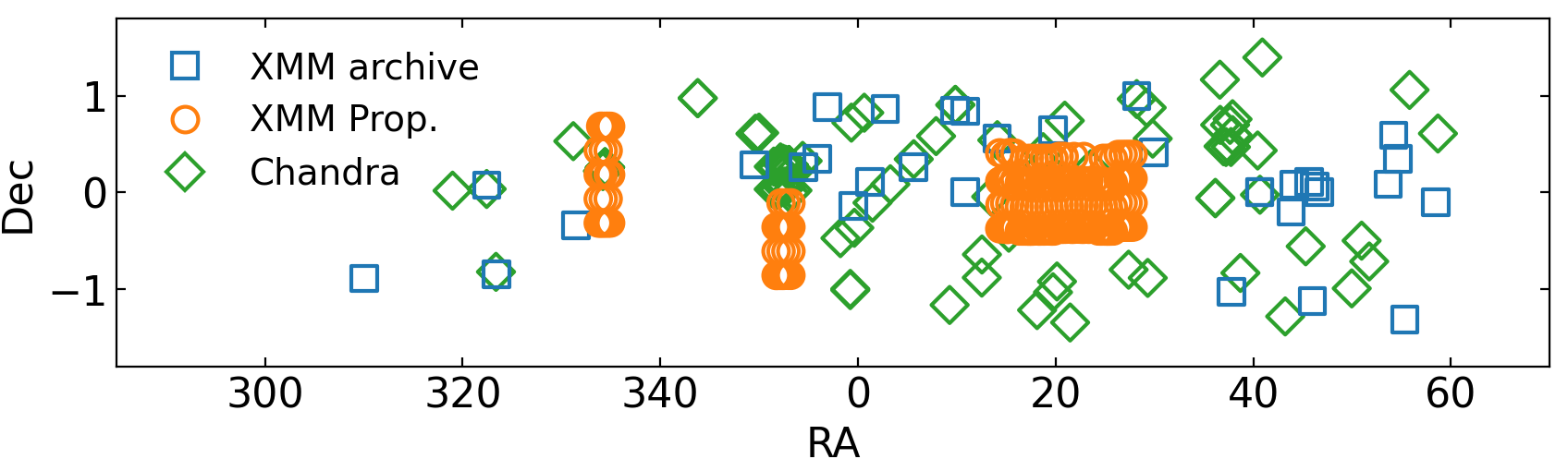}
    \caption{Stripe 82X coverage analyzed in this work: \textit{blue} squares and \textit{orange} circles represent \xmm archival and proprietary (AO10 and AO13) observations, respectively; while \textit{green} diamonds represent \chandra archival data.}
    \label{fig:s82_skycov}
\end{figure}

S82X comprises \xmm and \chandra observations from both proprietary (XMM AO10 and AO13, \citealt{lamassa13b,lamassa16}) and archival (\citealt{lamassa13a, lamassa16}) data for a total, non-overlapping area of $\sim$31.3 deg$^2$ (Fig. \ref{fig:s82_skycov}). The limiting fluxes are $8.7 \times 10^{-16}$, $4.7 \times 10^{-15}$, and $2.1 \times 10^{-15}$ erg s$^{-1}$ cm$^{-2}$ in the soft (0.5-2 keV), hard (2-10 keV), and full (0.5-10 keV) bands, respectively (\citealt{lamassa16}, hereafter \citetalias{lamassa16}). The broad band is 0.5–10 keV for \xmm observations and 0.5–7 keV for \chandra observations.
Table \ref{tab:obs_summary} summarizes the number of observations, covered areas, exposure times and number of sources. A detailed description of the \xmm campaign strategy, data reduction and analysis are given by \citet{lamassa13a,lamassa13b}, and \citetalias{lamassa16}.
The most recent version of the master catalog contains 6181 X-ray unique sources (\citealt{lamassa19_2}). Of these, 5150 ($\sim83\%$), 1520 ($\sim25\%$), and 5628 ($\sim91\%$) were detected in the soft, hard and full bands, respectively, with a detection significance of $>5\sigma$ for \xmm, and $>4.5\sigma$ for \chandra.
The total number of sources with redshift estimates is 5975 ($\sim97\%$), of which 3375 have spectroscopic redshifts (\citetalias{lamassa16}, \citealt{lamassa19_2}) and 5972 have photometric redshifts \citep{ananna17}. Multiwavelength counterpart matching was performed by running a maximum likelihood algorithm as discussed by \citetalias{lamassa16} and \citet{ananna17}.

\begin{table}[!t]
\small \center
\caption{Stripe 82X Observations}
\vspace{-0.05in}
\setlength\tabcolsep{4pt}
\begin{tabular}{ccccccc} 
\toprule
\textbf{Dataset\tablenotemark{a}}   & \textbf{Obs\tablenotemark{b}} & \textbf{Area\tablenotemark{c}} & \textbf{Exp\tablenotemark{d}}  &  \textbf{Srcs\tablenotemark{e}}  \\ 
- & - & (deg$^2$) & (ks) & - \\ \hline
Chandra archive      & 92        & 7.4         & 1842         & 969        \\
XMM archive          & 33        & 6.0         & 775          & 1599       \\
XMM AO10           & 2 (44)     & 4.6         & 240          & 751        \\
XMM AO13           & 7 (154)    & 15.6        & 980          & 2862       \\ \hline
Total              & 134 (323)  & 33.6 (31.3)  & 3837         & 6181       \\ \hline    
\end{tabular}
\begin{flushleft}
\tablenotetext{a}{Dataset analyzed.}
\tablenotetext{b}{Number of observations (pointings, for \xmm proprietary data).}
\tablenotetext{c}{Total (non-overlapping) area covered, in square degrees.}
\tablenotetext{d}{Total exposure time, in kiloseconds.}
\tablenotetext{e}{Number of sources: if a source is detected in multiple observations, the identification followed this priority (\citetalias{lamassa16}): XMM-AO13, XMM-AO10 + \xmm archive, \chandra archive. For example, if a source is detected both in \chandra and XMM-AO13, then it is identified as a XMM-AO13 source.}
\end{flushleft}
\label{tab:obs_summary} 
\end{table}

\section{X-ray analysis} \label{sec:analysis}
\subsection{Spectral extraction}\label{sec:analysis:spec}
The spectral extraction was performed with \xmm Standard Analysis System v1.3 (\textsc{SAS}\footnote{``Users Guide to the XMM-Newton Science Analysis System", Issue 16.0, 2021 (ESA: XMM-Newton SOC)}, \citealt{SAS}) and Chandra Interactive Analysis of Observations (\textsc{CIAO}, \citealt{ciao_paper}) v4.11, for \xmm and \chandra data, respectively. 
The overall procedure for the spectral extraction is the same for both telescopes.
For each source, we defined circular extraction regions by optimizing the signal-to-noise ratio (SNR), using a procedure similar to the \textsc{SAS} task \textsc{eregionalyse}\footnote{\url{https://xmm-tools.cosmos.esa.int/external/sas/current/doc/eregionanalyse.pdf}} (e.g., \citealt{ranalli13,peca21}).
A testing radius centered on the source centroid is varied until the maximum SNR is obtained. The SNR is computed as
\begin{equation}
    SNR = \frac{S}{\sqrt{S+BA}} ,
\end{equation}
where $S$ is the number of source counts from the varying source region, $B$ is the background counts from the corresponding background region, and $A$ is the ratio between the source and background areas. To avoid extremely small radii, we set a minimum radius of 16\arcsec\ or 0.5\arcsec, corresponding to the half-light widths at $E=1.5$ keV of the \xmm EPIC camera\footnote{\href{https://xmm-tools.cosmos.esa.int/external/xmm_user_support/documentation/uhb/}{XMM-Newton Users Handbook}} or the \chandra ACIS camera\footnote{\href{https://cxc.harvard.edu/proposer/POG/html/}{The Chandra Proposers’ Observatory Guide}}, respectively. There was no need to set a maximum radius since all the values were within or equal to the 90\% encircled energy radius for both telescopes.
The background regions were defined as annuli around the sources or---in source-crowded areas or in the case of sources near CCD gaps or edges---as circles close to the corresponding sources.
Background regions were defined to be 10 to 100 times larger than the corresponding source regions, and to avoid overlapping the source regions (considering the 90\% encircled energy at the source position).
Both the source and background spectra were binned to a minimum of 1 count bin$^{-1}$ with the \textsc{grppha} tool.

\subsection{Background handling}\label{sec:fittingprocedure}
The S82X observations are not deep enough (see Table \ref{tab:obs_summary}) to allow a good background sampling in local regions close to each source. Specifically, in our sample, there is not a minimum of one background count per spectral channel, which is required to apply a standard background subtraction technique (see Appendix B of the XSPEC manual\footnote{\href{https://heasarc.gsfc.nasa.gov/xanadu/xspec/manual/XSappendixStatistics.html}{XSPEC Manual}}). We instead decided to model the background and employ the c-stat statistics \citep{cash79} to deal with the low number of counts.
In order to extract a spectrum with enough statistics suitable to model the background, we proceeded as follows.
For \xmm, we selected one of the deepest observations in AO13 (0747440101), which is composed of 22 pointings with an average exposure time of $\sim$5 ks each. The detected sources in \citetalias{lamassa16} were removed from the observations considering circular regions with radii corresponding to the 90\% of the encircled energy (at 1.5 keV) at the source position. The background spectra were then extracted from each pointing in circular regions of radii 12$\arcmin$ centered at the aimpoint, and combined together with the SAS \textsc{epicspeccombine} tool. This procedure was repeated for the three PN, MOS1, and MOS2 cameras. 
For \chandra the procedure was similar. The selected deepest observations were 2252 ($\sim$71 ks) for ACIS-I, and 7241 ($\sim$49 ks) and 344 ($\sim$47 ks) for ACIS-S. We used a radius of 8$\arcmin$ for \chandra ACIS-I, while for \chandra ACIS-S we combined the spectra extracted from each of the six CCDs, with a radius of 4$\arcmin$. The stacking procedure was done with the CIAO \textsc{combine\_spectra} tool. 
The final background spectra have a number of counts between 50,000 and 150,000. They were modeled as described in Appendix \ref{app:bkg}, and the best-fit model was used to fit the local background for each source, leaving 
the background normalization free to account for possible local variations (e.g., \citealt{lanzuisi13}). 
The spectral fitting was performed using \textsc{XSPEC v12.10.1} \citep{xspec} through the PyXspec package\footnote{\href{https://heasarc.gsfc.nasa.gov/docs/xanadu/xspec/python/html/index.html}{PyXspec Documentation}}. Source and background spectra were fitted simultaneously.

\subsection{Spectral Models}
We fitted the spectra of each source using one or more of the models described below, based on their number of counts (see Section \ref{sec:analysis:thresholds}). 
Ordered from the simplest to the most complex:

\begin{figure*}[!tp]
    \center 
    \includegraphics[scale=0.48]{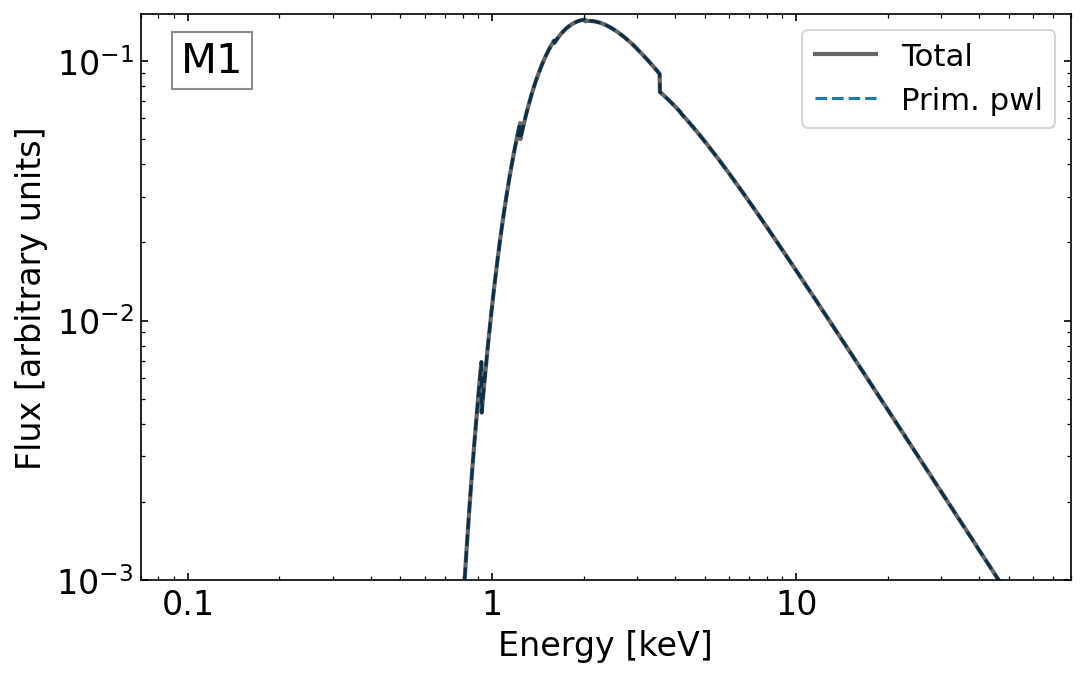}
    \includegraphics[scale=0.48]{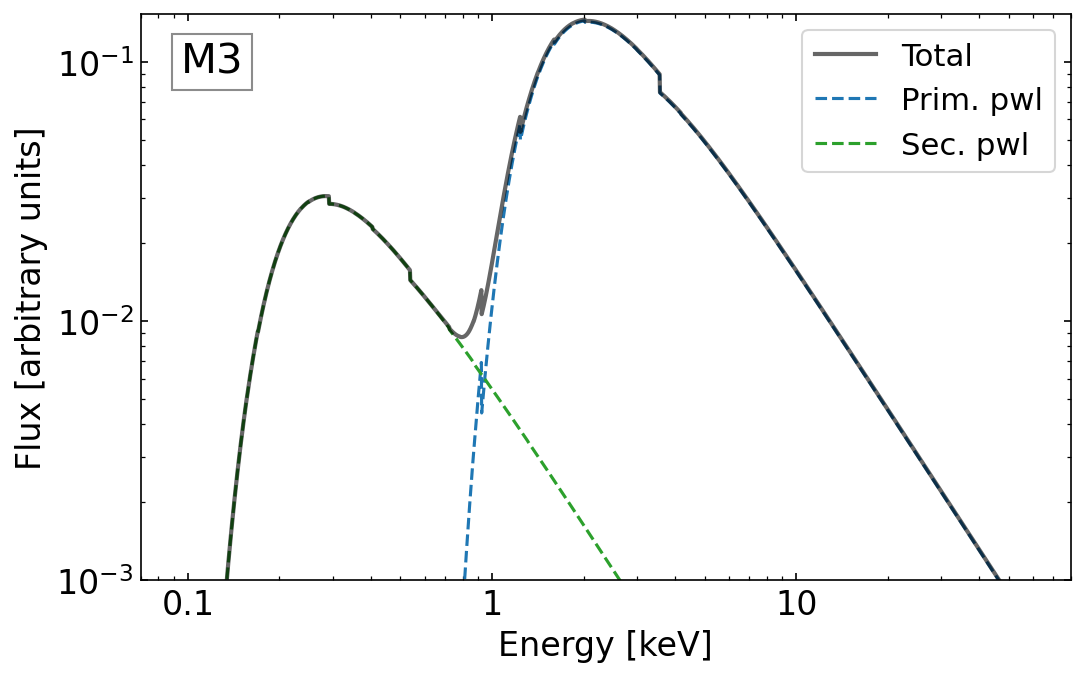}
    \includegraphics[scale=0.48]{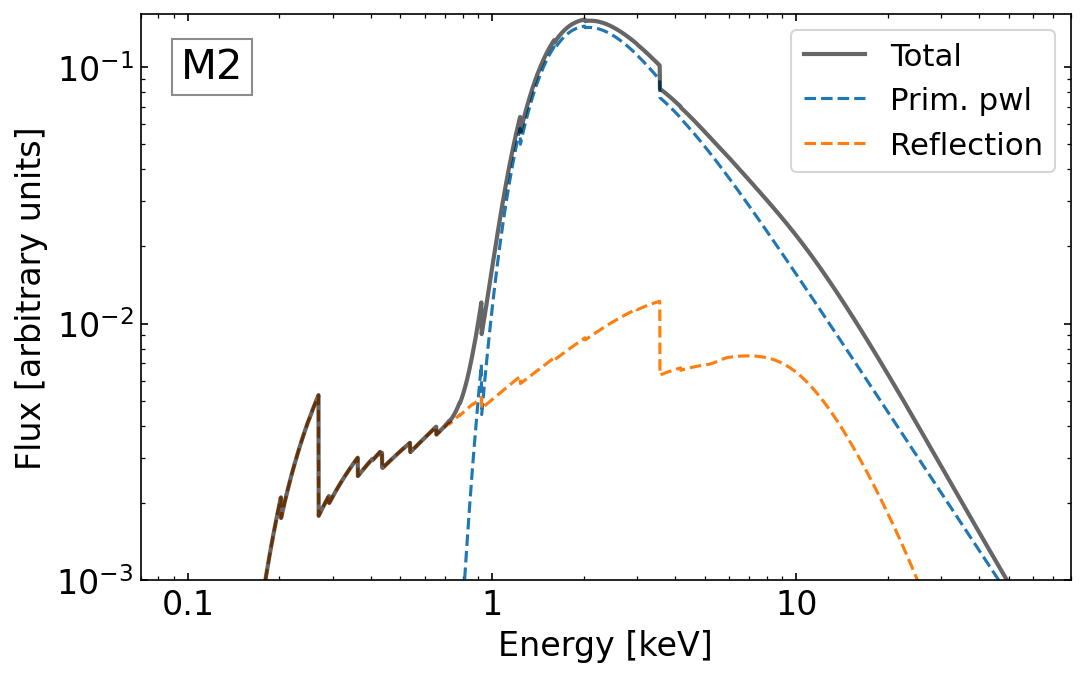}
    \includegraphics[scale=0.48]{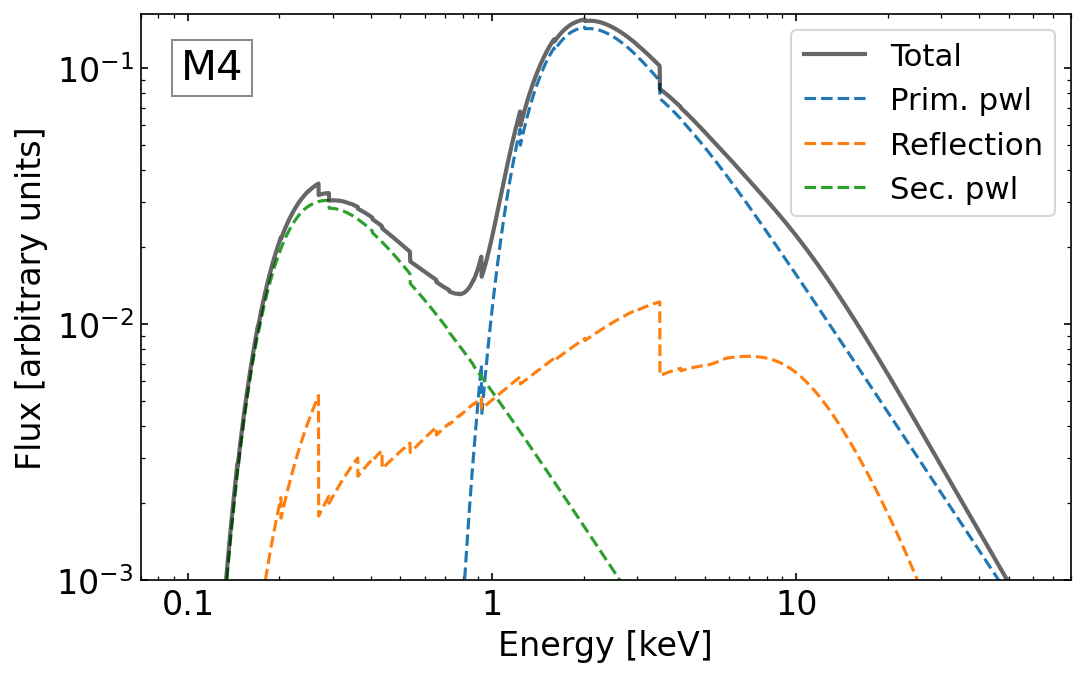}
    \caption{The four models used in the spectral analysis. M1: single absorbed power-law, M2: single absorbed power-law plus reflection, M3: double power-law model, and M4: double power-law model plus reflection. For this figure, the values $z=1$, $N_{\rm H}=10^{23}$ cm$^2$ and $\Gamma=1.9$ were assumed; Galactic absorption of 2$\times 10^{20}$ cm$^{-2}$ was applied; and the normalization of the secondary power-law was fixed at 2\% of the primary power-law. {\it Dashed lines} represent the main power-law (\textit{blue}), reflection (\textit{orange}), and secondary power-law (\textit{green}) component; while the \textit{black} solid lines show the total model. See text for model details.}
    \label{fig:models}
\end{figure*}

\begin{itemize}[leftmargin=*]
    \item M1: \textit{Single absorbed power-law} (XSPEC model \textsc{zphabs $\times$ cabs $\times$ zpowerlw}): 
    We modeled the effects of absorption on the primary X-ray emission (\textsc{zpowerlw}) by considering photoelectric absorption (\textsc{zphabs}) and Compton scattering (\textsc{cabs}). In particular, the latter is relevant for $\log N_{\rm H}/\mathrm{cm}^{-2}>22.5$ (e.g., \citealt{suchy12}), where part of the radiation is scattered out of the line of sight.
    The only parameter of the \textsc{cabs} component is the absorption $N_{\rm H}$, which was linked to the \textsc{zphabs} component.
    Even if its shape might not be the best description of the X-ray spectral shape (e.g., \citealt{mytorus}), the single power-law is a standard in the literature (e.g., \citealt{iwasawa12,marchesi16b}). Especially in the case of low photon statistics, where the use of many free parameters is not justified, the simple power-law is an effective model for comparing general physical properties such as $N_{\rm H}$ and $L_{\rm X}$ among AGN (e.g., \citealt{iwasawa20}).
    The free parameters for this model are then normalization, $N_{\rm H}$, and photon-index $\Gamma$.
    
    \item M2: \textit{Simple absorbed power-law plus reflection} (\textsc{zphabs $\times$ cabs $\times$ zpowerlw + pexrav}): The reflection component, produced by the reprocessing of the primary X-ray continuum by circumnuclear material, was introduced adding the \textsc{pexrav} model (\citealt{magdziarz95}) to M1. In \textsc{pexrav}, 
    both the photon index and the normalization were linked to the primary power-law. The reflection parameter $R$ was fixed to 1: since $R = \Omega/2\pi$, where $\Omega$ is the solid angle of the cold material visible from the hot corona, $R=1$ means the reflection is caused by an infinite slab illuminated by the isotropic corona emission. The other parameters were set to their default values. Due to the geometrical assumptions, \textsc{pexrav} can be considered a simplistic model (e.g., \citealt{yaqoob12, lamassa14, balokovic21}), but it is widely used 
    (e.g., \citealt{buchner14,ricci17}), especially in case of low photon statistics (e.g., \citealt{lanzuisi13}).
    The free parameters for this model are the same of M1.
    
    \item M3: \textit{Double absorbed power-law} (\textsc{zphabs $\times$ cabs $\times$ zpowerlw + const $\times$ zpowerlw}): A second unabsorbed power-law was added to M1, in order to cover a possible soft-excess 
    below $\sim$1-2 keV. It can be produced by different mechanisms, including scattered X-ray photons in circumnuclear material (e.g., \citealt{ueda07}), emission of thermal plasma due to star formation (e.g., \citealt{iwasawa11}), relativistic reflection (e.g., \citealt{vasudevan14}), Comptonization of optical/UV photons in plasma colder than the one in which X-ray photons originate (e.g, \citealt{boissay14}), or absorption by partially ionized material (e.g, \citealt{gierlinsky04}). 
    The photon index was linked to the primary power-law, 
    and the normalization was constrained to be $\leq$20\% of the primary component, as previously observed in X-ray surveys (e.g., \citealt{marchesi16b, ricci17}). The free parameters for this model are then the two power-law normalizations, $N_{\rm H}$, and photon-index $\Gamma$.
    
    \item M4: \textit{Double absorbed power-law plus reflection} (\textsc{zphabs $\times$ cabs $\times$ zpowerlw + const $\times$ zpowerlw + pexrav}): The reflection component was added to M3, with its parameters tuned as described for M2. The free parameters for this model are the same of M3.
\end{itemize}

Moreover, for each model:
\begin{itemize}[leftmargin=*]
    \item[-] The photon index was both left free to vary and fixed to a typical observed $\Gamma=1.8$ (e.g., \citealt{nandra94,piconcelli05}).
    
    \item[-] A calibration constant (\textsc{const}) was added before each model to consider possible normalization issues between \xmm and \chandra. This term was left free to vary when \xmm and \chandra spectra were fitted simultaneously, otherwise it was set to 1.
    
    \item[-] Fixed Galactic absorption (\textsc{phabs}) appropriate to the source position (\citealt{kalberla05}) was included.
    
    \item[-] The redshift parameter was frozen to the source redshift. We used spectroscopic redshift ($z_{spec}$) when available, and reliable photometric redshifts ($z_{phot}$) otherwise. We defined reliable $z_{phot}$ those redshifts with PDZ$\geq70\%$, where PDZ is defined as the probability that the true redshift is within $\pm0.1(1+z_{phot})$ of $z_{phot}$ (see \citet{ananna17} for details).
\end{itemize}


One may think that such simple models do not adequately reproduce the complex shape of the AGN spectrum. However, more sophisticated and physically motivated models, such as MYTORUS (\citealt{mytorus}) or \texttt{borus02} (\citealt{balokovic18}), require many more photons to be used effectively (e.g., \citealt{lamassa14, marchesi19, lamassa19}). Our sample has a median value of 38.5 net counts. With such low count statistics, parameters of sophisticated models need to be frozen to standard values to avoid possible degeneracies. As a consequence, their shape becomes similar to our models and produces consistent results (e.g., \citealt{brightman14}). It is worth mentioning that this is true for low obscuration levels up to $\log N_{\rm H}/\mathrm {cm}^{-2}\sim 24$ (e.g., \citealt{marchesi20}), which is what we found in our sample as shown in Section \ref{sec:results}.
Moreover, overly complex spectral shapes may lead to worse spectral fits: fluctuations in the spectrum, which are very common for low counts sources, may be misinterpreted by XSPEC as spectral features (\citealt{peca21}) causing inaccurate results. For these reasons, it makes more sense to use simple models in the case of low count statistics, when deriving main physical properties like $N_{\rm H}$ and $L_{\rm X}$. For completeness, we show a comparison between our results and the \texttt{borus02} model in Section \ref{sec:borus_comparison}.

\subsubsection{Counts threshold evaluation}\label{sec:analysis:thresholds}
The average photon statistics in the S82X requires an investigation of the minimum number of counts needed to obtain a reliable fit. We evaluated the fit performance with XSPEC spectral simulations.
For each model, we simulated spectra with typical AGN parameters: $\log N_{\rm H}/\rm{cm}^{-2}$ from 20 to 24.5 in logarithmic bins of 0.5, redshift from 0 to 5 in unitary bins, a fixed $\Gamma=1.8$, and a variable normalization in order to obtain a range of net counts between 0 and 100. 
For each parameter combination, 1000 spectra were simulated. We fitted the simulated spectra applying the same procedure described in \ref{sec:fittingprocedure}. 
To check the fit reliability as function of the simulated parameters (counts, $z$, $N_{\rm H}$), we computed the match percentage (\citealt{peca21}):
\begin{equation}
    \mathrm{MP
    } ({\mathrm cts}, z, N_{\rm H}, \Gamma) = \frac{N(X \pm \Delta X)}{N_{\rm sim}}(X) .
\end{equation}
This is the fraction of simulated spectra for which the fitted values of $X$ ($N_{\rm H}$ or $\Gamma$) are consistent with the simulated values within a given tolerance $\Delta X$ (90\% uncertainty) for specific values of the number of counts, redshift, $N_{\rm H}$, and spectral index. We chose a threshold of MP\%$\geq 50\%$. As shown by \cite{peca21}, this threshold is a fair compromise between having a large enough sample and spectral fit accuracy.
Since it is common to find $N_{\rm H}$ upper limits in X-ray surveys (e.g., \citealt{marchesi16b}), we considered them as good values, while we rejected $\Gamma$ solutions that were not constrained between boundaries.
Because $N_{\rm H}$ values are not known a priori as they are a free variable in the fit, the results were averaged over the $N_{\rm H}$ simulated values. The procedure was then repeated considering $\Gamma$ values in the range 1.6-2.1 (\citealt{nandra94,piconcelli05}) with a step of 0.1, but no significant changes in the match percentage were found.
In the simulations we used the response matrices from a subset of sources observed at different epochs (XMM-AO13, XMM-AO10, and \xmm and \chandra archives) to consider possible differences due to the effective area degradation during the years. We found a negligible effect on match percentage for bright sources and an overall scatter within $\sim$5\% for sources with $\lesssim$50 counts in both the telescopes. We averaged the match percentage obtained for each set of responses for \xmm and \chandra separately.

\begin{figure*}[!tp]
    \center 
    \includegraphics[scale=0.55]{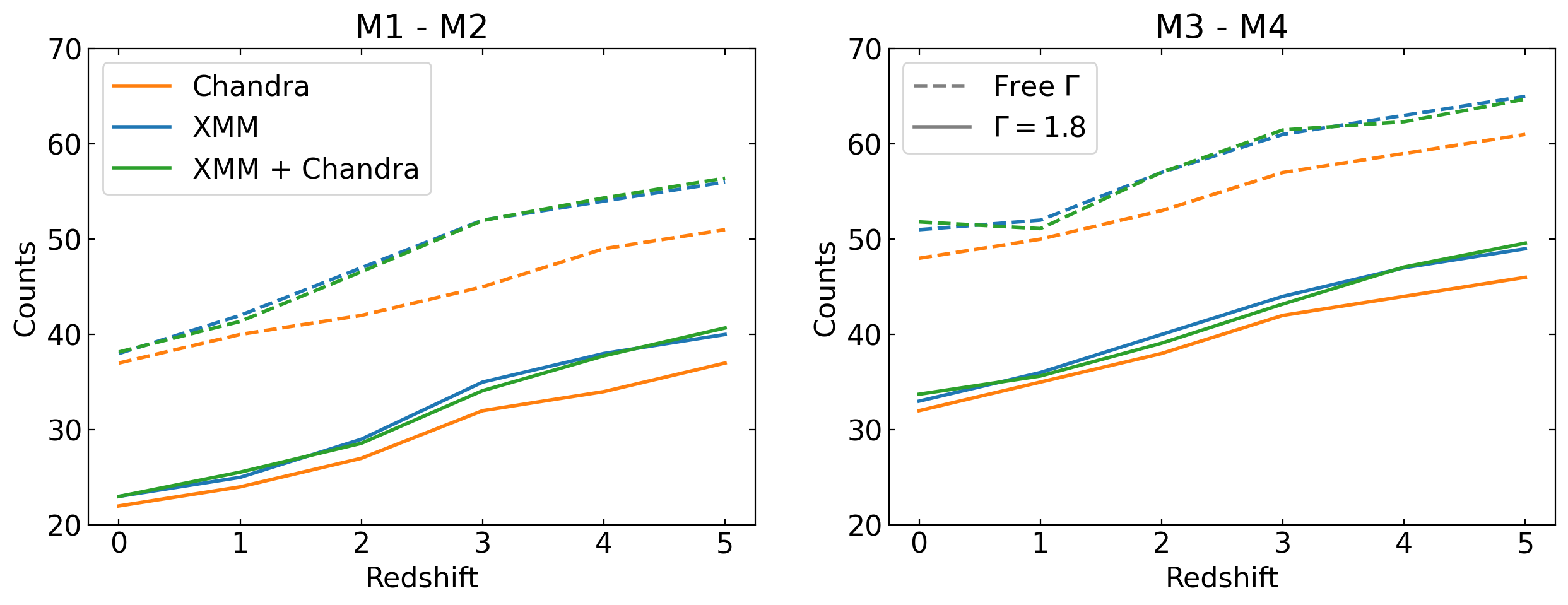}
    \caption{Counts thresholds determined for simple power-law models without and with reflection (M1-M2, respectively; \textit{left panel}) and double power-law models without and with reflection (M3-M4, respectively; \textit{right panel}). \textit{Blue} curves represent values obtained for \xmm data, \textit{orange} curves are those obtained for \chandra data, and \textit{green} curves represent \xmm and \chandra together. Solid curves show models where $\Gamma$ was frozen, while dashed curves indicate where $\Gamma$ was free to vary. In the \textit{left} (\textit{right}) panel, the curves are valid for both M1 and M2 (M3 and M4) because, even if a reflection component is added, all new parameters are either frozen to standard values or linked to other components.}
    \label{fig:cts_sim}
\end{figure*}

In Figure \ref{fig:cts_sim} we show the results of these simulations for models M1 to M4. As the model complexity increases (and with it the number of parameters), the minimum number of counts required for a reliable fit increases as well. For example, when fitting model M1 with $\Gamma=1.8$ (solid line, left panel), it is possible to get reliable results for sources down to $\sim$20 net counts; while using the same model with a free $\Gamma$ (dashed line, left panel), roughly 35-40 counts are required.
The same trends were obtained for model M3 (right panel), but with an increased minimum number of counts due to the more complex spectral shape.
For models M2 and M4, we used the same curves obtained for M1 and M3, respectively. In M2 and M4 we introduced a reflection component without new free parameters and, not surprisingly, the results are very similar to those of M1 and M3.
Other than the differences in the model complexity, there is also a trend with redshift. This is because as the redshift increases, the AGN spectral shape is shifted to lower energies until part of it lies outside the fitting range, resulting in a lower match percentage and, therefore, in a higher number of minimum counts. This effect is particularly important for obscured sources, whose spectral shape is absorbed in the soft band (e.g., \citealt{akylas06}).

When multi-instrument spectra were available for the same source, we fitted them simultaneously. We were then left with several fitting scenarios: from a single spectrum fitted alone to various \xmm and \chandra spectra fitted together. We opted to perform spectral simulations for the cases that happened most frequently in our sample, namely: i) three \xmm spectra (PN, MOS1, and MOS2), ii) three \xmm spectra plus a single \chandra spectrum, and iii) a single \chandra spectrum only.
As shown in Figure \ref{fig:cts_sim}, the curves found for cases i) and ii) are consistent with each other. For case iii), instead, we found a slight systematic improvement (up to $\sim$7 counts) compared to the other cases. This can be explained because of the different backgrounds in \chandra and \xmm. Having a lower background level, \chandra spectra alone need a slightly lower number of counts to get a reliable fit.
We also tested the above cases using both \chandra ACIS-I and ACIS-S spectra, without finding a significant change in the results.
Based on the computed curves (Figure \ref{fig:cts_sim}), we set our count thresholds as a function of $z$, number of counts, and model.

\subsection{Model comparison}\label{sec:analysis:modelcomparison}
When more than one model is used to fit a source, a statistical comparison among those models is required to select the best result.
Two tests often applied together are the Bayesian Information Criterion (BIC, \citealt{schwarz1978}) and the Akaike Information Criterion (AIC, \citealt{akaike1974}). Both are valid for nested and non-nested models. In general, due to a different penalization term (see e.g., \citealt{kass95}), BIC penalizes more complex models, i.e., with a larger number of parameters, compared to AIC (e.g., \citealt{feigelson12,chakraborty21}). We therefore decided to apply the AIC test to not penalize the double power-law models, which are more complex but more realistic and physically motivated than the single power-laws (e.g., \citealt{lanzuisi15, ricci17}). 

AIC is defined as
\begin{equation}\label{eq:aic}
    AIC = -2 \ln(L) + 2k ,
\end{equation}
where $L$ is the maximum likelihood from the fit and $k$ is the number of free parameters. In XSPEC, $L$ can be retrieved from the fit results, as $cstat=-2\times ln(L)$.
When the sample size is small, however, the AIC test (as expressed in Eq. \ref{eq:aic}) may over-fit, selecting models with too many free parameters (e.g., \citealt{liddle+07}). This issue can be solved by introducing a correction term in Eq. \ref{eq:aic} (e.g., \citealt{CAVANAUGH1997}):
\begin{equation}
    AIC_c = AIC + \frac{2k(k+1)}{N-k-1} ,
\end{equation}
where $N$ is the number of data points.
The AIC criterion states that when there is a set of competing models, the one with the lowest AIC value is preferred.
However, it is not the absolute AIC value that is important, but the AIC differences between different models (\citealt{burnham02}).
In our case, when a source has enough counts to be fitted with a set of models, the AIC differences are computed over the models in the set:
\begin{equation}
    \Delta AIC_{c,i} = AIC_{c,i} - AIC_{c,min} ,
\end{equation}
where $AIC_{c,min}$ is the minimum value in the set. 
If more than one model falls in the range $0\leq \Delta AIC_{c,i} < 2$, then the evidence that these models are competitive for being the best model is considered ``substantial" (e.g., \citealt{burnham02}). When this case is verified, we take the more complex model as the best model since it has more diagnostic power on the physical nature of the AGN.  

For completeness, it is worth mentioning that the BIC criterion ($BIC=-2 \ln(L) + k \ln(N)$) was also tested. As expected, for the majority (91\%) of the sources we obtained the same results of the AIC criterion. For the remaining 9\%, the BIC prefers simpler models, but never with a ``strong" (i.e., $\Delta BIC>6$, \citealt{kass95}) evidence against the more complex models chosen by the AIC. For these cases, the competitive models are then roughly equally consistent with the data, and there is not statistical evidence to strongly favor one over the other, further justifying our choice to pick the more physical motivated one.
In this regard, it should be noted that tests such as AIC and BIC compare models from a purely statistical perspective, and thus a physical interpretation in the choice of the best model should also be taken into account.

\section{Results} \label{sec:results}
In this section we report the main results derived from the spectral fitting.
We narrowed the sample to include only sources with $z_{spec}$, reliable $z_{phot}$ (PDZ$\geq70\%$), and with enough counts. Spectroscopic classified stars were also removed. The final sample contains 2937 candidate AGN ($\sim 48\%$ of the total sample). 
Since X-ray sources with $\log L_{\rm X,2\text{-}10keV}/\mathrm{erg\,\, s}^{-1}<42$ may be either low luminosity AGN or non-AGN (e.g., \citetalias{buchner15}, \citealt{padovani17}), we performed the spectral analysis for all the 2937 sources and discussed our results with and without the 185 ($\sim6\%$) sources with $\log L_{\rm X}/\mathrm{erg\,\, s}^{-1}<42$.
The results are released within the catalog described in Appendix \ref{app:master_cat}.

\subsection{Fit results}
\begin{figure}[!tp]
    \center 
    \includegraphics[scale=0.5]{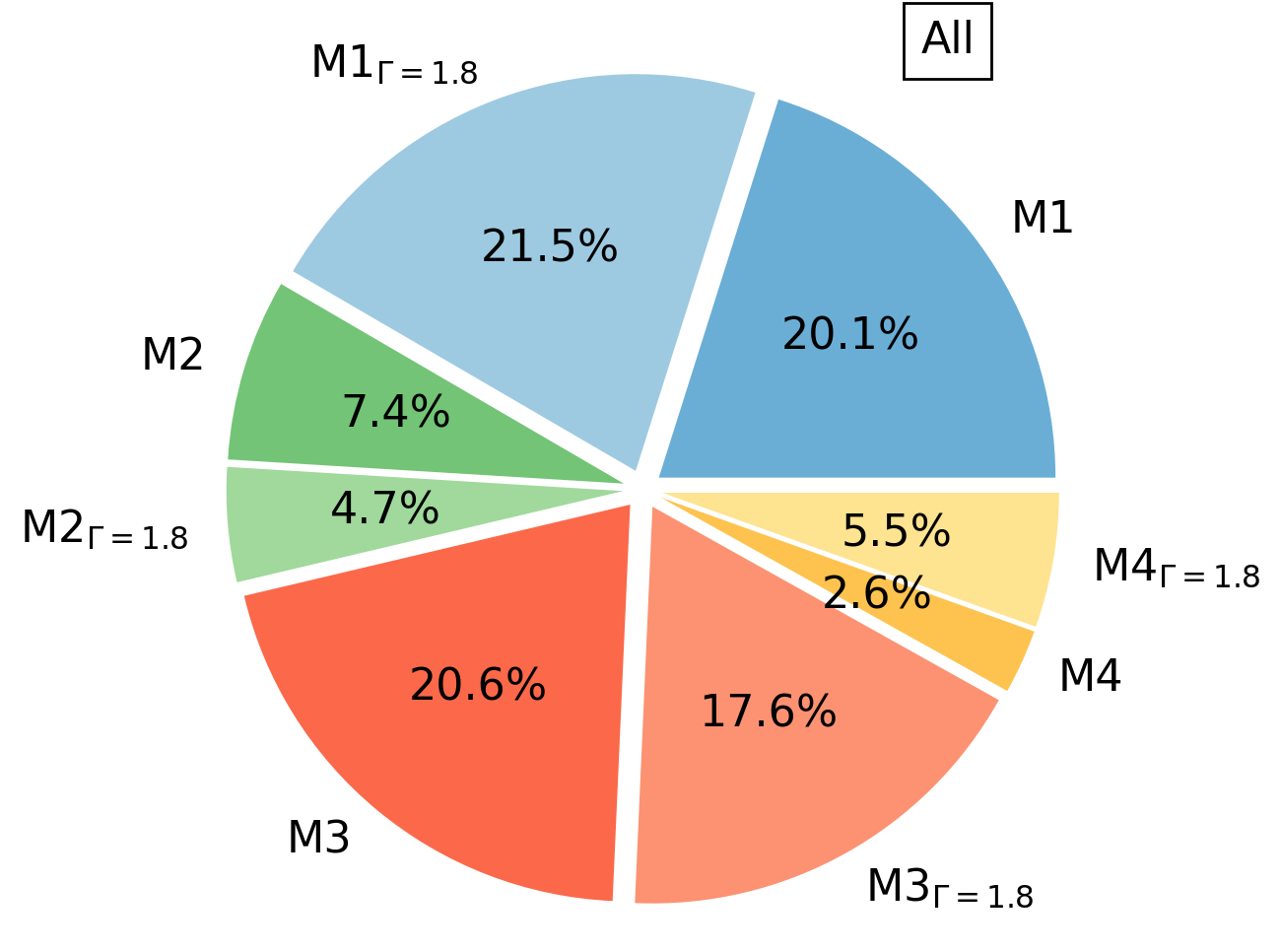}\\
    \vspace{1cm}
    \includegraphics[scale=0.5]{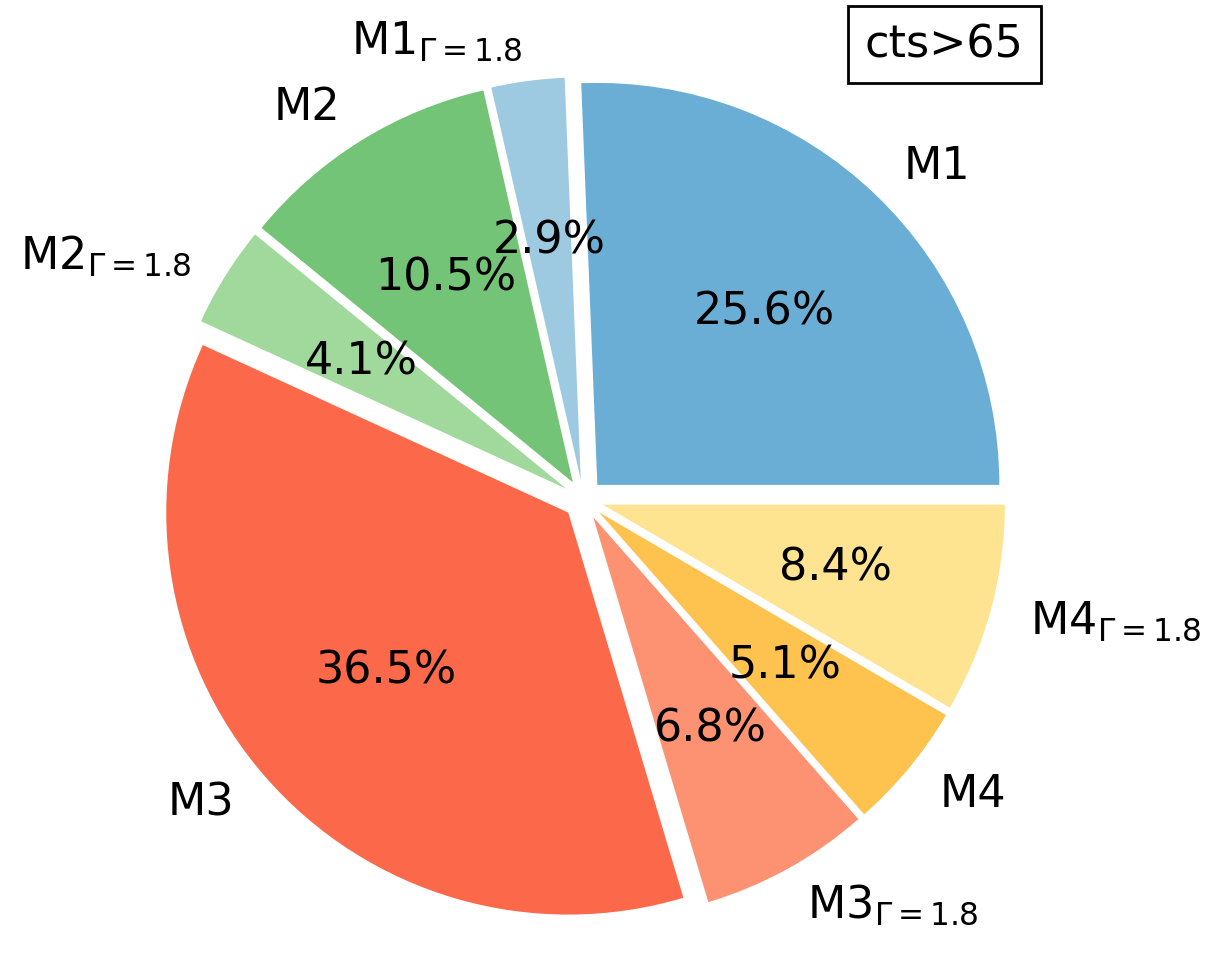}
    \caption{Pie charts of the selected best-fit models for all the sources (\textit{top panel}) and for sources with net counts $>65$ (\textit{bottom panel}). 
    For each model we show the results with a free $\Gamma$ (models M1 to M4) and frozen $\Gamma$ (models M1 to M4 with $\Gamma=1.8$ as subscript). Percentages are shown in each pie slice.}
    \label{fig:best_models_pie}
\end{figure}

In Figure \ref{fig:best_models_pie} we show the results of model selection described in Section \ref{sec:analysis:modelcomparison}.
Considering the full sample (top panel), around 80\% of the sources preferred models without the reflection component (M1 and M3): even if there is the same number of free parameters between models with and without reflection, the $\Delta AIC$ criterion prefers simpler spectral shapes.
There is not a significant difference between models with free and frozen $\Gamma$, and single power-law models (M1 and M2) are preferred ($\sim54\%$) over double power-law models (M3 and M4). However, there is a bias towards simpler models in this pie chart, since spectral models were fitted as a function of the number of counts. 
Considering, instead, only sources with $>65$ counts (bottom panel), i.e. where all the models were tested according to our count thresholds, we can see that while models without reflection (M1 and M3) are still preferred ($\sim72\%$), now models with a fixed $\Gamma$ are disfavoured\footnote{M4 is an exception, but the percentages are similar.} and double power-law models (M3 and M4) are preferred ($\sim57\%$) over the single power-law models (M1 and M2). 
In particular, the last trend is confirmed for sources with $>250$ counts, where double power-law models are preferred over single power-law models in  $\sim64\%$ of the cases. 
We can conclude that, in general, the secondary power-law component becomes statistically more evident as the number of counts increases. In other words, since more complex spectral models have more parameters than simple ones, more counts are needed for them to be statistically chosen as the best representation of observed spectra.

\begin{figure*}[!tp]
    \center 
    \includegraphics[scale=0.46]{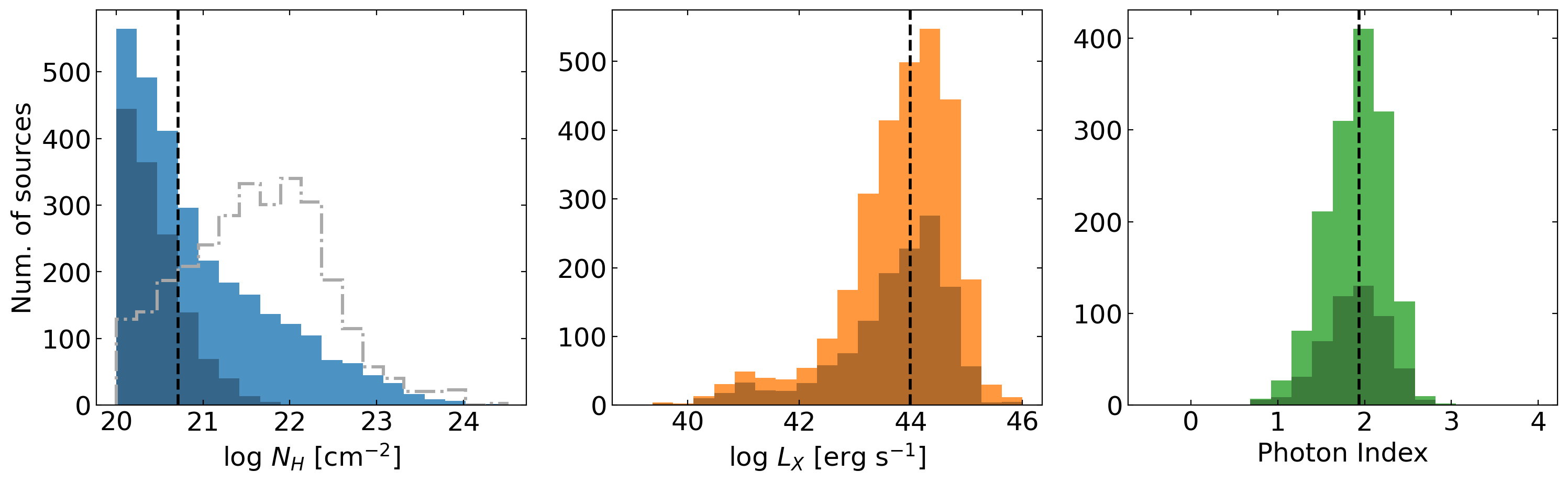}
    \caption{ Distributions of key parameters from spectral analysis. {\it Left to right}: $N_{\rm H}$ ({\it blue}), intrinsic 2-10 keV luminosity ({\it orange}), and photon index ({\it green}). The \textit{grey dash-dotted line} in the left panel represents the $N_{\rm H}$ distribution considering upper limits. For each distribution, the median values (see text) are shown with a dashed vertical line. Darker colors indicate sources where the derived $N_{\rm H}$ has only upper limits. In the photon index distribution, all the sources fitted with a fixed $\Gamma=1.8$ were removed.}
    \label{fig:results1}
\end{figure*}

\begin{figure}[!tp]
    \center 
    \includegraphics[scale=0.385]{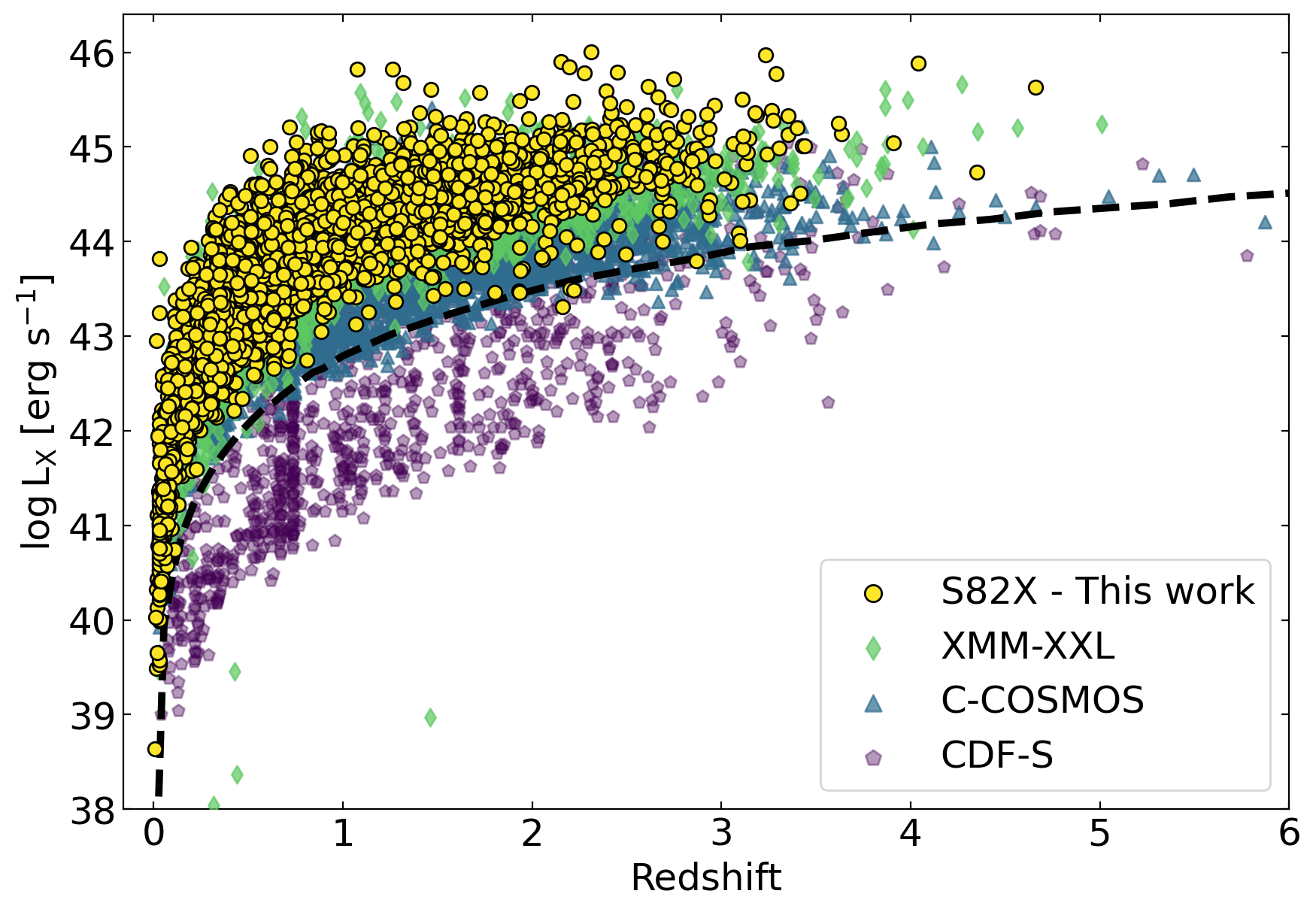}
    \caption{Intrinsic, de-absorbed 2-10 keV luminosity as a function of redshift derived in this work (\textit{yellow circles}) compared to XMM-XXL (\textit{green diamonds}; \citealt{pierre16}, \citealt{liu16}), C-COSMOS (\textit{blue triangles}; \citealt{civano16}, \citealt{marchesi16b}), and CDF-S (\textit{purple pentagons}; \citealt{luo17}, \citealt{liu17}) surveys. The dotted black line represents the S82X survey sensitivity limit from \citetalias{lamassa16}.}
    \label{fig:lx_z_nh}
\end{figure}

Figure \ref{fig:results1} shows the observed distributions of column density (left panel), X-ray luminosity (absorption-corrected, 2-10 keV, rest-frame; middle panel), and photon index (right panel); whose median values are
$\log N_{\rm H}/\mathrm {cm}^{-2}=20.7_{-0.5}^{+1.2}$ ($21.6_{-0.9}^{+0.7}$ considering upper limits), $\log L_{\rm X}/\mathrm{erg\,\, s}^{-1}=44.0_{-1.0}^{+0.7}$, and $\Gamma=1.94_{-0.39}^{+0.31}$, respectively. The uncertainties correspond to the 16$^{th}$ and 84$^{th}$ percentiles.
These results are consistent with the spectral analysis of the XMM-XXL North field ($\Gamma=1.98_{-0.05}^{+0.09}$, $\log N_{\rm H}/\mathrm {cm}^{-2}=20.9_{-0.3}^{+0.8}$, and $\log L_{\rm X}/\mathrm{erg\,\, s}^{-1}=44.0_{-1.0}^{+0.6}$, by \citet{liu16}), which has similar depth and area as S82X.
Moreover, we find that 420 ($\sim14\%$) of the detected S82X AGN are obscured ($\log N_{\rm H}/\mathrm {cm}^{-2} > 22$), similarly to the $\sim12\%$ of XMM-XXL.
It is worth noticing that these numbers also agree with those obtained by \citet{mountrichas20}, which performed a X-ray analysis on infrared selected AGN in S82X.
Obscuration and X-ray luminosity distributions vary as a function of depth and covered area (e.g., \citetalias{aird15,ananna19}). Indeed, analyzing the smaller, deeper C-COSMOS field (2.2 deg$^2$, 4.6 Ms), \citet{marchesi16b} found mean $\log L_{\rm X}/\mathrm{erg\,\, s}^{-1}=43.7_{-0.7}^{+0.5}$ and $\log N_{\rm H}/\mathrm {cm}^{-2}=22.7_{-0.5}^{+0.4}$, while in the still smaller and deeper CDF-South field (CDF-S, $\sim0.11$ deg$^2$, 7 Ms), \citet{liu17} found median $\log L_{\rm X}/\mathrm{erg\,\, s}^{-1}=43.5_{-0.6}^{+0.7}$ and $\log N_{\rm H}/\mathrm {cm}^{-2}=23.0_{-1.0}^{+0.9}$. 
As expected, large-volume surveys like S82X or XMM-XXL are more sensitive to luminous, unobscured AGN, while deep pencil-beam surveys detect more obscured and faint objects.
Figure \ref{fig:lx_z_nh} shows this concept in the luminosity-redshift plane: while S82X shares the high luminosity region with XMM-XXL, smaller area surveys cover lower luminosity ranges.
In particular, CDF-S and C-COSMOS fields combined have only three sources at $\log L_{\rm X}/\mathrm{erg\,\, s}^{-1}>45$ and $z>3$, while S82X and XMM-XXL have 21 and 13 sources, respectively. This strengthens the statement that large area surveys are needed to detect and study high redshift and high luminosity AGN.

\begin{figure}[!tp]
    \center \vspace{1mm}
    \includegraphics[scale=0.43]{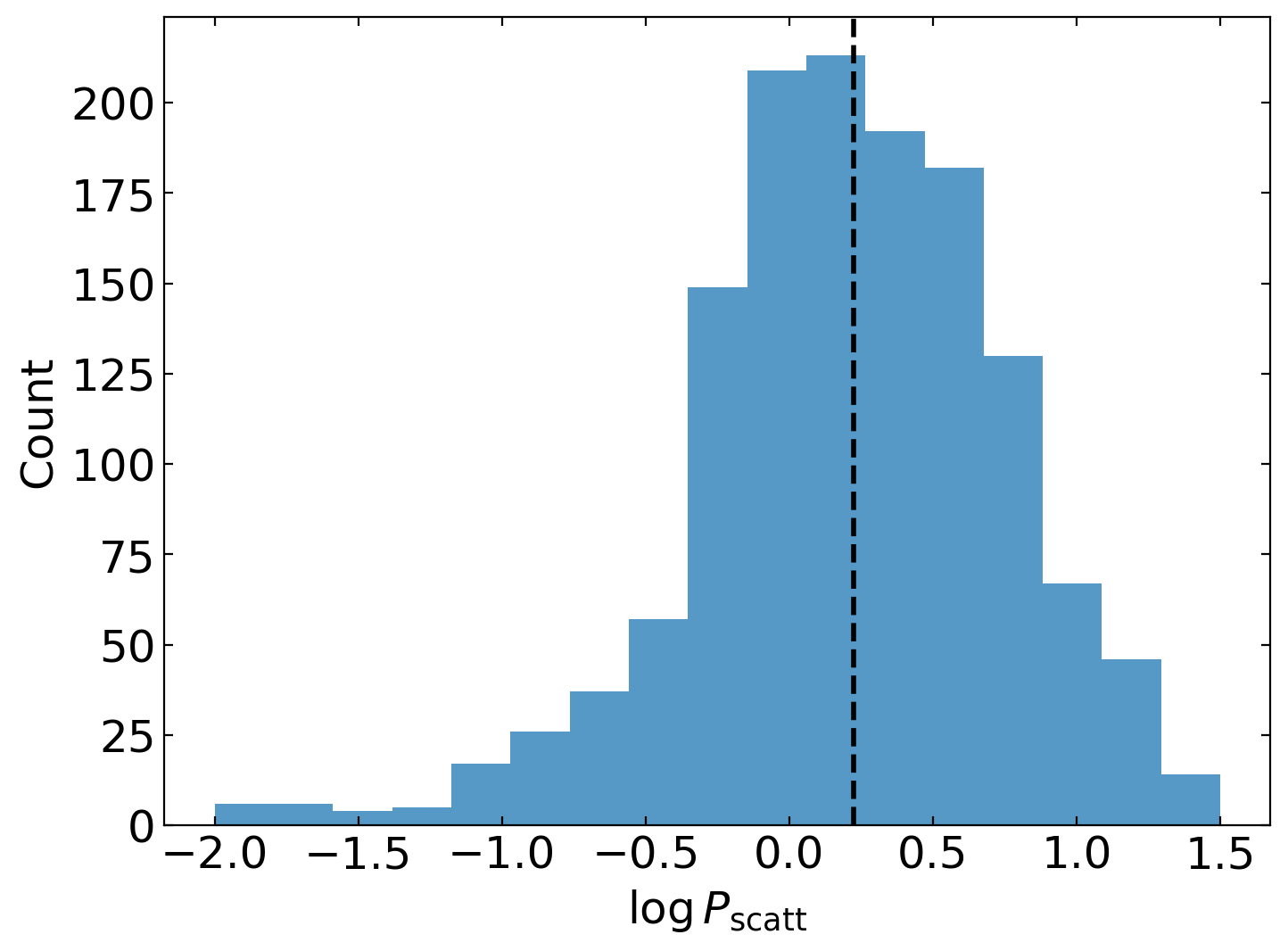}
    \caption{Distribution of the $P_{\rm scatt}$ parameter, which represents the percentage of emission from the secondary component relative to the primary power-law. The \textit{dashed} line indicates the median value, $\log P_{\rm scatt} = 0.2\pm 0.5$, which indicates that most sources show a significant soft-excess and should be fitted with models that account for it.
    }
    \label{fig:fscatt}
\end{figure}

Our photon index values are in agreement with many other surveys (e.g., CDF-S, \citealt{liu17}; C-COSMOS, \citealt{marchesi16b}; Swift/BAT, \citealt{ricci17}; J1030, Signorini et al., in prep), suggesting that, on average, the primary power-law slope is not correlated with luminosity or absorption.
Figure \ref{fig:fscatt} shows the distribution of the $P_{\rm scatt}$ parameter, which represents the percentage of emission in the secondary power-law (model M3 and M4) compared to the primary component. As discussed before, we observed a significant soft-excess component in $\sim57\%$ of the sources, in agreement with results from \citet{ricci17} in the local Universe. The median value
is $\log P_{\rm scatt} = 0.2\pm 0.5$, also consistent with 
Ricci et al. ($0.1_{-0.7}^{+0.6}$). 
This consistency suggests that AGN should routinely be fitted with models that include a secondary power-law or other component that account for the soft-excess. 

By excluding possible non-AGN sources with $\log L_{\rm X}/\mathrm{erg\,\, s}^{-1}<42$, we found no significant impact on the model selection analysis and consistent median values on the derived physical parameters: $\log N_{\rm H}/\mathrm {cm}^{-2}=20.8_{-0.5}^{+1.2}$ ($21.7_{-0.9}^{+0.7}$ considering upper limits), $\log L_{\rm X}/\mathrm{erg\,\, s}^{-1}=44.1_{-0.8}^{+0.6}$, with no changes for $\Gamma$ and $\log P_{\rm scatt}$.


\subsection{Comparison to a more sophisticated torus model}
\label{sec:borus_comparison}
In this analysis, we used simple models due to the limits set by low photon statistics. While these models allow a reliable parameter estimation, more complex and physically motivated shapes might better describe the actual AGN spectrum and its features.
The spectrum of an unobscured AGN is relatively easy to model, since the main power-law component dominates the overall spectral shape. Some features, such as emission lines and/or absorbers of various nature (e.g., \citealt{blustin05}) may be present, but require high photon statistics to be detectable. On the other hand, the spectral shape of obscured AGN is driven by obscuration, and it may be modeled even with a low number of counts (e.g., \citealt{iwasawa12,iwasawa20,peca21}). For these reasons, we tested our results for obscured sources with a more detailed model.

\begin{figure}[!tp]
    \center 
    \includegraphics[scale=0.45]{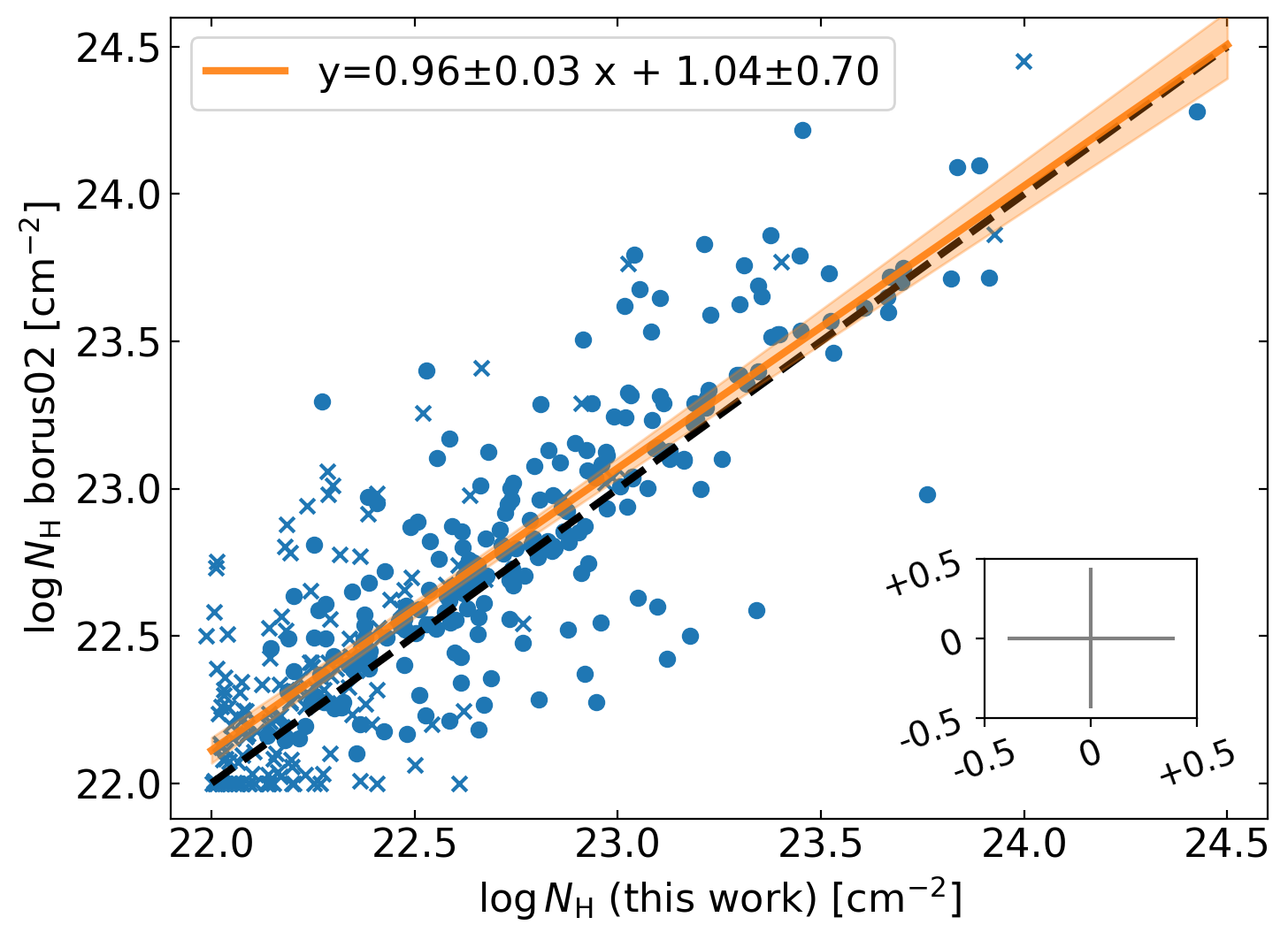}
    \caption{Comparison between log$N_{\rm H}$ values obtained with the \texttt{borus02} model and models M1 to M4. Dots represent sources where both upper and lower limits were constrained, while crosses indicate sources where only upper and/or lower limits were found. The one-to-one relation is shown with a \textit{black dashed line}. The best-fit line (\textit{orange}) and its 2$\sigma$ uncertainty (\textit{shaded orange}) are also shown. The insert box shows the average error bars at the 90\% confidence level.}
    \label{fig:borus}
\end{figure}

\texttt{borus02} (\citealt{balokovic18}) is a self-consistent model that considers absorption, scattering, and emission lines. Assuming a toroidal geometry and a varying covering factor, it allows to separate the average torus column density from the line-of-sight column density, which are different when the torus is not homogeneous.
However, as discussed before, for low photon statistics spectra the reliability of the results strongly depends on the number of free parameters. We therefore assumed a fixed covering factor $f_c=0.5$ and viewing angle $cos\theta_{obs}=0.5$, and we linked the different $N_{\rm H}$ values together. Iron abundance was set to solar.
Since this model allows $\log N_{\rm H}$ values in the range 22-25.5, we tested it only on sources with log$N_{\rm H}>22$ cm$^{-2}$.
To make a fair comparison, we applied the same constraints used in models M1 to M4. If a source is best fitted with M1 or M2, we then turned off the secondary power-law in \texttt{borus02}, while we left its normalization to be $\leq$ 20\% of the primary power-law for sources best-fitted with M3 or M4. The same logic was applied for free or frozen $\Gamma=1.8$. With these restrictions, our \texttt{borus02} model does not add any new free parameter relative to our previous models. The comparison of derived $N_{\rm H}$ values is shown in Figure \ref{fig:borus}, along with the best-fit relation derived with linear regression modeling.
Overall, we found a good agreement between the two approaches, even if the $N_{\rm H}$ values derived with \texttt{borus02} for $\log N_{\rm H}/\mathrm {cm}^{-2}<23.5$ are slightly higher, but still consistent within the uncertainties (see insert box).
At the 90\% confidence level for individual spectra, only $\sim6\%$ (26/420) of AGN have $N_{\rm H}$ values inconsistent with the one-to-one relation. 
This degree of agreement indicates that use of complex and more realistic models, at least in the present case of low count statistics, is not necessary.

\section{AGN cosmic evolution} \label{sec:evol}
The evolution of AGN, and of the fraction that are obscured, has been studied by many works (e.g., \citealt{ueda03}, \citealt{treister06}, \citealt{aird2010}, \citetalias{ueda14}, \citetalias{buchner15, aird15}, \citealt{vito18}, \citetalias{ananna19}); however, it remains an open question in observational astrophysics.
Here we analyze the S82X sample, which includes more AGN at high luminosities and redshifts than do smaller, deeper surveys, to derive the XLF and the evolution of the intrinsic obscured fraction.
Below we describe how we built intrinsic distributions by correcting for observational biases (Section \ref{sec:obs_evol:corr}), and how we derived the obscured fraction (Section \ref{sec:osc_frac_section}) and the XLFs for both obscured and unobscured AGN (Section \ref{sec:xlf}).
The results derived in this section were obtained for the 2752 sources ($\sim45\%$) with $\log L_{\rm X}/\mathrm{erg\,\, s}^{-1}>42$, to avoid possible non-AGN sources.

\subsection{Corrections for selection effects}\label{sec:obs_evol:corr}

\subsubsection{Malmquist bias} \label{sec:malmquist_bias}
 The Malmquist bias---i.e., the loss of faint sources due the flux limit of a survey---affects not only intrinsically faint sources, but also those that are heavily obscured. This is because sources with the same flux and redshift may be either more obscured and intrinsically more luminous, or less obscured and less luminous.
Therefore, we corrected for a luminosity-, $N_{\rm H}$-, and redshift-dependent bias, as follows.

\begin{figure}[!tp]
    \center 
    \includegraphics[scale=0.45]{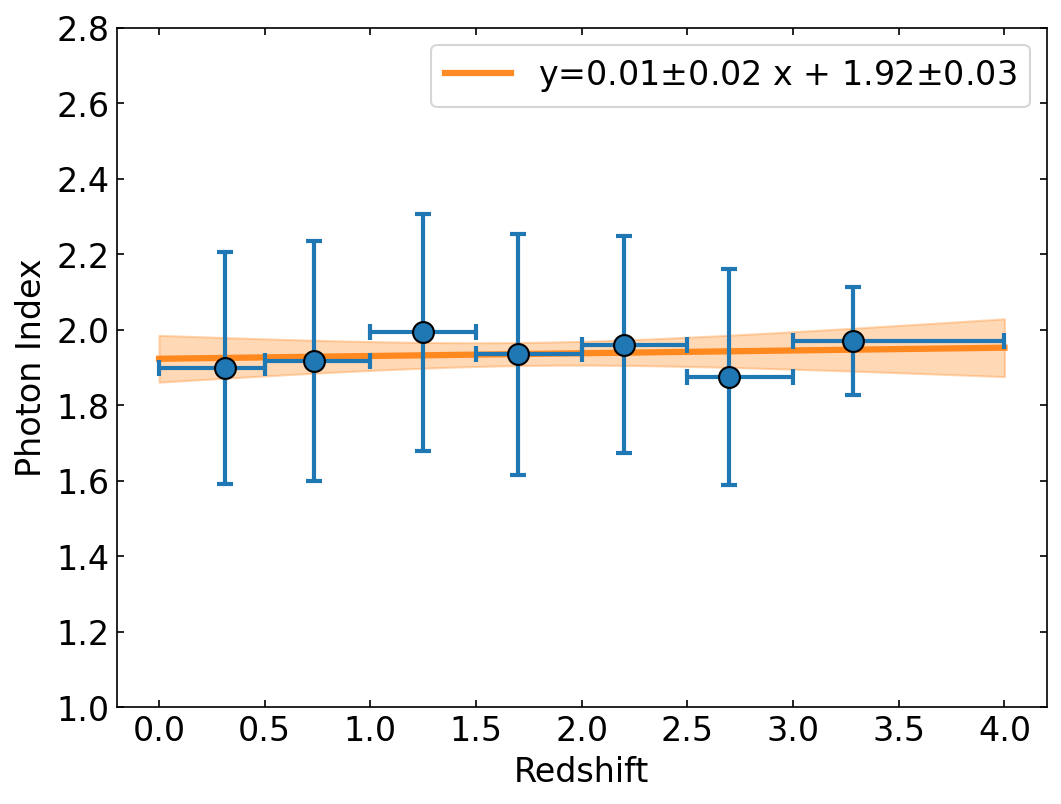}
    \caption{Photon index appears constant as a function of redshift. Values on the x-axis are centered at the redshift median value of each bin. Error bars on the y-axis represent one standard deviation in each bin, while on the x-axis represent the bin width. The best-fit line and its uncertainty at 2$\sigma$ are also shown (\textit{orange}).}
    \label{fig:gamma_evol}
\end{figure}

First, we used XSPEC to simulate AGN spectra as a function of $L_{\rm X}$, $z$, and $N_{\rm H}$, for models M1-M4. For each parameter combination and model, 1000 spectra were created. The adopted parameter space was similar to what is commonly observed for X-ray AGN: $\log L_{\rm X}/\mathrm{erg\,\, s}^{-1}$ in the range 42-46.5 with a step of 0.5, $\log N_{\rm H}/\mathrm {cm}^{-2}$ in the range 20.5-24.5 with a step of 0.5, and redshifts from 0 to 4 with a step of 0.5.
We did not assume a single $\Gamma$ value as usually done in this kind of simulation; instead, we randomly sampled the observed $\Gamma$ distribution (Figure \ref{fig:results1}, right panel). 
This has the advantage of not fixing a standard value, which is not what is observed in any X-ray survey. We assumed that the observed distribution is similar to the distribution of objects that are missing due to the survey sensitivity, which is justified since the $\Gamma$ distributions in deeper surveys are similar to what we found for S82X.
We assumed $\Gamma$ does not evolve with redshift since we do not observe such evolution in the S82X sample (see Fig. \ref{fig:gamma_evol}), nor is such evolution seen in other samples (e.g., \citealt{shemmer06,vito19}).
Response matrices for both \xmm\ and \chandra\ were applied, appropriate to the exposure times and observation epochs in our dataset.

\begin{figure}[!tp]
    \center 
    \includegraphics[scale=0.45]{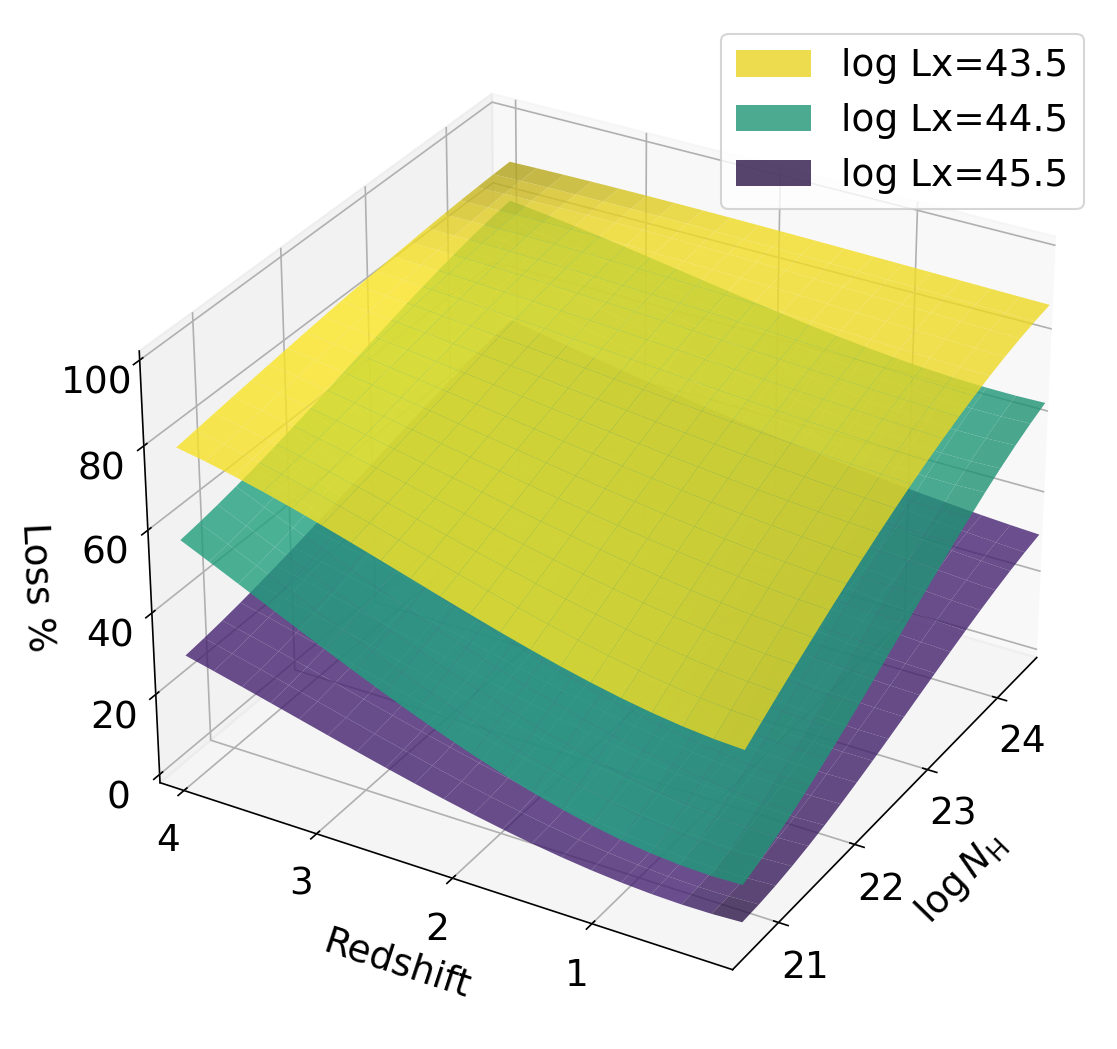}
    \caption{Loss surfaces for three luminosity values: $\log L_{\rm X}/\mathrm{erg\,\, s}^{-1}$ = 43.5 (\textit{yellow}), 44.5 (\textit{green}), and 45.5 (\textit{purple}), as a function of redshift and absorption. See text for the fully explored parameter space. At each $N_{\rm H}$, and $z$, the corresponding $L_{\rm X}$ surface gives the fraction of sources lost in the S82X due to observational limits. Surfaces are smoothed for graphic purposes.}
    \label{fig:sim_all}
\end{figure}

We then defined the loss-sensitivity function based on the number of simulated sources that satisfied our counts thresholds, $N_{cts>th}$ (thresholds were discussed in Section \ref{sec:analysis:thresholds}), relative to the total number of sources ($N_{tot}$) simulated for each combination of $z$, $N_{\rm H}$, and $L_{\rm X}$:
\begin{equation}\label{eq:loss}
    {\rm Loss} (z, N_{\rm H}, L_{\rm X}) = 1 - \frac{N_{cts>th}(z, N_{\rm H}, L_{\rm X})}{N_{tot}(z, N_{\rm H}, L_{\rm X})} .
\end{equation}
Equation \ref{eq:loss} represents the fraction of sources lost in S82X due to Malmquist bias. 
The same procedure was applied for \xmm and \chandra separately (see Appendix \ref{app:sens_sim}), and the results were combined together with a weighted average of 0.85 for \xmm and 0.15 for \chandra, reflecting that the final S82X sample consists of 15\% of sources detected with \chandra only.
The fractional loss as a function of $N_{\rm H}$ and $z$ can be represented with iso-luminosity surfaces, as shown in Figure \ref{fig:sim_all}\footnote{This figure can be downloaded as a \texttt{numpy} 3D array from GitHub: 
\href{https://github.com/alessandropeca/S82X_sensitivity_correction}{https://github.com/alessandropeca/S82X\_Correction}} (which includes also the corrections discussed in Section \ref{sec:skycov_bias} and \ref{sec:edd_bias}). We lose up to $\sim85\%$ of sources with $\log L_{\rm X}/\mathrm{erg\,\, s}^{-1}=43.5$ and $\log N_{\rm H}/\mathrm {cm}^{-2}>23$ at $z>3$, while only $\lesssim10\%$ when those are unobscured, luminous ($\log L_{\rm X}/\mathrm{erg\,\, s}^{-1}=45.5$), and at $z<1$.

We were then able to estimate how many objects were lost in the analysis, and then derive the intrinsic number of sources by simply applying the correction to the observed data.
Specifically, we corrected our sample on $L_X$, $N_H$, and $z$ bins according to the resolution of the simulation or, where not possible due to lack of sources, on larger bins.
It is worth mentioning that we can correct only for the parameter space we detected. Which is to say, we can not correct for sources with, for example, $\log N_{\rm H}/\mathrm {cm}^{-2}>24$ and $\log L_{\rm X}/\mathrm{erg\,\, s}^{-1} >46$ since they were not available in our sample.

\subsubsection{Sky coverage bias} \label{sec:skycov_bias}
\begin{figure}[!tp]
    \center 
    \includegraphics[scale=0.45]{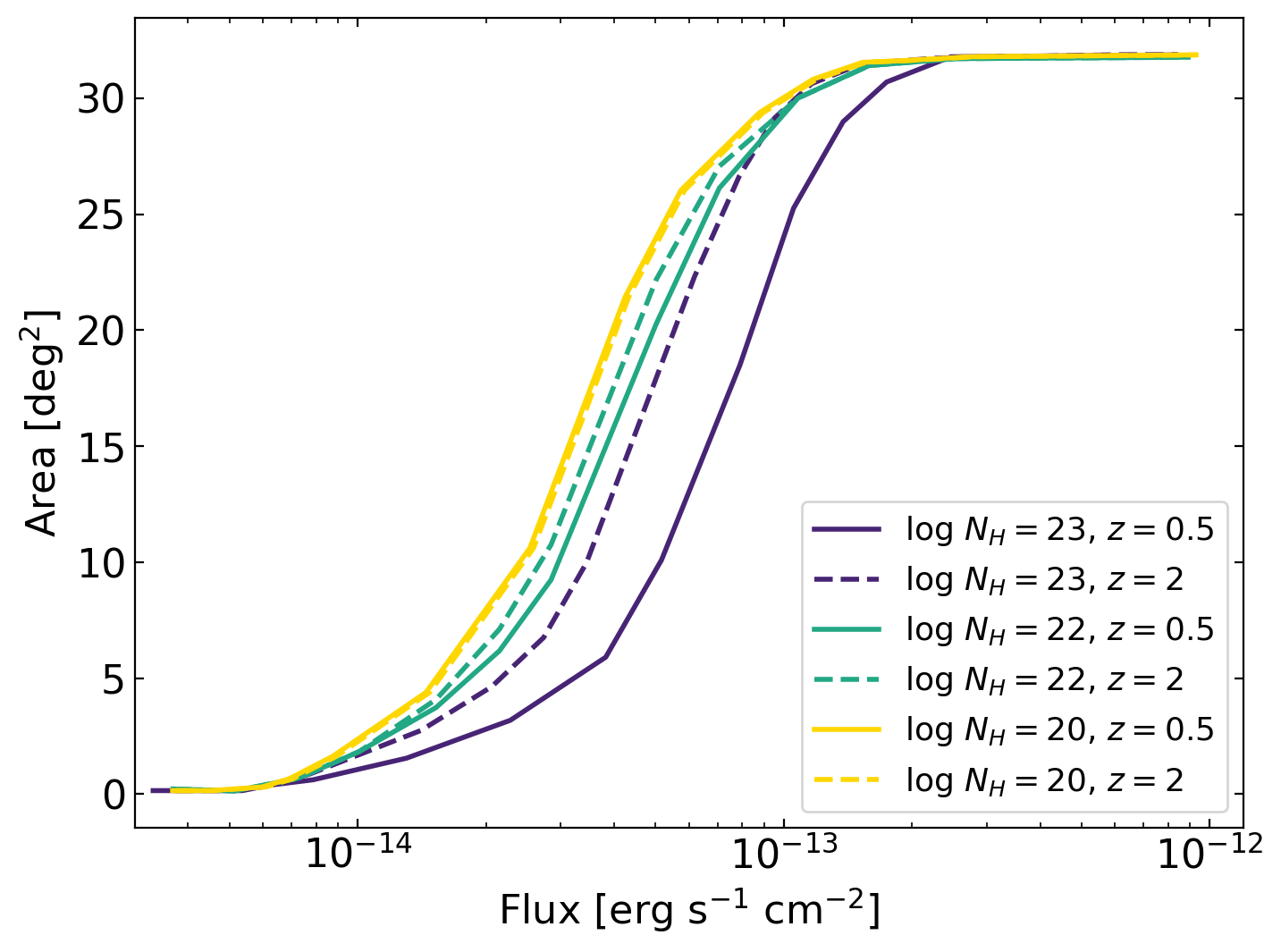}
    \caption{Sky coverage curves for $\log N_{\rm H}/\mathrm {cm}^{-2}=20$ (\textit{gold}), 22 (\textit{green}), and 23 (\textit{purple}). Solid and dashed lines represent sensitivities at $z=0.5$ and 2, respectively. The sky coverage bias impacts negatively on high $N_{\rm H}$ sources, while the increasing redshift mitigates it because of a favorable K correction.}
    \label{fig:sens_curves}
\end{figure}

The sky coverage of an X-ray survey is defined as the sky covered area as a function of the limiting flux. Thus, sources with different flux are detectable in different sky areas. This introduces a bias due to the AGN spectral shape, since it strongly depends on $N_{\rm H}$ and redshift. In particular, at a fixed redshift and luminosity, the sky coverage is biased against the detection of high-$N_{\rm H}$ AGN, which have lower fluxes than unobscured ones. This means that different AGN populations have different sky coverages as a function of $N_{\rm H}$ and $z$.
We simulated 1000 sources for each set of parameters, as explained in Section \ref{sec:malmquist_bias}, to quantify this effect. Then, we applied our count thresholds and derived the sensitivity curves. An example is shown in Figure \ref{fig:sens_curves}.

It is essential to notice that the sky coverage bias is independent of the Malmquist bias. Even if they both affect the detection of sources with high $N_{\rm H}$, the sky coverage bias compensates for the fact that different areas are sensitive to different observed fluxes, while the Malmquist bias depends on the intrinsic properties ($z$, $L_{\rm X}$, and $N_{\rm H}$) of the sources.
We corrected for the sky coverage bias by weighting each source by the reciprocal of its sky coverage at the corresponding observed flux (e.g., \citealt{liu17,yuan20}). The weights were then included in the simulations performed in Section \ref{sec:malmquist_bias}.

\subsubsection{Eddington bias}\label{sec:edd_bias}
\begin{figure}[!tp]
    \center 
    \includegraphics[scale=0.48]{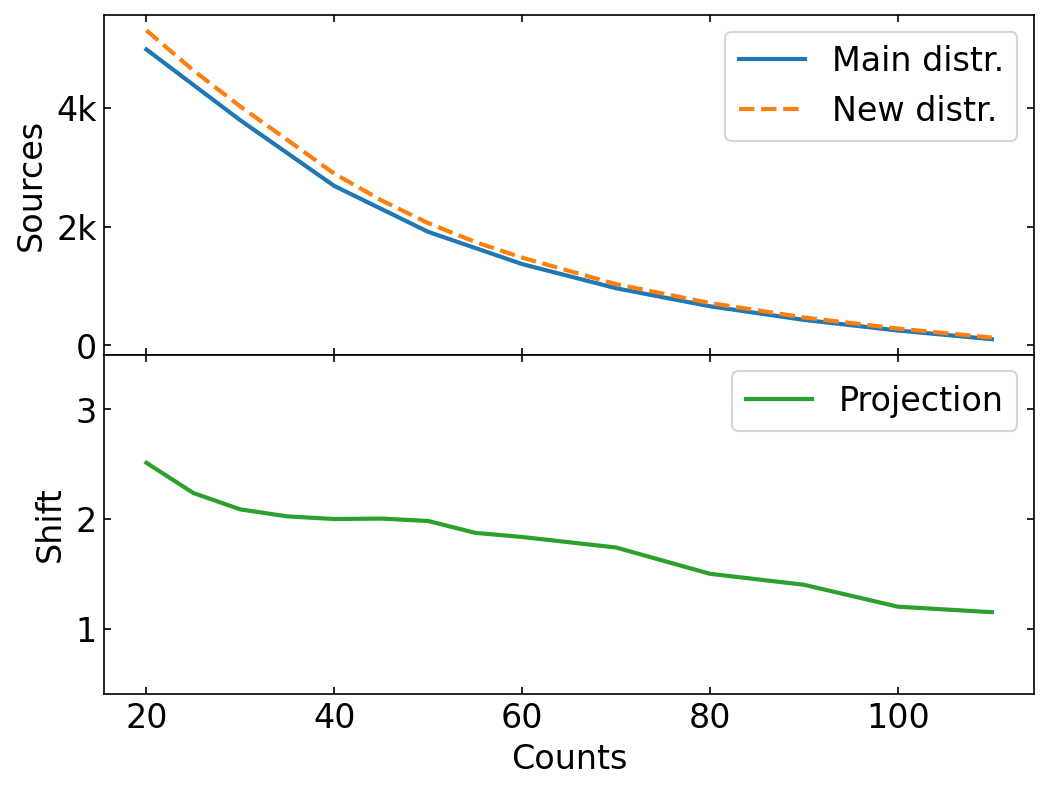}
    \caption{\textit{Upper panel:} Smoothed, inverse cumulative counts distribution of the main catalog (\textit{solid blue}) and of the simulated sample obtained by adding Poissonian fluctuations (\textit{dashed orange}). \textit{Bottom panel:} The projected horizontal shift between the two curves (\textit{solid green}) was assumed to be a good approximation of the Eddington bias (see text). The curves are cut between 20 counts, below which no sources were selected, and 110 counts, after which the slope of the shift projection becomes steady.}
    \label{fig:edd_bias}
\end{figure}

Another effect to be considered is the Eddington bias (\citealt{eddington1913}). Since faint sources are more numerous than brighter ones, statistical fluctuations may boost their source counts, therefore leading to an overestimation of sources when count thresholds are used to select samples.
We considered this effect by applying an additive correction similar to the one described in \citet{liu17}:
for each source in the main \citetalias{lamassa16} catalog, 
we generated 1000 copies with counts drawn from a Poissonian distribution with mean number of counts corresponding to the value of the parent object.
The resulting distribution for all replicated sources shows an excess of sources at low counts (Figure \ref{fig:edd_bias}), reflecting the Eddington bias. The horizontal shift between the two distributions, 
shown in the lower panel, 
was assumed as pseudo-correction. 
In particular, we corrected for the Eddington bias by applying this pseudo-correction to our selection curves, obtaining effective counts thresholds that were used in the bias corrections described in Section \ref{sec:skycov_bias} and \ref{sec:malmquist_bias}. As shown by the bottom panel of Figure \ref{fig:edd_bias}, 
the Eddington bias affects the sample selection by only $\sim$2 net counts or less, and is therefore almost negligible in our analysis.

\subsubsection{Redshift completeness}
To derive the properties of the intrinsic AGN populations, we must consider the redshift completeness of the sample, i.e., how many X-ray sources have a well constrained redshift estimate. S82X is almost 100\% complete in redshifts; however, including only photometric redshifts with high probability (PDZ$\geq70\%$), the redshift completeness drops to $\sim75\%$.
To verify that the remaining $\sim25\%$ did not impact our results,
we analyzed two test samples with redshift completeness of 90\% and of 95\%.
Within the uncertainties, the obscured fractions and the derived XLF 
in those test samples were consistent with the 75\% complete sample we analyze below (Section \ref{sec:osc_frac_section} and \ref{sec:xlf}), with no systematic differences in the best-fit parameters. 
In addition, we performed a Kolmogorov-Smirnov (K-S) test between the 95\% and 75\% samples on the corresponding $N_{\rm H}$, $L_{\rm X}$, and $z$ distributions. The obtained $p$-values are 0.45, 0.27, and 0.33, respectively, indicating that the samples are not significantly different in those quantities. 
We conclude that sources missing due to redshift incompleteness do not introduce a bias in our results. 
In support of this conclusion, we note that low-quality photometric redshifts (PDZ$<70\%$) are not systematically inaccurate (e.g., all low or all high) if compared to their spectroscopic values.

\subsection{Obscured fraction evolution} \label{sec:osc_frac_section}
\begin{figure*}[!t]
    \center 
    \includegraphics[scale=0.47]{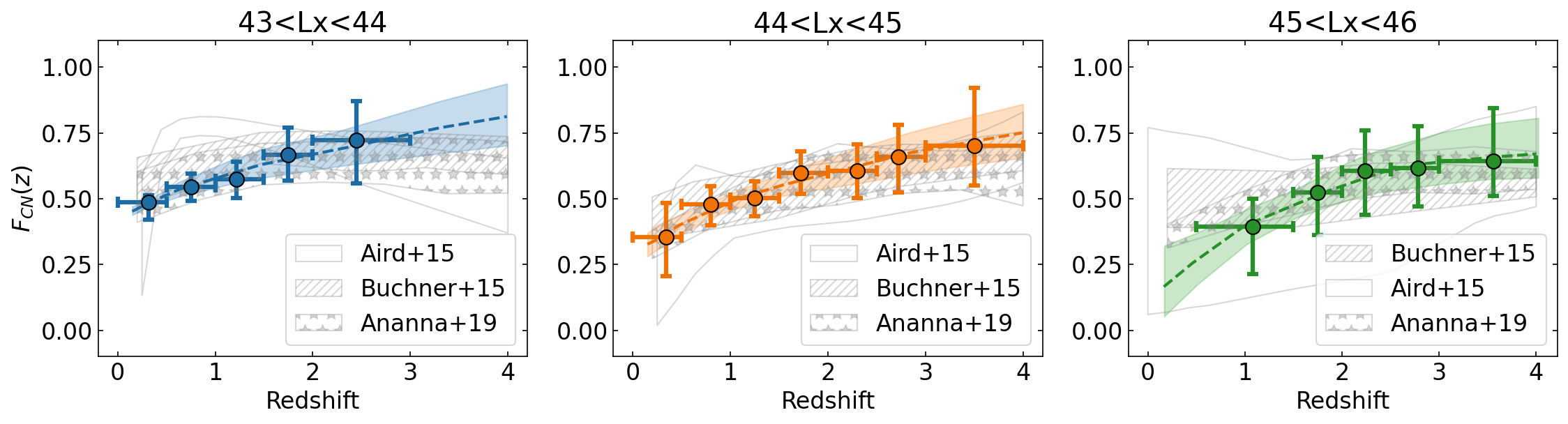}
    \caption{Obscured fraction evolution ($22\leq \log N_{\rm H}/\mathrm{cm}^{-2} < 24$), corrected for observational biases, as a function of redshift for three $\log L_{\rm X}/\mathrm{erg\,\, s}^{-1}$ bins: 
    43 to 44 (\textit{left}), 44 to 45 (\textit{middle}), and 45 to 46 (\textit{right}). The derived points overlaps the models from \citetalias{ananna19} 
    (\textit{shaded stars}), \citetalias{buchner15} (\textit{shaded slashed}), and \citetalias{aird15} (\textit{empty shaded}). Error bars were obtained with 
    a bootstrapping procedure (see text). Our models are shown with colored dashed lines in each panel, and the 1$\sigma$ confidence levels with colored 
    shaded areas. In the x-axis, each point is centered at the median redshift value of the corresponding bin.}
    \label{fig:nhfrac_l}
\end{figure*}

\begin{figure*}[!t]
    \center 
    \includegraphics[scale=0.5]{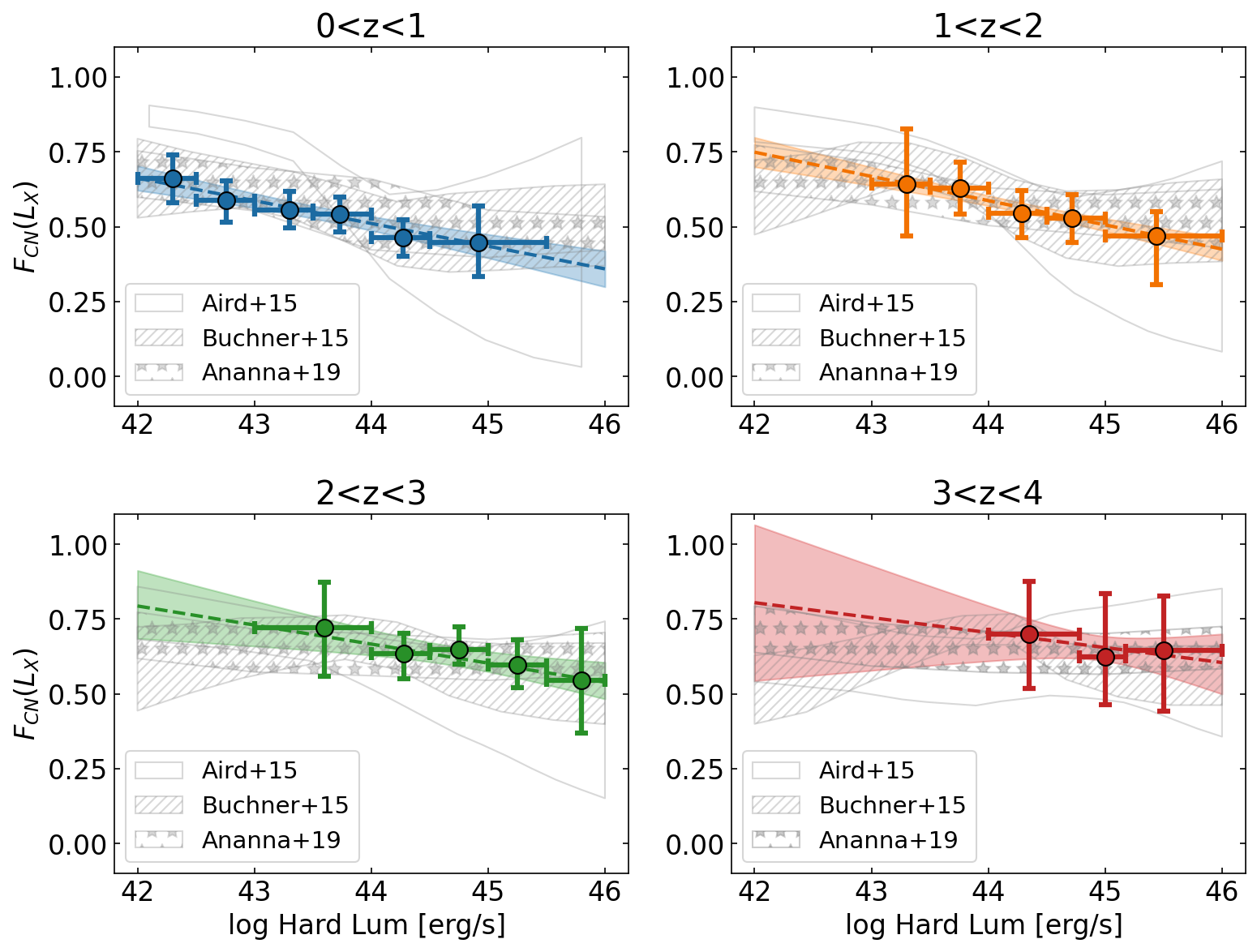}
    \caption{Same obscured fraction evolution represented in Figure \ref{fig:nhfrac_l}, but as a function of intrinsic $L_{\rm X}$ in four redshift bins: 0 to 1 (\textit{upper left}), 1 to 2 (\textit{upper right}), 2 to 3 (\textit{lower left}), and 3 to 4 (\textit{lower right}). The derived points overlap the models from \citetalias{ananna19} (\textit{shaded stars}), \citetalias{buchner15} (\textit{shaded dashed}), and \citetalias{aird15} (empty shaded). 
    Error bars were obtained with a bootstrapping procedure (see text). On the x-axis, each point is centered at the median luminosity value of the corresponding bin.}
    \label{fig:nhfrac_z}
\end{figure*}

The obscured fraction is key to understand how obscured and unobscured AGN populations evolve through cosmic time. Moreover, 
complete knowledge of obscured accretion is essential to understanding AGN intrinsic luminosities and the overall accretion history of supermassive black holes, as well as how the two are linked (e.g., \citealt{hopkins06}).
Once corrected for observational biases, the obscured AGN fraction was calculated as a function of redshift (Figure \ref{fig:nhfrac_l}) and $L_{\rm X}$ (Figure \ref{fig:nhfrac_z}). We defined the obscured fraction as the number of obscured, Compton-thin ($22 \leq \log N_{\rm H}/\mathrm{cm}^{-2} < 24$) AGN divided by the total number of AGN. Since we identified only two Compton-thick ($\log N_{\rm H}/\mathrm {cm}^{-2}\gtrsim24$) sources, we focus here on the unobscured and Compton-thin populations.
To avoid large uncertainties, we binned our data such that each bin contained at least 8 objects in both the numerator and denominator, with the exception of the red panel in Figure \ref{fig:nhfrac_z}, where we used 6 objects as a minimum in order to have at least three points. 
Error bars in both the figures were computed with a bootstrapping procedure (e.g., \citealt{vito18}): the $N_{\rm H}$ distribution derived from the spectral analysis was used to generate a new random $N_{\rm H}$ distribution, with the same total number of sources and allowing for repetition. Then, a new obscured fraction was computed and the procedure was repeated 1000 times. At this point we had 1000 different values for each $N_{\rm H}$ bin, from which confidence intervals were computed.

First, we built the obscured, Compton-thin fraction ($F_{CN}$) as a function of redshift for three luminosity bins, from $\log L_{\rm X}/\mathrm{erg\,\, s}^{-1}=$43 to 46 erg
(Figure \ref{fig:nhfrac_l}). We assumed a broken power-law model in the following form:
\begin{equation}
    F_{CN}(z) = 
    \begin{cases}
        A (z/z_*)^{-\alpha_1} &\text{if $z\leq z_*$}\\
        A (z/z_*)^{-\alpha_2} &\text{if $z> z_*$}
    \end{cases}
\end{equation}
where $A$ is the normalization constant, and $z_*$ is the characteristic redshift between the two slopes $\alpha_1$ and $\alpha_2$. To fit the data points to the assumed model we used \textsc{emcee} (\citealt{foreman13}), a solid Markov chain Monte Carlo (MCMC) implementation of the Goodman-Weare algorithm (\citealt{goodman10}). The best-fit parameters are shown in Table \ref{table:fit_res} (top). 
We find that the obscured fraction increases with redshift at all luminosities.
Formally, $z_*$ and $\alpha_1$ increase with luminosity, while no trend was found for $A$ and $\alpha_2$. 
Second, we built the obscured fraction as a function of $L_{\rm X}$, for four redshift bins from 0 to 4 (Figure \ref{fig:nhfrac_z}). We assumed a simple line model, since we did not see any more complex trend in the data points, and applied the same fitting procedure described above. The best-fit results are shown in Table \ref{table:fit_res} (bottom). No trend was found in the best-fit parameters; however, the normalization in the selected luminosity range increases with the redshift.
Our results are consistent with previous works (\citealt{treister06}, \citealt{hasinger08}, \citetalias{aird15, buchner15}; and \citetalias{ananna19}) within the uncertainties (see also Section \ref{sec:comparison}).

Overall, there is a strong positive redshift evolution, with the Compton-thin fraction increasing as a function of redshift at all luminosities. Moreover, the $z_*$ parameter increases with the luminosity bins, indicating a luminosity-dependent evolution. Indeed, we found an anti-correlation with the luminosity, at all redshifts.
However, the uncertainties at high redshift and low luminosity, as well as those at low redshift and high luminosity, prevent us from further considerations on the luminosity-dependent evolution.
Considering $\log L_{\rm X}/\mathrm{erg\,\, s}^{-1}>43$, and integrating over the redshift range 0-4, we obtained an obscured Compton-thin fraction of $57\pm6\%$, while exploring the high $L_{\rm X}$ ($\log L_{\rm X}/\mathrm{erg\,\, s}^{-1}>45.5$) and high redshift ($z>3$) regime we obtained an obscured fraction of $64\pm12\%$.
This is consistent with what was found by e.g., \citetalias{buchner15} and \citetalias{ananna19}.

\begin{table}[!t]
\centering
\caption{Fit Parameters for Evolution of Obscured Fraction\tablenotemark{a}}
\setlength\tabcolsep{4pt}
\begin{tabular}{ccccc}
\toprule
\multicolumn{5}{c}{{Obscured Fraction. vs. Redshift}}                                         \\ \hline
Bin                        & $A$     & $z_*$     & $-\alpha_1$ & $-\alpha_2$ \\ \hline
$43< L_{\rm X} \leq44$ & 0.58$_{-0.05}^{+0.04}$ & 1.10$_{-0.06}^{+0.05}$ & 0.14$_{-0.04}^{+0.04}$ & 0.25$_{-0.05}^{+0.05}$ \\
$44< L_{\rm X} \leq45$ & 0.52$_{-0.04}^{+0.03}$ & 1.32$_{-0.05}^{+0.04}$ & 0.25$_{-0.04}^{+0.03}$ & 0.34$_{-0.05}^{+0.04}$ \\
$45< L_{\rm X} \leq46$ & 0.60$_{-0.07}^{+0.06}$ & 2.24$_{-0.07}^{+0.06}$ & 0.55$_{-0.06}^{+0.07}$ & 0.19$_{-0.05}^{+0.06}$ \\ 
\multicolumn{5}{c}{} \\ \hline
\multicolumn{5}{c}{{Obscured Fraction vs. Luminosity}}                                     \\ \hline
Bin                             & \multicolumn{2}{c}{$m$} & \multicolumn{2}{c}{$q$}   \\ \hline
$0< z \leq 1$  & \multicolumn{2}{c}{$-0.076\pm0.009$}  & \multicolumn{2}{c}{$3.85\pm0.04$}   \\
$1< z \leq 2$  & \multicolumn{2}{c}{$-0.081\pm0.011$}  & \multicolumn{2}{c}{$4.14\pm0.07$}   \\
$2< z \leq 3$  & \multicolumn{2}{c}{$-0.064\pm0.020$}  & \multicolumn{2}{c}{$3.48\pm0.09$}   \\
$3< z \leq 4$  & \multicolumn{2}{c}{$-0.05\pm0.04$}  & \multicolumn{2}{c}{$2.91\pm0.19$}   \\ \hline
\end{tabular}
\begin{flushleft}
\tablenotetext{a}{Best-fit parameters for evolution with redshift (top): normalization, $A$; characteristic redshift, $z_*$; and two power-law indices, $\alpha_1$ and $\alpha_2$ (see Fig. \ref{fig:nhfrac_l}). For dependence on luminosity (bottom): best-fit parameters are slope, $m$, and intercept, $q$ (see Fig. \ref{fig:nhfrac_z}). Uncertainties correspond to the 1$\sigma$ confidence level.}
\end{flushleft}
\label{table:fit_res}
\end{table}

\subsection{Luminosity function}\label{sec:xlf}
The X-ray luminosity function is an important tracer of AGN evolution and is key to understand the accretion history of supermassive black holes. We built it using the corrected, (i.e., intrinsic) distributions of $N_{\rm H}$ and $L_{\rm X}$ derived from our analysis. 
One of the commonly used methods to build the XLF is the $1/V_{a}$ estimator (e.g., \citealt{miyaji00}), which is a generalization of the $1/V_{max}$ method (\citealt{schmidt1968}). The intuitive idea behind this method is to construct a binned differential luminosity function (LF) in redshift intervals. However, it may produce biases at low fluxes close to the data sensitivity and when the luminosity bins are large (see e.g., \citealt{page2000, miyaji01}). 
Several methods were designed to address the limitations of the $1/V_a$ approach (e.g., \citealt{lynden1971,miyaji01,ananna22}); we used a method similar to \citet{page2000}, which works with Poissonian statistics, so there is no minimum number of objects required in each bin.

\begin{figure*}[!t]
    \center 
    \includegraphics[scale=0.39]{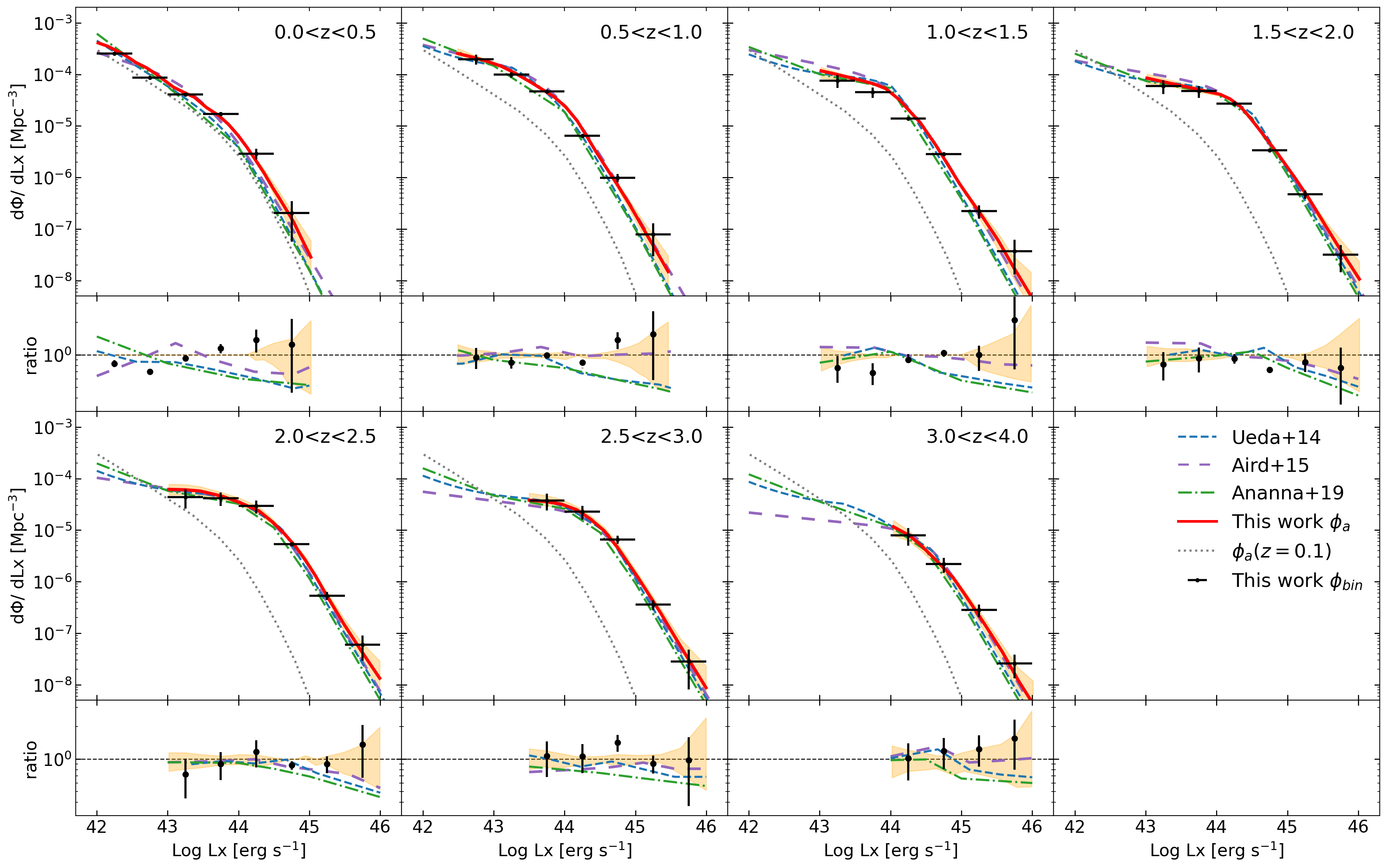}
    \caption{ Derived XLFs for seven redshift bins. 
    \textit{Black points} represent the binned XLFs ($\phi_{bin}$, \citealt{page2000}), while \textit{red solid lines} show the KDE results ($\phi_a$, \citetalias{yuan22}) estimated at the redshift bin centers.
    For both methods, we calculate the XLF only where there are sufficient sources. 
    \textit{Shaded orange regions} represent the 3$\sigma$ uncertainties of the kernel density estimation. As a comparison, we show the XLFs derived by \citetalias{ueda14} (\textit{dashed blue}), \citetalias{ananna19} (\textit{dash-dotted orange}), and \citetalias{aird15} (\textit{long dashed purple}). All the XLFs were computed for $\log N_{\rm H}/\mathrm {cm}^{-2} < 24$ AGN. The residuals, in the form of the ratio between each XLF and the KDE $\phi_a$, are shown at the bottom of each panel. The $z=0.1$ XLF (\textit{dotted black}) derived with the KDE method is also plotted in each panel for reference.}
    \label{fig:XLF}
\end{figure*}

Our derived binned XLF ($\phi_{bin}$) is shown in Figure \ref{fig:XLF} (\textit{black points}), for 7 redshift bins from $z=0$ to 4. 
The derived binned LF is consistent, within the uncertainties, with the XLFs from \citetalias{ueda14}, \citetalias{aird15}, and \citetalias{ananna19}.
To fit the XLF, we applied the kernel density estimation (KDE) method described by \citet{yuan20,yuan22}. 
This nonparametric approach takes advantage of the mathematics beyond the KDE (\citealt{Wasserman06}), a well-established procedure to estimate continuous density functions (e.g., \citealt{chen17, davies18}).
This method does not require any model assumptions, since it generates the XLF relying only on the available data.
In particular, we apply the adaptive KDE version of the method, which automatically adapts the bandwidth of each individual kernel avoiding possible biases due to data inhomogeneity. The choice of the adaptive KDE is justified by applying a K–S test as described by \citealt{yuan22} (\citetalias{yuan22}, hereafter).
The analytical form of the adaptive KDE XLF, $\phi_a$ (for the full mathematical treatment, see \citetalias{yuan22}) is
\begin{equation}
    \phi_a(z,L) = \frac{n(Z_2-Z_1)\hat{f}_a(x,y|h_{10},h_{20},\beta)}{(z-Z_1)(Z_2-z)\Omega dV/dz} ,
\end{equation}
where [$Z_1,Z_2$] is the dataset redshift range, $n$ is the number of objects in such range, $\Omega$ is the covered solid angle, $dV/dz$ is the the differential comoving volume per unit of solid angle, and $\hat{f}_a(x,y|h_{10},h_{20},\beta)$ is the density of $(x,y)$, which corresponds to the $(z,L)$ pair in the KDE parameter space. 
This function, $\hat{f}_a$, defined in equation A1 in \citetalias{yuan22}, depends on three parameters: the two bandwidths $h_{10}$ and $h_{20}$, which define the window width of the density estimation, and the sensitivity parameter, $\beta$, which allows for a flexible and adaptive kernel width. 
\citetalias{yuan22} provide a Bayesian MCMC routine to determine the posterior distributions of the bandwidths and the $\beta$ parameter, and then the uncertainty estimation on the XLF. 

\begin{figure}[!t]
    \center 
    \includegraphics[scale=0.45]{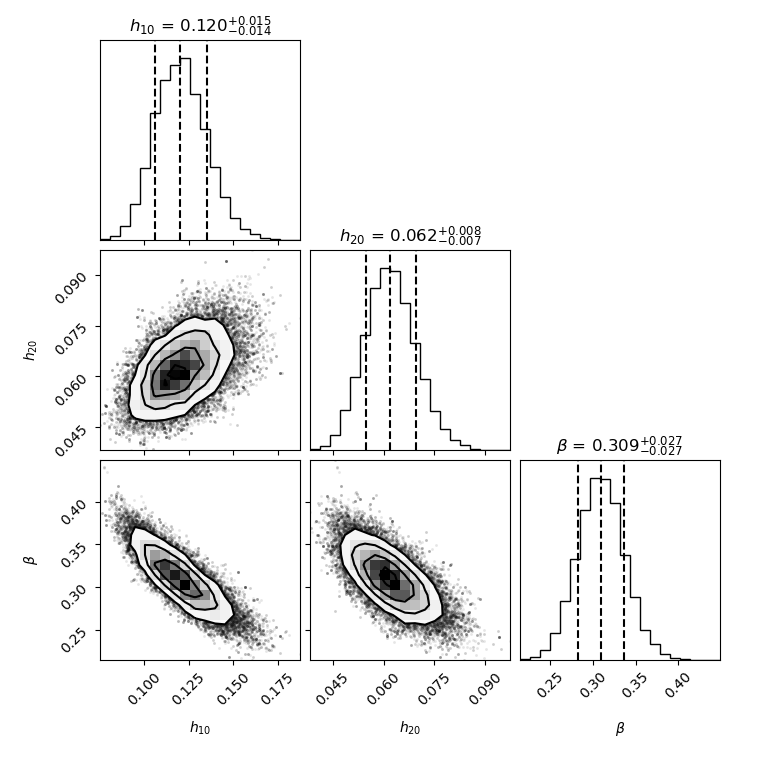}
    \caption{Posterior distributions of the bandwidths ($h_{10}$, $h_{20}$) and the $\beta$ parameter of the adaptive KDE XLF, obtained with the routine provided by \citetalias{yuan22}. Uncertainties correspond to the 16$^{th}$ and 84$^{th}$ percentiles, while the best fit value is the median. All the parameters are well constrained: $h_{10}=0.120_{-0.014}^{+0.015}$, $h_{20}=0.062_{-0.007}^{+0.008}$, and $\beta=0.309\pm0.027$, meaning a good kernel density estimation of the XLF.}
    \label{fig:XLF_posterior}
\end{figure}

The derived XLF is shown in Figure \ref{fig:XLF} (\textit{solid red lines} and \textit{orange contours}).
In principle, the KDE method allows extrapolation of the XLF beyond the luminosity limits of the observations; however, the resulting high luminosity tail may overestimate the true XLF, especially when using the adaptive method (\citealt{yuan20}), so our XLFs are reported only for observed luminosities.
As shown by the posterior error estimations in Figure \ref{fig:XLF_posterior}, the derived best-fit parameters are well constrained, meaning a good kernel estimation (\citetalias{yuan22}).

The XLF shows a strong luminosity evolution, since as the redshift increases the AGN density increases at high luminosity. 
We also see a density evolution, since the ratio between the XLF and the XLF at $z=0.1$ varies with $z$. In particular, the AGN density increases up to $z\sim2$, then declines at $z\gtrsim 2.5$. These trends are in agreement with what was found previously by \citetalias{ueda14}, \citetalias{aird15}, and \citetalias{ananna19}. It is worth mentioning the differences among these works: \citetalias{ueda14} and \citetalias{ananna19} analyzed sources in the 0.5-195 keV range, while \citetalias{aird15} analyzed the 0.5-7 keV band only (see also \citetalias{ananna19} for a detailed comparison), similar to the 0.5-10 keV range used in this work. Moreover, while they used different combinations of deep and wide surveys, we used the S82X data only.
At the bottom of each panel in Figure \ref{fig:XLF}, we plot the ratio between our XLF and the others (\citetalias{ueda14, aird15, ananna19}), showing they are in generally good agreement. 
Even though there is a slight increase of AGN densities at high luminosities, the uncertainties of both our XLF and the others (omitted in Figure \ref{fig:XLF} for clarity) become larger at these regimes, making them consistent within the errors and preventing us from further considerations.
The nonparametric approach does not impose any particular shape on the XLF, but a break followed by significant steepening (a ``knee" in the XLF), is required by the data. 
In practice, AGN at the break luminosity dominate the X-ray emission produced by the full population.
This break luminosity evolves to higher luminosities with increasing redshift.

\begin{figure*}[!t]
    \center 
    \includegraphics[scale=0.55]{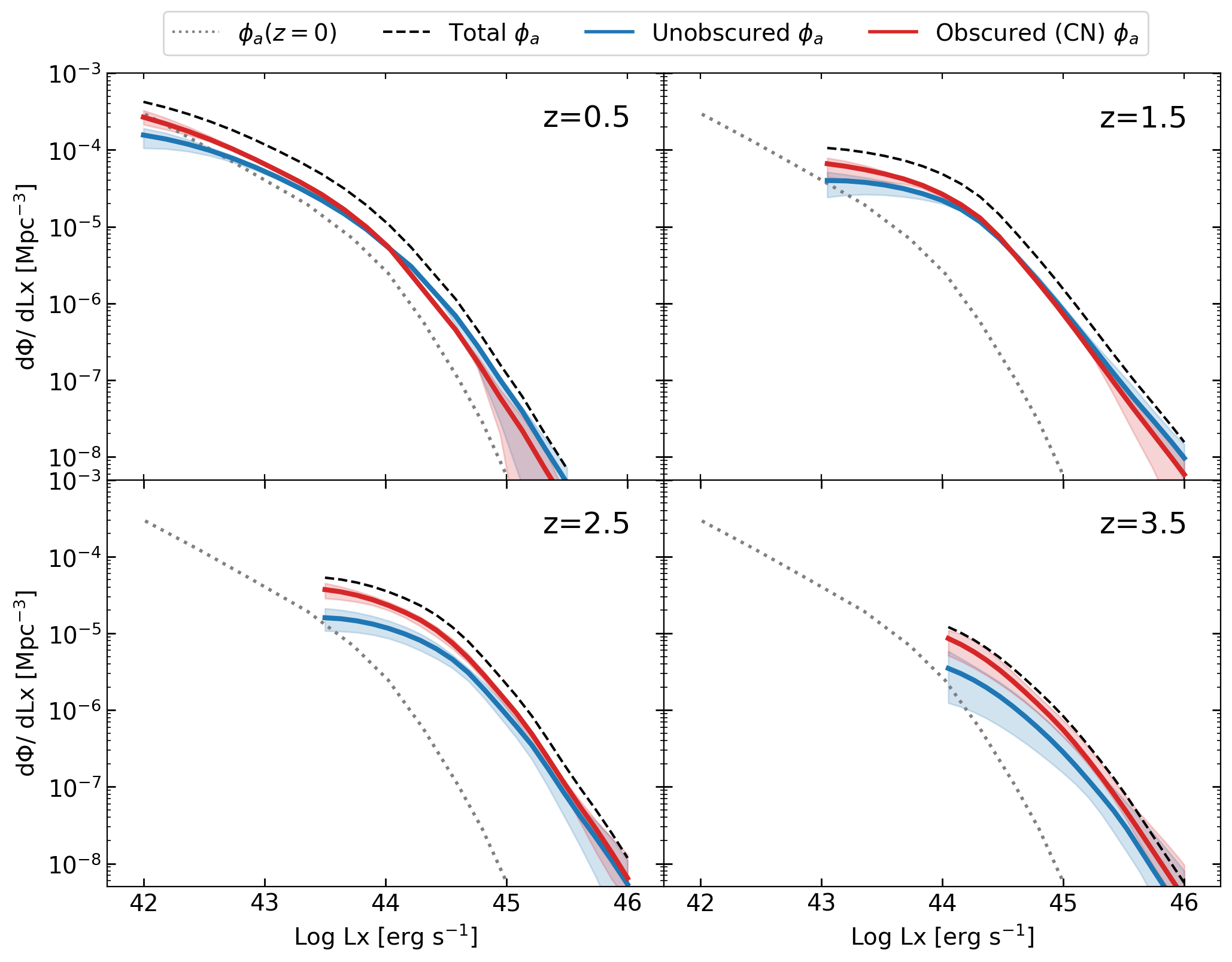}
    \caption{Total (\textit{dashed black}), unobscured ($\log N_{\rm H}/\mathrm {cm}^{-2}<22$, \textit{solid blue}) and obscured ($22\leq \log N_{\rm H}/\mathrm{cm}^{-2}<24$, \textit{solid red}) XLFs derived with the KDE method (\citealt{yuan22}) for the corrected samples at $z=$ 0.5, 1.5, 2.5, and 3.5. Shaded regions represent the 3$\sigma$ uncertainties of the kernel density estimation. XLFs are cut off where we lack sources. The total XLF at $z=0.1$ (\textit{dotted black line}) is also shown. The best-fit parameters are $h_{10}=0.192_{-0.028}^{+0.031}$, $h_{20}=0.047_{-0.012}^{+0.013}$, and $\beta=0.304_{-0.045}^{+0.044}$ for obscured AGN, and $h_{10}=0.170_{-0.025}^{+0.028}$, $h_{20}=0.071_{-0.009}^{+0.011}$, and $\beta=0.258_{-0.049}^{+0.049}$ for unobscured AGN.}
    \label{fig:XLF_sep}
\end{figure*}

We also derived the XLF separately for obscured (CN) and unobscured AGN. 
The results are shown in Figure \ref{fig:XLF_sep}. 
Unobscured AGN slightly dominate at high luminosity ($\log L_{\rm X}/\mathrm{erg\,\, s}^{-1}\gtrsim$ 44-45) up to $z\sim1.5$-2, with the obscured population dominating at lower luminosities.
At higher redshifts ($z>2$), instead, the obscured population dominates at all luminosities. 
This is similar to what was found by \citetalias{aird15}. 
In general, both unobscured and obscured AGN experience strong luminosity evolution,
as well as 
density evolution. 
Table \ref{tab:xlf_numbers} gives number densities for the total, unobscured, and obscured XLFs.

\begin{figure}[!t]
    \center 
    \includegraphics[scale=0.45]{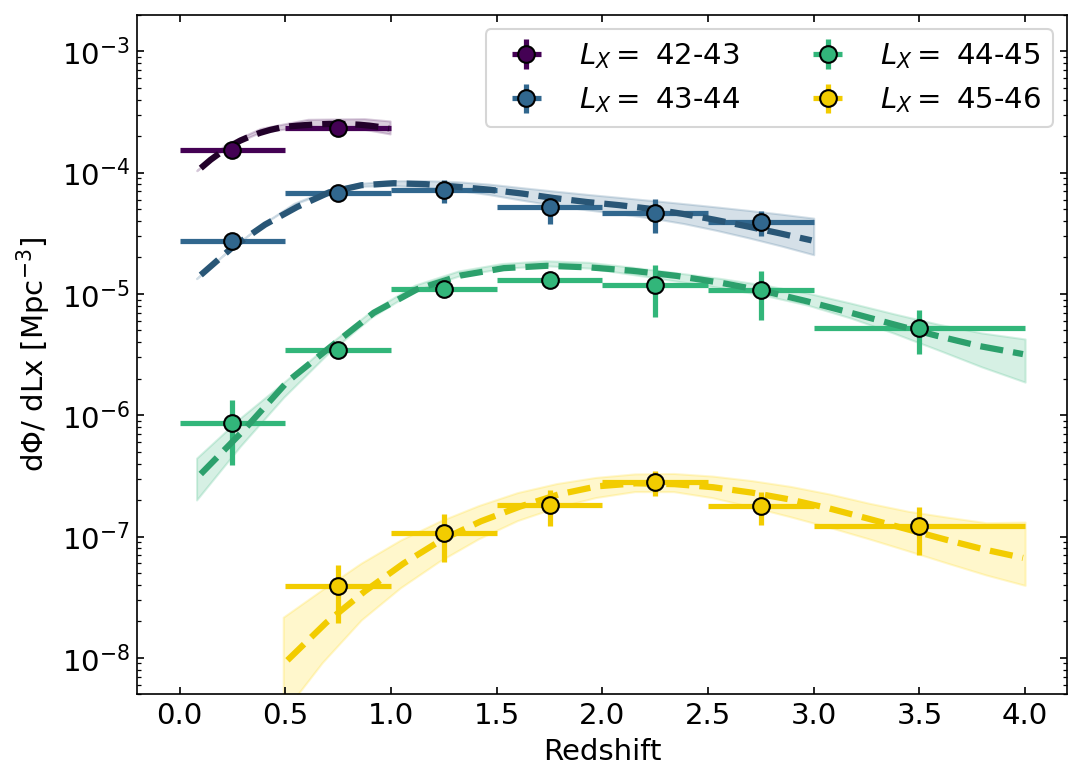}
    \includegraphics[scale=0.45]{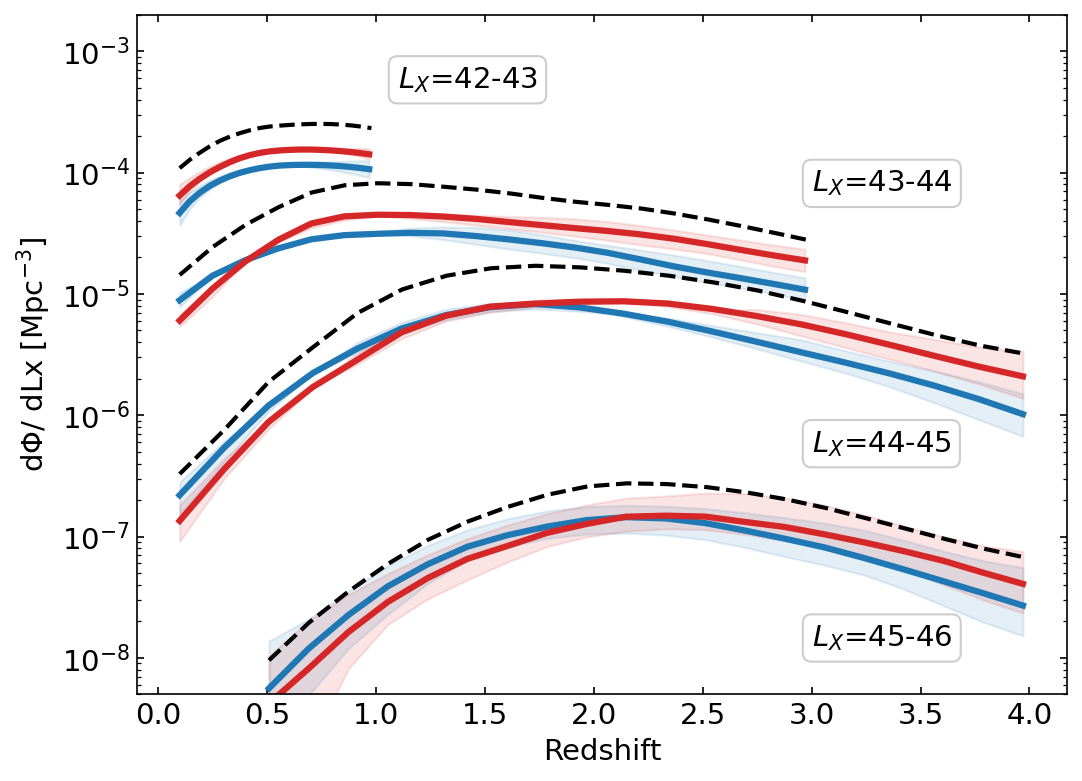}
    \caption{\textit{Top panel}: Total AGN space density as a function of redshift and for $L_{\rm X}$ bins described in the legend. Dashed lines represent the derived XLF ($\phi_a$), while the points correspond to the binned XLF ($\phi_{bin}$), both integrated over corresponding the luminosity bin.
    \textit{Bottom panel}: The \textit{black dashed lines} are the same total space densities as in the top panel, each divided into unobscured (\textit{blue}) and obscured (\textit{red}) AGN populations. Shaded regions are the 3$\sigma$ uncertainties of the KDE estimation.}
    \label{fig:XLF_downsizing}
\end{figure}

In Figure \ref{fig:XLF_downsizing} (top panel), we show the spatial AGN density as a function of redshift, for different luminosity bins. As the luminosity increases, the peak of the density moves towards higher redshifts. This pattern describes the so-called downsizing scenario (e.g., \citealt{cowie03,ueda03}), where the AGN density is driven by a combination of merger-driven mechanisms for high luminosity AGN, and secular processes dominated by minor mergers and disk instabilities for intermediate and low luminosity AGN (e.g., \citealt{treister12,miyaji15}).
As we did for the XLF, we split the total AGN space density into unobscured and obscured populations (Figure \ref{fig:XLF_downsizing}, bottom panel). Except for the $\log L_{\rm X}/\mathrm{erg\,\, s}^{-1}$=42-43 bin, which is not covered adequately by our relatively shallow data, it is clear how the combination of the underlying populations models the total AGN density shape.
In particular, the distributions of the two populations peak at different redshifts, with the obscured (unobscured) one at higher (lower) redshifts. This reflects the fact that obscured AGN become more dominant as the redshift increases.

\begin{table*}[!t]
\centering
\begin{tabular}{ccccc}
\toprule
$\log L_{\rm X}$     &  $z$  &  $\phi_a$                   &  $\phi_{unobs}$             &  $\phi_{obs}$   \\
(erg s$^{-1}$) &    -   & (Mpc$^{-3}$)                &  (Mpc$^{-3}$)               & (Mpc$^{-3}$)    \\
\hline
42-43  &  0.1  &  $1.11_{-0.02}^{+0.03}\cdot 10^{-4}$  &  $4.29_{-1.00}^{+0.73}\cdot 10^{-5}$  &  $7.13_{-0.51}^{+0.88}\cdot 10^{-5}$ \\
42-43  &  0.5  &  $2.39_{-0.04}^{+0.21}\cdot 10^{-4}$  &  $9.40_{-0.09}^{+0.16}\cdot 10^{-5}$  &  $1.51_{-0.01}^{+0.02}\cdot 10^{-4}$ \\
42-43  &  1.0  &  $2.32_{-0.24}^{+0.34}\cdot 10^{-4}$  &  $9.18_{-1.59}^{+2.00}\cdot 10^{-5}$  &  $1.45_{-0.13}^{+0.18}\cdot 10^{-4}$ \\
43-44  &  0.1  &  $1.49_{-0.05}^{+0.01}\cdot 10^{-5}$  &  $8.17_{-1.05}^{+1.42}\cdot 10^{-6}$  &  $6.64_{-0.65}^{+0.72}\cdot 10^{-6}$ \\
43-44  &  0.5  &  $4.66_{-0.03}^{+0.41}\cdot 10^{-5}$  &  $2.13_{-0.02}^{+0.02}\cdot 10^{-5}$  &  $2.56_{-0.03}^{+0.04}\cdot 10^{-5}$ \\
43-44  &  1.0  &  $8.16_{-0.31}^{+0.41}\cdot 10^{-5}$  &  $3.39_{-0.01}^{+0.05}\cdot 10^{-5}$  &  $4.77_{-0.07}^{+0.03}\cdot 10^{-5}$ \\
43-44  &  1.5  &  $7.05_{-0.67}^{+0.88}\cdot 10^{-5}$  &  $2.71_{-0.45}^{+0.56}\cdot 10^{-5}$  &  $4.51_{-0.47}^{+0.33}\cdot 10^{-5}$ \\
43-44  &  2.0  &  $5.67_{-0.80}^{+0.82}\cdot 10^{-5}$  &  $1.90_{-0.42}^{+0.33}\cdot 10^{-5}$  &  $4.09_{-0.51}^{+0.68}\cdot 10^{-5}$ \\
43-44  &  2.5  &  $4.15_{-0.71}^{+1.15}\cdot 10^{-5}$  &  $1.27_{-0.27}^{+0.38}\cdot 10^{-5}$  &  $3.17_{-0.44}^{+0.49}\cdot 10^{-5}$ \\
43-44  &  3.0  &  $2.92_{-0.71}^{+1.30}\cdot 10^{-5}$  &  $9.26_{-2.05}^{+2.70}\cdot 10^{-6}$  &  $2.15_{-0.43}^{+0.44}\cdot 10^{-5}$ \\
44-45  &  0.1  &  $3.37_{-1.04}^{+1.37}\cdot 10^{-7}$  &  $2.41_{-0.47}^{+0.76}\cdot 10^{-7}$  &  $1.04_{-0.33}^{+0.61}\cdot 10^{-7}$ \\
44-45  &  0.5  &  $1.92_{-0.27}^{+0.19}\cdot 10^{-6}$  &  $1.07_{-0.10}^{+0.06}\cdot 10^{-6}$  &  $8.49_{-0.39}^{+0.45}\cdot 10^{-7}$ \\
44-45  &  1.0  &  $8.47_{-0.23}^{+0.74}\cdot 10^{-6}$  &  $4.18_{-0.38}^{+0.32}\cdot 10^{-6}$  &  $4.31_{-0.41}^{+0.32}\cdot 10^{-6}$ \\
44-45  &  1.5  &  $1.54_{-0.34}^{+0.17}\cdot 10^{-5}$  &  $7.07_{-0.62}^{+0.65}\cdot 10^{-6}$  &  $8.54_{-0.80}^{+0.29}\cdot 10^{-6}$ \\
44-45  &  2.0  &  $1.62_{-0.06}^{+0.13}\cdot 10^{-5}$  &  $6.89_{-0.25}^{+0.23}\cdot 10^{-6}$  &  $9.57_{-0.41}^{+0.29}\cdot 10^{-6}$ \\
44-45  &  2.5  &  $1.28_{-0.11}^{+0.09}\cdot 10^{-5}$  &  $4.70_{-0.49}^{+0.58}\cdot 10^{-6}$  &  $8.50_{-0.63}^{+0.41}\cdot 10^{-6}$ \\
44-45  &  3.0  &  $8.55_{-0.60}^{+1.42}\cdot 10^{-6}$  &  $2.91_{-0.51}^{+0.85}\cdot 10^{-6}$  &  $5.93_{-1.10}^{+0.59}\cdot 10^{-6}$ \\
44-45  &  3.5  &  $5.19_{-1.03}^{+1.06}\cdot 10^{-6}$  &  $1.74_{-0.49}^{+0.56}\cdot 10^{-6}$  &  $3.63_{-0.81}^{+0.58}\cdot 10^{-6}$ \\
44-45  &  4.0  &  $3.02_{-1.24e}^{+1.13}\cdot 10^{-6}$ &  $8.98_{-3.35}^{+4.74}\cdot 10^{-7}$  &  $2.24_{-0.72}^{+0.89}\cdot 10^{-6}$ \\
45-46  &  0.5  &  $9.87_{-6.20}^{+12.3}\cdot 10^{-9}$  &  $6.43_{-4.62}^{+8.55}\cdot 10^{-9}$  &  $3.53_{-3.42}^{+5.03}\cdot 10^{-9}$ \\
45-46  &  1.0  &  $5.46_{-2.15}^{+3.02}\cdot 10^{-8}$  &  $3.33_{-1.21}^{+1.56}\cdot 10^{-8}$  &  $2.16_{-0.72}^{+1.12}\cdot 10^{-8}$ \\
45-46  &  1.5  &  $1.65_{-0.51}^{+0.39}\cdot 10^{-7}$  &  $8.44_{-1.22}^{+3.43}\cdot 10^{-8}$  &  $8.04_{-1.69}^{+2.49}\cdot 10^{-8}$ \\
45-46  &  2.0  &  $2.70_{-0.54}^{+0.44}\cdot 10^{-7}$  &  $1.28_{-0.33}^{+0.39}\cdot 10^{-7}$  &  $1.44_{-0.29}^{+0.46}\cdot 10^{-7}$ \\
45-46  &  2.5  &  $2.63_{-0.40}^{+0.38}\cdot 10^{-7}$  &  $1.14_{-0.24}^{+0.37}\cdot 10^{-7}$  &  $1.54_{-0.19}^{+0.65}\cdot 10^{-7}$ \\
45-46  &  3.0  &  $1.96_{-0.64}^{+0.33}\cdot 10^{-7}$  &  $7.64_{-2.04}^{+4.56}\cdot 10^{-8}$  &  $1.26_{-0.16}^{+0.41}\cdot 10^{-7}$ \\
45-46  &  3.5  &  $1.24_{-0.45}^{+0.40}\cdot 10^{-7}$  &  $3.46_{-1.46}^{+3.87}\cdot 10^{-8}$  &  $6.18_{-1.50}^{+4.87}\cdot 10^{-8}$ \\
45-46  &  4.0  &  $7.14_{-3.20}^{+5.96}\cdot 10^{-8}$  &  $1.96_{-1.13}^{+3.01}\cdot 10^{-8}$  &  $3.55_{-1.65}^{+3.66}\cdot 10^{-8}$ \\
\hline
\end{tabular}
\caption{Numerical values for the derived XLFs: total ($\phi_a$), unobscured ($\log N_{\rm H}/\mathrm {cm}^{-2}<22$, $\phi_{unobs}$), and obscured ($22\leq \log N_{\rm H}/\mathrm{cm}^{-2}<24$, $\phi_{obs}$), for different luminosity and redshift bins. Uncertainties are reported at 3$\sigma$ confidence level on the kernel density estimation.}
\label{tab:xlf_numbers}
\end{table*}

\subsection{Black hole accretion density}\label{sec:bhad}
AGN evolution can be further analyzed by computing the black hole accretion density (BHAD, or $\Psi_{\rm BHAD}$), which encapsulates the SMBH growth history (e.g., \citealt{soltan82}). The BHAD is defined as
\begin{equation}\label{eq:bhad}
    \Psi_{\rm BHAD}(z) = \int \frac{(1-\epsilon)}{\epsilon c^2} L_{\rm bol} \phi (L_{\rm bol},z) d\log L_{\rm bol}\,\, \mathrm{[cgs]} ,
\end{equation}
where $\epsilon$ is the radiative efficiency (here assumed to be $\epsilon=0.1$; e.g., \citealt{hopkins07,vito18}), $L_{\rm bol}$ is the bolometric luminosity, and $\phi(L_{\rm bol},z)$ is the bolometric luminosity function at a given redshift.
We derived $L_{\rm bol}$ and $\phi(L_{\rm bol},z)$ from their X-ray analogues by applying the bolometric correction 
found by \citet{duras20}. Due to the large scatter in 
these values
(e.g., \citealt{lusso12, duras20}), we also applied the bolometric correction derived by \citet{hopkins07}, which is higher in the range $\log L_{\rm X}/\mathrm{erg\,\, s}^{-1}\sim$42-45, but lower at higher luminosities. The 
differences in these bolometric corrections
dominate the uncertainties in the BHAD.
Equation (\ref{eq:bhad}) is ideally integrated 
over all luminosities 
(e.g., \citealt{delvecchio14}). Practically, we selected the $\log L_{\rm bol}$ range corresponding to $\log L_{\rm X}/\mathrm{erg\,\, s}^{-1}$=42-47.
Since our derived XLF do not cover entirely this range, we extrapolated our results as follows. Motivated by the fact that the main contribution to the AGN accretion rate is produced by the break luminosity of the XLF, which we constrained at all redshifts, and that XLF models have typically double power-law shapes (e.g., \citealt{vito14}, \citetalias{aird15}), we extended our XLF by maintaining constant the average slope before and after the break luminosity. 
Since at $z\gtrsim3$ our XLF is poorly constrained at low luminosities, we assumed the \citetalias{aird15} slope, whose XLF best overlaps with ours (see Figure \ref{fig:XLF}).

\begin{figure*}[!t]
    \center 
    \includegraphics[scale=0.5]{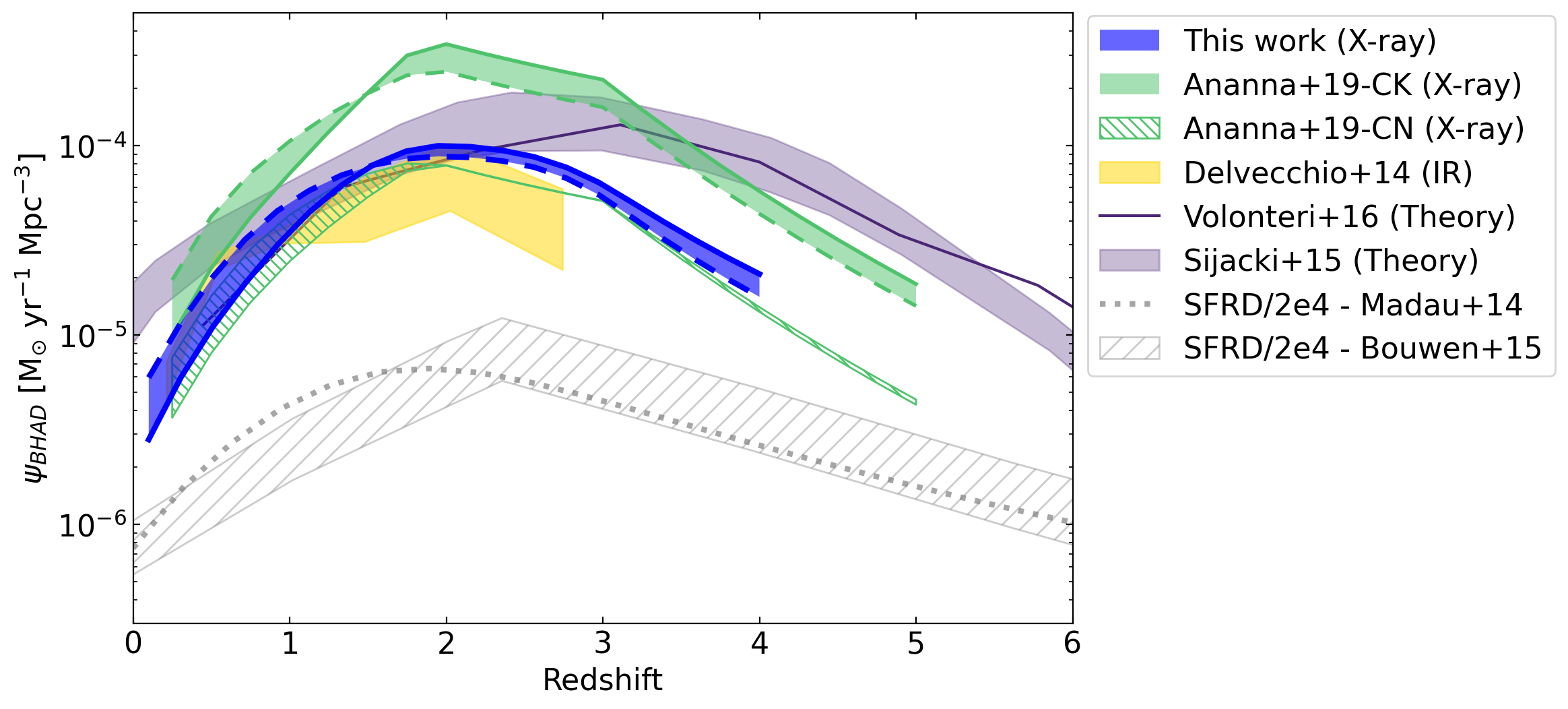}
    \caption{Evolution of the black hole accretion density (BHAD) derived in this work (\textit{blue curves}), compared with \citetalias{ananna19} (\textit{green curves}, with and without Compton-thick AGN), and the theoretical works of \citet{sijacki15} (\textit{shaded purple}) and \citet{volonteri16} (\textit{purple line}). The continuous and dashed lines correspond to the BHADs derived with the bolometric corrections of \citet{duras20} and \citep{hopkins07}, respectively.
    The \citet{delvecchio14} BHAD, obtained from infrared data, is also shown (\textit{shaded yellow}). For comparison, we added the \citet{madau14} and \citet{bouwens15} star formation rate densities (scaled by a factor $2\times 10^4$ for clarity, \textit{grey dotted line} and \textit{grey dashed area}, respectively).} \label{fig:bhad}
\end{figure*}

Our derived BHAD is shown in Figure \ref{fig:bhad}, along with several other estimates for comparison. 
First, we show the BHAD derived from the \citetalias{ananna19} XLF because it was based on the most comprehensive data available, including hard X-ray surveys from \textit{NuSTAR} and \textit{Swift}-BAT, which are sensitive up to $\sim195$ keV and thus identify highly obscured sources that otherwise could be missed at lower energies (e.g., \citealt{gilli07,treister09,ballantyne11}). Moreover, the \citetalias{ananna19} model simultaneously reproduces number counts from the largest surveys over different areas and depths, as well as constraints from the cosmic X-ray background, and is therefore the most up-to-date population synthesis model for AGN. In any case, due to the similarities between our XLF and the other X-ray XLFs, their BHADs are similar to ours.
Considering Compton-thin AGN only (\textit{hatched green}), the BHAD derived from \citetalias{ananna19} agrees well with our curve (\textit{solid blue}). The complete \citetalias{ananna19} BHAD (\textit{solid green}) is a factor $\sim$2-3 higher, as expected because of the substantial number of Compton-thick AGN, which are not well sampled in our survey nor other samples (e.g., \citetalias{ueda14}, \citetalias{aird15}).

The yellow region in Figure \ref{fig:bhad} was derived by \citet{delvecchio14} from \textit{Herschel} infrared observations of AGN, using decomposition of  UV-optical-infrared spectral energy distributions (SEDs) to isolate the emission from black hole accretion. However, this procedure may miss some AGN when the SED is dominated by stellar emission (e.g., \citealt{delvecchio17}), especially if the AGN radiation is buried behind obscuring material (e.g., \citealt{delmoro13, hatcher21}).
Hence, it is not surprising that the \citet{delvecchio14} infrared curve lies near or below our curve;
indeed, this comparison suggests that the SED-decomposition approach could miss the majority of Compton-thick AGN, or at least a significant fraction of them.

Theoretical simulations of black hole growth from \citet{sijacki15} and \citet{volonteri16} are also shown in Figure \ref{fig:bhad} (\textit{purple curves}). 
These models are similar to the \citetalias{ananna19} curve, although slightly different in shape, and they lie well above our curve, particularly for $z\gtrsim2$.
The discrepancy between X-rays and theoretical simulations is known in the literature (e.g., \citealt{vito18, barchiesi21}), and it is generally explained with the challenges in efficiently detecting heavily obscured, Compton-thick sources in X-ray surveys (e.g., \citealt{hickox18}).
These tensions suggest that Compton-thick AGN may have a substantial role in the supermassive black holes accretion history, especially at high redshift, even though their contribution is still not well understood. 

We also compare our results with the star formation rate densities (SFRD) found by \citet{madau14} and \citet{bouwens15} (in \textit{grey},  scaled down by a factor $2\times10^4$ to avoid confusion with the other curves). As for the BHAD, there is a peak around redshift $z\sim2$. The similar evolution of the BHAD and SFRD is considered evidence in favor of the SMBH host galaxy co-evolution scenario (e.g., see \citealt{vito18} or \citealt{hickox18} for a review). In detail, however, these quantities seem to have a different evolution, especially at high redshifts. In particular, the slope of the BHAD at $z>2$ is steeper than that of the SFRD, implying that black holes grow faster, on shorter time scales, compared to their host galaxies
(e.g., \citetalias{aird15}).
However, the large uncertainties in deriving the BHAD and the SFRD from different observations, simulations, and methods, leave the discussion open.

\subsection{On the evolutionary picture} 
Our analysis shows how the total AGN space density evolves over cosmic time, and how its shape depends on the interplay between unobscured and obscured AGN. 
We now discuss models that might explain the observed trends with luminosity, obscuration, and redshift.

One of the most invoked models to explain the decreasing obscured fraction with increasing X-ray luminosity is the receding torus model (e.g., \citealt{lawrence91,simpson05}). In this model, a decrease in the covering factor of the dusty torus (e.g., \citealt{buchner15,matt19}) is caused by increasing radiation pressure, photoionization of gas clouds, dust sublimation, or some combination thereof (e.g., \citealt{honig07,akylas08, mateos17}). The receding torus model therefore links the distribution of line-of-sight column densities and intrinsic luminosity. 
However, it is unclear how to connect this model to the redshift evolution, since it predicts less obscuration at higher redshifts, where AGN are on average more luminous, whereas the opposite trend is observed.

Another common picture is that luminous AGN are triggered by mergers (e.g., \citealt{sanders88,treister12}), which supply gas to the black hole (\citealt{hopkins06,somerville08}) and, at least initially, produce high levels of obscuration (e.g., \citealt{sanders88,blecha18}).
During this stage, the SMBHs grow close to the Eddington limit, generating a combination of radiation and kinetic pressure capable of blowing out gas and dust around the central engine, leading to a largely unobscured phase (\citealt{sanders88}, \citealt{treister10}) during which the AGN shines 
until the accretion processes consumes the gas reservoir (e.g., \citealt{hopkins08}).
It is therefore likely that high-luminosity AGN generate efficient feedback mechanisms capable of depleting and consuming the gas supply (e.g., \citetalias{buchner15}), linking the X-ray luminosity to the accretion processes (e.g., \citealt{hopkins07}, \citetalias{ueda14}).
In this scenario, one expects the fraction of unobscured sources to increase with luminosity, in any redshift bin, as observed (Fig. \ref{fig:XLF_sep}). 
At the same time, for fixed luminosity, the fraction of unobscured sources should decrease with increasing redshift because mergers, and obscuration in general (e.g., \citealt{carilli13}), become more common at high redshift (e.g., \citealt{ricci17_2}).
Both trends are observed in our results (bottom panel of Fig. \ref{fig:XLF_downsizing}).

\citet{hopkins05} proposed an interpretation of the AGN luminosity function in which the bright end corresponds to AGN at the maximum of their accretion history, while the faint end traces AGN during small accretion events either at the beginning or the end of their activity peak.
The bulk of AGN accretion is then produced by the ``knee" of the LF, where the density of AGN at their activity peak is maximized.
Both the XLF density evolution and the BHAD derived in this work show a peak at $z\sim2$.
At these redshifts, obscured AGN dominate for $\log L_{\rm X}/\mathrm{erg\,\, s}^{-1}>43$ (Fig. \ref{fig:XLF_sep}), meaning the majority of black hole accretion occurs in an obscured stage. 

\subsection{Comparison to Other Works}\label{sec:comparison}
In this work, we analyzed the X-ray spectra of AGN from the S82X sample, then derived the intrinsic distributions and luminosity functions of X-ray-selected AGN by correcting
for observational biases (Section \ref{sec:obs_evol:corr}). In particular, we used simulations to correct for Malmquist bias (loss of faint sources at the flux limit), sensitivity bias (from the range of spectral shapes), and Eddington bias (statistical fluctuations near the flux limit). 
We did not assume any functional form for the XLF, $N_{\rm H}$ distributions, or X-ray spectra, allowing these to be dictated by the data (and bias corrections).
Because we used only observations from the S82X survey, we avoided the challenge of correcting for different survey sensitivities.

We were able to measure the evolving XLF and place strong constraints on the obscured fraction up to $z=4$ because of the many S82X AGN with $z>3$ and $\log L_{\rm X}/\mathrm{erg\,\, s}^{-1} >45.5$. 
Previous results based on smaller, deeper surveys, which lack sources in this redshift and luminosity regime, are usually extrapolated or less well constrained. 
Our results agree well with those of \citetalias{ueda14}, \citetalias{buchner15}, and \citetalias{ananna19}; \citetalias{aird15} predicts more obscured AGN at $z\lesssim1$, and with different trends at $\log L_{\rm X}/\mathrm{erg\,\, s}^{-1}$ 43-44, but is consistent at all other redshifts and luminosities. It is worth noticing that \citealt{georgakakis17} derived a slightly lower fraction of obscured AGN at $z\sim1$ and $\log L_{\rm X}/\mathrm{erg\,\, s}^{-1}$ 44-45 in XMM-XLL, but consistent with ours within the errors.
Moreover, our derived obscured AGN fractions are consistent, within the uncertainties, also with \citealt{treister06} and \citealt{hasinger08}, which derived the obscured fraction of AGN relying on optical classification and using a combination of optical spectra and X-ray photometry, respectively.

The effects of our non-parametric approach can be seen in a less evident ``knee" in the XLF and by a slope that can change with redshift and luminosity, instead of having a fixed number of slopes dictated by the model (Figure \ref{fig:XLF}). 
We also have sufficient obscured AGN to determine the XLF independently for Compton-thin and unobscured AGN separately (Fig. \ref{fig:XLF_sep}).
Other XLFs at high redshift ($z>3$) and high luminosity ($\log L_{\rm X}/\mathrm{erg\,\, s}^{-1} > 45.5$) have usually been obtained using mainly unobscured sources, 
using assumptions or extrapolations to derive information on the underlying obscured population 
(e.g., \citealt{hickox18}). Commonly adopted solutions are: i) extrapolating local obscured AGN distributions up to high redshift (e.g., \citetalias{ueda14}), ii) using only a few objects to constrain the obscured fractions (e.g., \citetalias{buchner15}), or iii) including additional information such as constraints from the X-ray background 
(e.g., \citealt{gilli07}, \citetalias{ananna19}). 
However, these approaches generally fail, if taken individually, 
to satisfy all current available data (see comparisons by \citetalias{ananna19}). 

\subsection{Compton-thick AGN}\label{sec:c_thick}
\begin{figure}[!t]
    \center 
    \includegraphics[scale=0.47]{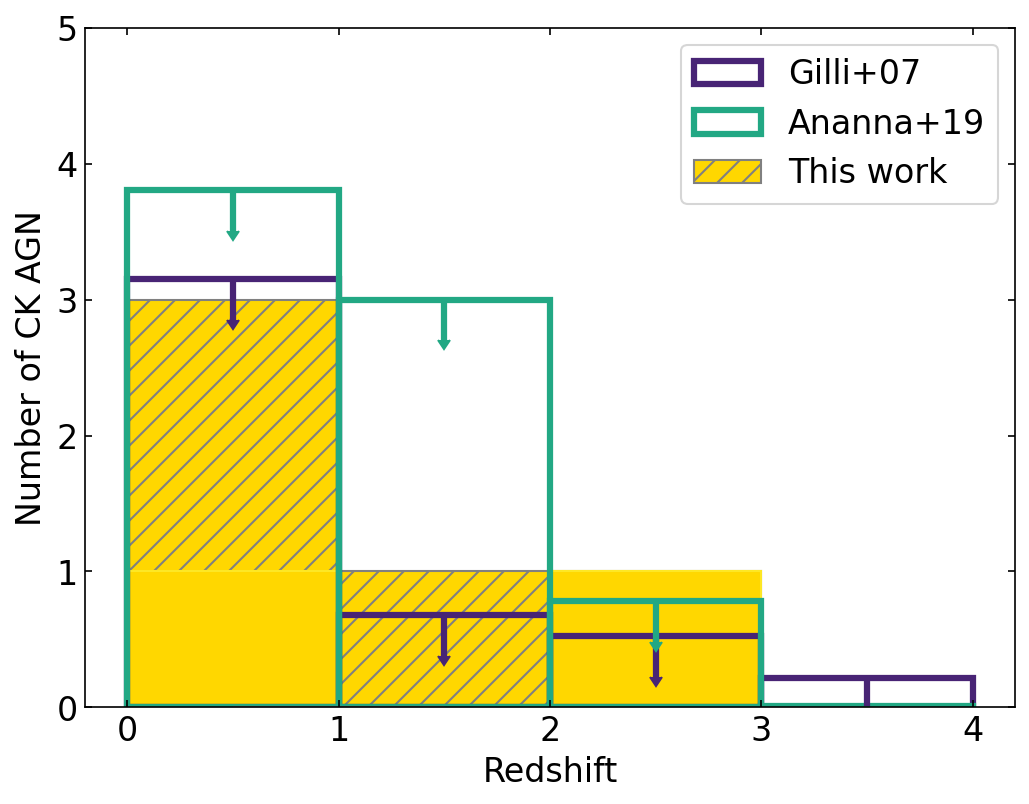}
    \caption{The number of Compton-thick AGN ($\log N_{\rm H}/\mathrm {cm}^{-2}=$ 24-26) identified in this work (\textit{gold}) is in good agreement with the predicted number from \citealt{gilli07} (\textit{purple line}) and \citealt{ananna19} (\textit{green line}), albeit with limited statistics. {\it Hatched regions} correspond to $N_{\rm H}$ upper limits above $\log N_{\rm H}/\mathrm {cm}^{-2}=$ 24. Model predictions are upper limits because of our quality cut on the photometric redshifts.}
    \label{fig:cthick_prediction}
\end{figure}

We were able to place strong constraints on unobscured and Compton-thin AGN, but the role of highly obscured, Compton-thick AGN is difficult to analyze with the S82X sample alone (see also Section \ref{sec:bhad}).
We found just two Compton-thick objects plus three other candidates (which have $N_{\rm H}$ upper limits above $10^{24}$ cm$^{-2}$).
Figure \ref{fig:cthick_prediction} compares these numbers with the number of expected Compton-thick AGN from the \citet{gilli07} and \citetalias{ananna19} models, taking into account the S82X limiting flux and the cuts we applied on the number of counts, but not the selection in photometric redshifts. In particular, the latter may be crucial in recovering the full number of these highly obscured sources, since they are usually faint and difficult to detect in optical/infrared bands. 
Thus, these predictions have to be considered as upper limits. 

The total number of predicted Compton-thick AGN in the redshift range 0-4 is $\sim$4.5 and $\sim$7.6 for the two models, respectively. Given the limited statistics, these numbers are consistent with what we found, although three of five sources have only $N_{\rm H}$ upper limits, and thus their interpretation as Compton-thick sources is uncertain. 
This result suggests that to investigate the intrinsically luminous, Compton-thick population we need much larger numbers of AGN, either from wider areas or deeper observations. This is especially true at $z\gtrsim2$, where the number of 
detected highly obscured AGN should drop dramatically due to their faintness (e.g., \citealt{treister04}).

\section{Future prospects} \label{sec:future_perspectives}

Population studies with high luminosity and high redshift AGN strongly depend on the number of objects available in such extreme regimes. In this work, we used the available S82X data from \citetalias{lamassa16} and \citealt{lamassa19_2}, which allowed us to investigate AGN up to $z=4$ and $\log L_{\rm X}/\mathrm{erg\,\, s}^{-1}= 46$ for both unobscured and obscured (Compton-thin) AGN populations. To overcome these limits, larger samples are needed.
Since \citetalias{lamassa16}, there are new \xmm and \chandra archival observations in the Stripe 82 field, such that the total non-overlapping area reaches $\sim 50$ deg$^2$. Taking into account also overlapping observations, we estimate to almost double the current number of objects and to increase the overall depth by a factor of $\sim2$. This will be a significant improvement over our current results, especially for the Compton-thick AGN population. 
We note that the study of obscured AGN populations in particular will benefit from the \textit{Herschel} and \textit{Spitzer} infrared data available in the S82X, which helps determine photometric redshifts (e.g., \citealt{salvato09, ananna17, peca21}) and improves analysis of spectral energy distributions (Auge et al., in prep.).
Moreover, we are continuing spectroscopic follow-up of S82X sources (\citealt{lamassa19_2}), meaning that the number of spectroscopic redshifts will grow in the next years, adding more sources to the X-ray spectral analysis sample.

Another approach is to include other X-ray surveys in the analysis (e.g., \citetalias{ueda14,aird15,buchner15,ananna19}).
Deep X-ray surveys can reveal heavily obscured sources, but typically cover very small volumes, and thus are not sensitive to rare objects at high luminosity (e.g., \citealt{marchesi16b,liu17}). 
Even the XMM-XXL field (XXL North, $\sim 25$ deg$^2$) has only 5 objects at $\log L_{\rm X}/\mathrm{erg\,\, s}^{-1}>45.5$, all of them classified as broad-line AGN from optical spectra (\citealt{liu16}), and S82X AGN ($\sim 31$ deg$^2$) do not exceed $\log L_{\rm X}/\mathrm{erg\,\, s}^{-1}=46$.
Much larger surveys, such as
eFEDS (\citealt{brunner22}) and the forthcoming all-sky \textit{eROSITA} survey (\citealt{merloni13}), will be needed to detect rare and intrinsically luminous objects. 
Even then, \textit{eROSITA} has less effective area than {\it XMM-Newton} above 3 keV, and higher background than \chandra, so it is relatively less sensitive to obscured AGN.
For example, only 245 sources---$\sim 1$\% of the current eFEDS catalog---
are detected in the hard band (2.3-5 keV). 
Of course, since the \textit{eROSITA} final all-sky catalog (eRASS:8, \citealt{predehl21}) will have millions of extragalactic sources, even the obscured ones will be available in large numbers, enough to allow population studies. However, the corrections for obscured AGN missed by the soft-band selection will be large. 

Moreover, data above 10 keV are fundamental to detecting heavily obscured sources and Compton-thick AGN 
(e.g., \citealt{ricci17}; \citetalias{ananna19}; \citealt{marchesi19, koss22}). The combination of these very hard-band X-ray observations and simulations (e.g., \citealt{balokovic21}) to correct for observational biases is crucial to constrain the fraction of Compton-thick AGN and, therefore, to shed light on the black hole accretion in the young Universe and on the onset of the black hole-galaxy co-evolution.

Future planned X-ray telescopes will certainly improve our current knowledge of AGN populations. For example, the planned \textit{AXIS} mission (\citealt{mushotzky19}) is designed to have an effective area of roughly 7 (2) and 25 (5) times those of the current \xmm and \chandra capabilities at 1 (6) keV, respectively, with 
sub-arcsecond angular resolution over a $24\arcmin \times 24\arcmin$ field of view. This will be an improvement of a factor of $\sim$100 with respect to \chandra, whose PSF degrades rapidly outside $2\arcmin$.
\citet{marchesi20} predicted that \textit{AXIS} would detect more than 200,000 AGN in the 0.5-7 keV band, including $>7000$ at $z>3$ and tens at $z>6$, from a combination of deep, intermediate, and wide surveys.
This is just one example of possible next-generation telescopes---like \textit{Athena} \citep{barret19} 
and \textit{Lynx} \citep{gaskin19}---
which will be essential to better constraining
the obscured fraction and XLF at the highest redshifts and luminosities, and to unveiling a new window for AGN population studies in the Universe up to $z\sim8$ (\citealt{marchesi20}).

\section{Summary} \label{sec:summary}

We exploited the full extent of X-ray and multiband data in the S82X field by analyzing the X-ray spectra of the 2937 sources with a solid redshift estimate (\citetalias{lamassa16}, \citealt{lamassa19_2}).
Using simulations, we established thresholds for the minimum number of detected counts needed to get a reliable fit.
We considered simple power-law models as well as more complex shapes, including soft-excess and reflection components. The best fits were then identified using the AIC criterion.
    
We derived the evolving AGN X-ray luminosity function, correcting for observational biases through extensive simulations, and without assuming any particular functional form. In addition to the total XLF, we derived separate XLFs for obscured and unobscured AGN populations, and we computed the obscured fraction of AGN as a function of both redshift and luminosity.
S82X AGN with high luminosities ($L_{\rm X} > 45.5$ erg s$^{-1}$) and redshifts ($z>3$) add new statistical weight not available from smaller volume surveys.

Our XLF shows a ``knee" imposed by the data, which represents where AGN dominate the luminosity at a particular redshift.
The shape of the total XLF exhibits both luminosity and density evolution: AGN are more luminous at higher redshift and, for fixed luminosity, have higher densities at lower redshift. The unobscured and obscured XLFs, whose combination constitutes the total XLF in changing contributions, reveal that obscured AGN dominate at all luminosities at $z>2$.
The fraction of  AGN that are obscured increases with redshift, and decreases with increasing luminosity.

We used the XLF to compute the black hole accretion density as a function of redshift. Although our BHAD has a similar shape compared to other works based on Compton-thin AGN, it lies below theoretical predictions.
Including Compton-thick AGN roughly doubles the emission and largely resolves the disagreement with theory, 
even if the slightly different shapes of the curves remain to be explained.
This result suggests that Compton-thick AGN have an essential role in the growth of supermassive black holes, and that their contribution exceeds previous estimates (e.g., \citetalias{ueda14,aird15}).
The BHAD for X-ray-selected AGN lies above estimates from far-infrared-selected AGN, implying that, at least with present flux limits, infrared surveys may miss Compton-thick AGN.

The patterns found for the XLFs, obscured fractions, and BHAD 
confirm the cosmic downsizing of black hole growth.  
The similar evolution of the black hole and star formation densities supports the idea of co-evolution between SMBH and host-galaxy, but suggesting different time scales during their evolution.


Current and upcoming X-ray surveys with larger volumes will increase the statistics of AGN at high luminosity, high obscuration, and high redshift beyond the work presented here.
This includes both  future surveys with {\it eROSITA} and proposed X-ray observatories,
and newer archival X-ray observations in S82X, which have the advantage of invaluable ancillary multiwavelength data, especially in the infrared.

\acknowledgments
We acknowledge the anonymous referee for the valuable comments that improved the quality of the paper.
AP and NC acknowledge
Chandra X-ray Observatory grant AR2-23010X. CMU acknowledges support from National Aeronautics and Space Administration via ADAP grant 80NSSC18K0418.
ET acknowledges support from FONDECYT Regular 1190818 and 1200495, ANID grants CATA-Basal AFB-170002, ACE210002, and FB210003, and Millennium Nucleus NCN19\_058.
MB acknowledges support from the YCAA Prize Postdoctoral Fellowship.
AP acknowledges D. Costanzo for all the support over the years.

\,


\software{PyXspec \citep{pyxspec}, 
          XSPEC \citep{xspec},
          CIAO \citep{ciao_paper},
          SAS \citep{SAS},
          astropy \citep{astropy},
          Matplotlib \citep{matplotlib},
          TOPCAT \citep{topcat},
          kdeLF \citep{yuan22}.
          }

\appendix

\section{Background modeling} \label{app:bkg}
\begin{figure*}[!tp]
    \center 
    \includegraphics[scale=0.33]{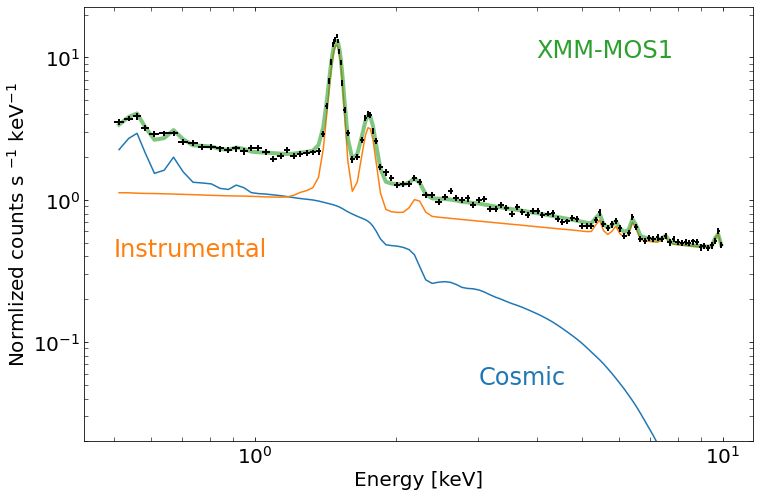}
    \includegraphics[scale=0.33]{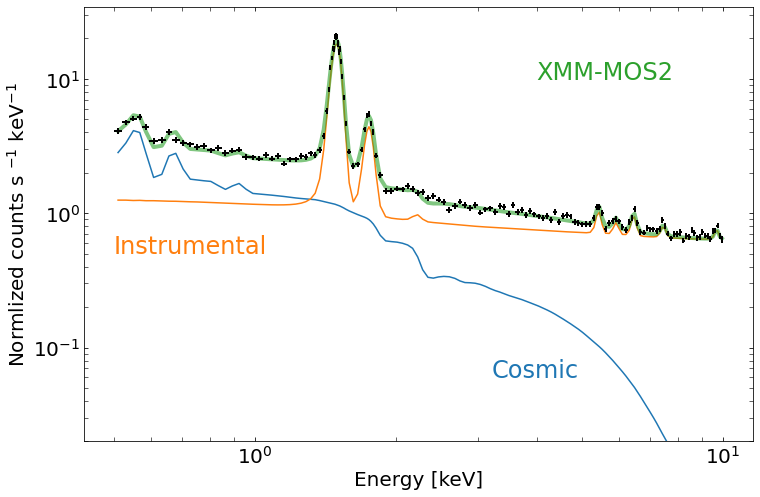}
    \includegraphics[scale=0.33]{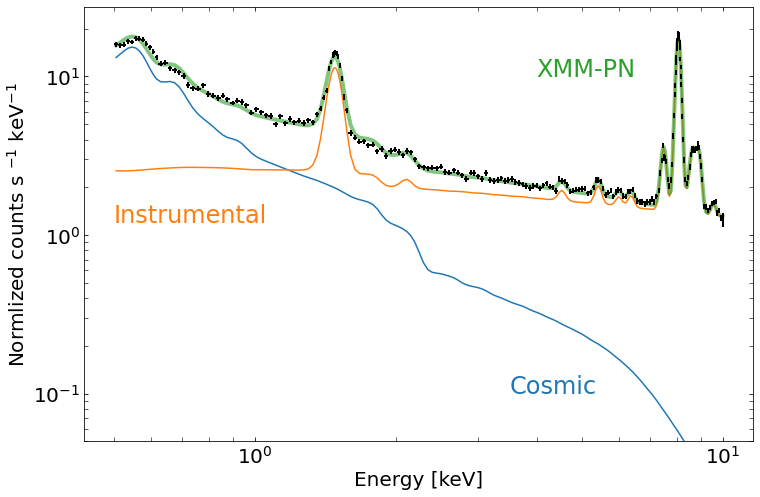}
    \includegraphics[scale=0.33]{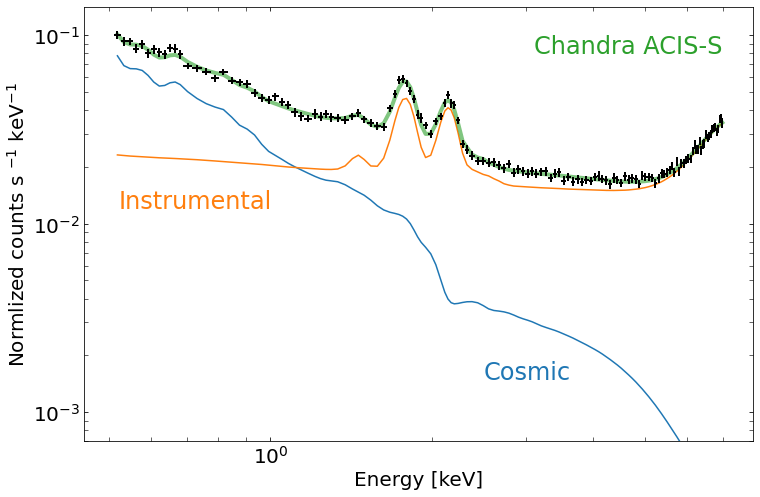}
    \includegraphics[scale=0.33]{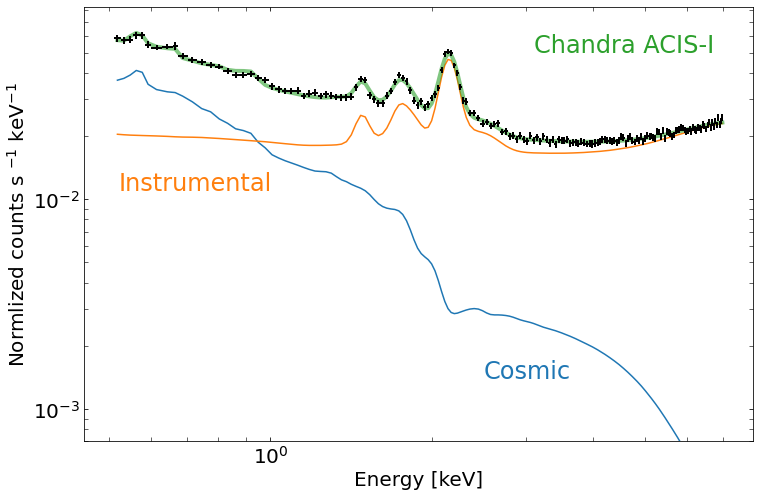}
    \caption{Background spectra used in the background modeling procedure. From the \textit{top-left panel}: \xmm MOS1, MOS2, PN, \chandra ACIS-S, and ACIS-I. In each panel, the best-fit model (\textit{thick green line}) was obtained with a cosmic (\textit{blue line}) plus an instrumental (\textit{orange line}) components, as discussed in the text.}
    \label{fig:bkg_spectra}
\end{figure*}

In this section, we describe the models adopted to fit the backgrounds in the \xmm and \chandra data.
The observed X-ray background is composed of two main components: cosmic and instrumental (blue and orange curves in Fig. \ref{fig:bkg_spectra}, respectively). The cosmic background 
is produced by the sum of Galactic and extragalactic components: 
\begin{itemize}
	\item Two thermal components (XSPEC model \textsc{apec}; e.g., \citealt{lanzuisi15}) representing a local hot bubble ($kT\sim0.04$ keV) and interstellar gas in the Galactic disk ($kT\sim0.12$ keV) 
	\citep{kuntz00};
	\item The extragalactic X-ray background produced by unresolved emission from discrete sources (e.g., \citealt{comastri95}), which can be modeled with a $\Gamma \sim 1.4$ power-law, modified by Galactic absorption (e.g., \citealt{deluca04}).
\end{itemize} 

The instrumental background is the sum of three main components:
\begin{itemize}
	\item Several emission lines produced by the telescope instrumentation, which we modeled with a combination of Gaussian lines (see Table \ref{tab:background});
	\item Residuals from the filtering of quiescent soft protons (QSP, e.g., \citealt{baldi12});
	\item The cosmic ray-induced continuum (NXB, \citealt{leccardi08}).
\end{itemize}
The QSP and NXB were modeled with two broken power-laws for \xmm (e.g., \citealt{leccardi08}), and with a single broken power-law plus a broad Gaussian line above $\sim$ 5 keV for \chandra. For \chandra only, we also added a broad Gaussian line between 1 and 3 keV (\citealt{fiore12}), which may be a mother-daughter artifact produced during the Charge Transfer Inefficiency (CTI) correction (\citealt{bartalucci14}). Table \ref{tab:background} summarizes these components. 

It is worth noticing that for both \xmm and \chandra, we repeated the above modeling using backgrounds extracted from two different epochs: from the oldest observations to 2008, and from 2010 to the most recent, to check for possible differences due to the effective area degradation of the telescopes. Once the background best-fit was obtained, we fitted a sub-sample of 1000 random sources using the derived backgrounds. Since we found that the derived fluxes, $N_{\rm H}$, and luminosities were consistent within each other, we modeled the background using the set of observations described in Section \ref{sec:fittingprocedure} .

\begin{table*}[!t]
\hspace{-2cm}
\footnotesize
\begin{tabular}{|c|c|c|c|c|c|c|}
\hline
\multirow{2}{*}{Instrument} & Component             & \multirow{2}{*}{Best-fit} & Component             & \multirow{2}{*}{Best-fit} & Component             & Best-fit                                                              \\
                                          & (Model)                    &                                       &  (Model)                   &                                      & (Model)                   &          (keV)                                                                            \\ \hline
\multirow{3}{*}{MOS1}       & \multirow{11}{*}{QSP}                   & \multirow{11}{*}{$E_b = 5$ keV}           & \multirow{11}{*}{NXB}     & $E_b=7$ keV            & \multirow{11}{*}{Em. lines}             &              \\
                            & \multirow{11}{*}{\textsc{(brkpwl})}          & \multirow{11}{*}{$\Gamma_1=0.4$}           & \multirow{11}{*}{(\textsc{brkpwl})}              & $\Gamma_1=0.22$          & \multirow{11}{*}{(\textsc{Gauss})}       &  1.49 (Al K$\alpha$), 1.74 (Si K$\alpha$),        \\
                            &                       &  \multirow{11}{*}{$\Gamma_2=0.8$}           &                       & $\Gamma_2=0.05$           &                                                                                                        & 2.21 (Au M$\alpha$, M$\beta$), 5.42 (Cr K$\alpha$),     \\ \cline{1-1} \cline{5-5}
\multirow{3}{*}{MOS2}       &                       & \multirow{8}{*}{}         &                       & $E_b=3$ keV              &                                                                                                                                  &  5.90 (Mn K$\alpha$), 6.40 (Fe K$\alpha$),       \\
                            &                       &                           &                       & $\Gamma_1=0.32$         &                              &               7,48 (Ni K$\alpha$), 9.7 (Au L$\alpha$)                                                                        \\
                            &                       &                           &                       & $\Gamma_2=0.2$            &                       &                                                                                      \\ \cline{1-1} \cline{5-5} \cline{7-7} 
\multirow{5}{*}{PN}         &                       &                           &                       & \multirow{4}{*}{$E_b=2$ keV}              &                       & 1.49 (Al K$\alpha$), 2.10 (Au M$\alpha$),                                            \\
                            &                       &                           &                       & \multirow{4}{*}{$\Gamma_1=0.8$}           &                       & 4.51 (Ti K$\alpha$), 5.42 (Cr K$\alpha$),                                            \\
                            &                       &                           &                       & \multirow{4}{*}{$\Gamma_2=0.4$}           &                       & 5.90 (Mn K$\alpha$), 6.40 (Fe K$\alpha$),                                            \\
                            &                       &                           &                       &                           &                       & 7.48 (Ni K$\alpha$), 8.04 (Cu K$\alpha$),                                            \\
                            &                       &                           &                       &                           &                       & 8.63 (Zn K$\alpha$), 8.9 (Zn K$\alpha$),                                             \\
\multicolumn{1}{|l|}{}      & \multicolumn{1}{l|}{} & \multicolumn{1}{l|}{}     & \multicolumn{1}{l|}{} & \multicolumn{1}{l|}{}     & \multicolumn{1}{l|}{} & 9.57 (Zn K$\beta$)                                                                   \\ \hline

\multicolumn{5}{c}{} \\ \hline
\multirow{2}{*}{Instrument} &      \multicolumn{2}{c|}{Component}             & \multicolumn{2}{|c|}{\multirow{2}{*}{Best-fit}}      & Component     & Best-fit                                                       \\
                                           &      \multicolumn{2}{c|}{(Model)}                    & \multicolumn{2}{|c|} {}                                        & (model)            &  (keV)                                                                                         \\ \hline
\multirow{3}{*}{ACIS-I}        &     \multicolumn{2}{c|}{}                                & \multicolumn{2}{c|}{$E_b=3$ keV}                     &    \multirow{7}{*}{Em. lines}                      &                           \\
\multicolumn{1}{|l|}{}           & \multicolumn{2}{c|}{}                                    & \multicolumn{2}{c|}{$\Gamma_1=0.22$}            &     \multirow{7}{*}{(\textsc{Gauss})}                    &  \\
\multicolumn{1}{|l|}{}           & \multicolumn{2}{l|}{}                                      & \multicolumn{2}{c|}{$\Gamma_2=0.18$}           &                         & \multicolumn{1}{l|}{}                                                                \\
\multicolumn{1}{|l|}{}           & \multicolumn{2}{c|}{QSP + NXB}                   & \multicolumn{2}{c|}{$E_{line}=7.5$ keV}          &                         & 1.49 (Al K$\alpha$), 1.74 (Si K$\alpha$),                                                              \\ \cline{1-1} \cline{4-5}
\multirow{3}{*}{ACIS-S}      & \multicolumn{2}{c|}{(\textsc{brkpwl + Gauss})}          & \multicolumn{2}{c|}{$E_b=3$ keV}                    &                         & \multicolumn{1}{l|}{2.15 (Au M$\alpha$, M$\beta$), 1-3 (Au-Mg)$^*$}                                                               \\
                                           & \multicolumn{2}{c|}{}                                    & \multicolumn{2}{c|}{$\Gamma_1=0.20$}            &                         & \multicolumn{1}{c|}{}                                                                 \\
                                           & \multicolumn{2}{c|}{}                                    & \multicolumn{2}{c|}{$\Gamma_2=0.18$}            &                         & \multicolumn{1}{c|}{}                                                                 \\
                                           & \multicolumn{2}{c|}{}                                    & \multicolumn{2}{c|}{$E_{line}=7.5$ keV}            &                         & \multicolumn{1}{c|}{}                                                                 \\ \hline
\end{tabular}
\label{tab:bkgmodel}
\caption{Best-fit parameters for the instrumental background component. From the left: instrument; background component (used model), and best-fit parameters. The assumed models were: broken power-law (\textsc{brkpwl}), where $E_b$ is the break energy and $\Gamma_1$ and $\Gamma_2$ the two slopes; and Gaussian lines (\textsc{Gauss}), whose energies were frozen at their nominal values (\citealt{freyberg04,leccardi08,fiore12}). $^*$Mother-daughter system (see \citealt{bartalucci14} for details).}
\label{tab:background}
\end{table*}

\section{Column description} \label{app:master_cat}
The results from the spectral analysis are included in the catalog released with this paper. The details of the table columns are given below.
\begin{itemize}
    \item[-] \textit{[1] Source ID}: Source ID from \citetalias{lamassa16}.
    \item[-] \textit{[2] Obs ID}: Observation ID from \citetalias{lamassa16}.
    \item[-] \textit{[3] Src Exp}: Effective exposure time in seconds. If the source was observed by multiple telescopes, it is the sum of the effective times.
    \item[-] \textit{[4-12] Counts}: Source counts in the full, soft, and hard bands, respectively. Uncertainties are computed according to \cite{gehrels86} at the 1$\sigma$ confidence level.
    \item[-] \textit{[13-15] Hardness ratio}: defined as $HR=\frac{H-S}{H+S}$, where $H$ and $S$ are the count rates in the hard and soft bands, respectively, computed with the BEHR tool \citep{park06}. Uncertainties are at the 1$\sigma$ confidence level.
    \item[-] \textit{[16-24] Flux}: Fluxes in the full, soft, and hard bands, respectively, in units of erg s$^{-1}$ cm$^{-2}$, derived from the spectral analysis. Errors are at the 90\% confidence level.
    \item[-] \textit{[25-33] Obs Lum}: Observed X-ray luminosity, in erg s$^{-1}$, in the full, soft, and hard bands, respectively, derived from the spectral analysis. Errors are at the 90\% confidence level.
    \item[-] \textit{[34-42] Lum}: Intrinsic (rest-frame) and de-absorbed X-ray luminosity [erg s$^{-1}$] in the full, soft, and hard band, respectively, derived from the spectral analysis. Errors are at the 90\% confidence level.
    \item[-] \textit{[43-45] NH}: Obscuring column density, in units of cm$^{-2}$, derived from the spectral analysis. Errors are at the 90\% confidence level.
    \item[-] \textit{[46-49] Gamma}: Photon index, $\Gamma$, derived from the spectral analysis. Errors are at the 90\% confidence level.
    \item[-] \textit{[49] Scattering fraction}: Ratio between the secondary and the primary power-law components, derived for double power-law models.
    \item[-] \textit{[50] Redshift}: Best redshift from \citetalias{lamassa16} and \citealt{lamassa19_2}: $z_{spec}$ if available, $z_{phot}$ otherwise.
    \item[-] \textit{[51] Redshift flag}: ``1" for spectroscopic redshifts, and ``2" for photometric redshifts.
    \item[-] \textit{[52] Model}: Best-fit model used for deriving the results: ``1" for single power-law (M1), ``2" for single power-law plus reflection (M2), ``3" for double power-law (M3), and ``4" for double power-law plus reflection (M4). See details in the text. 
\end{itemize}

\section{Flux and Luminosity comparison with L16} \label{app:fluxcomparison}
\begin{figure}[!tp]
    \center 
    \includegraphics[scale=0.45]{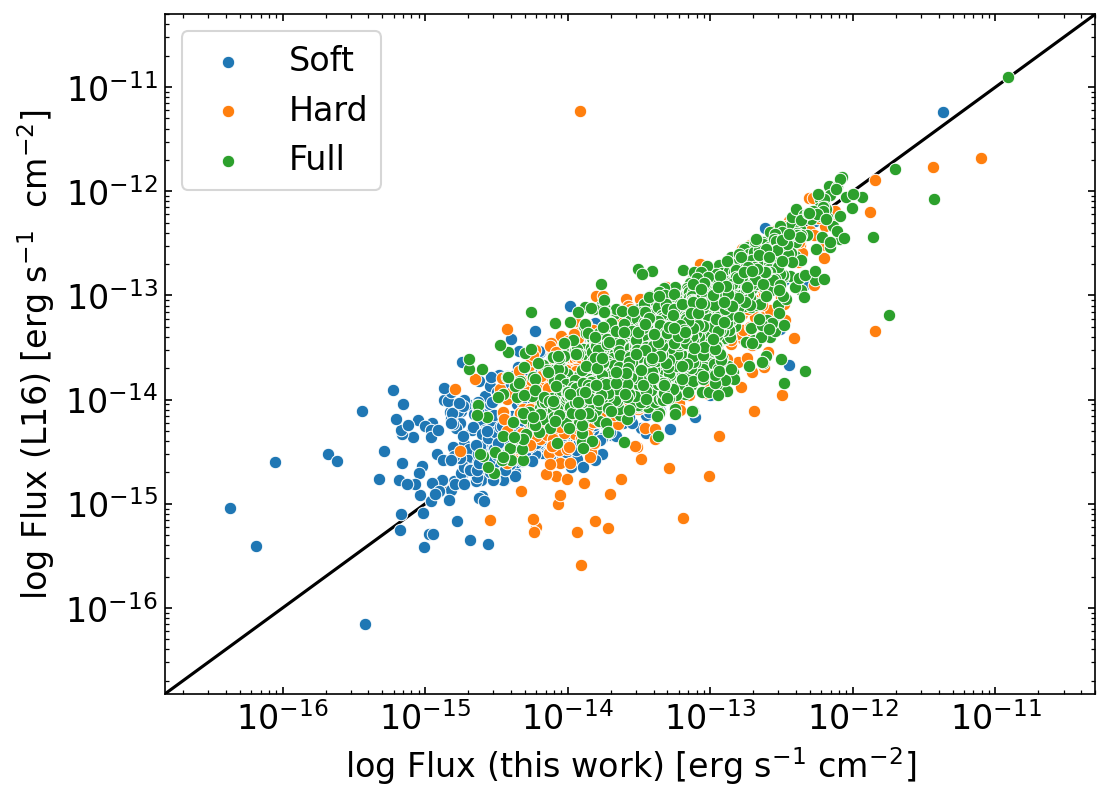}
    \includegraphics[scale=0.45]{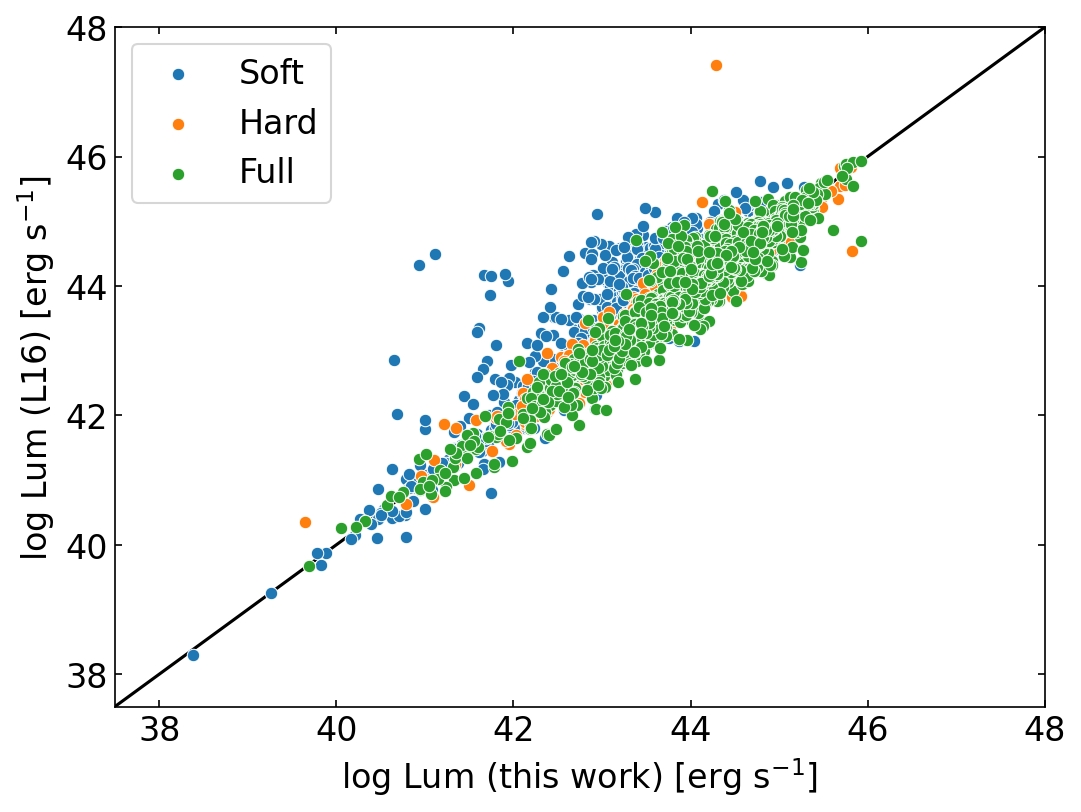}
    \caption{Comparison between derived fluxes (\textit{top panel}) and observed luminosities (\textit{bottom panel}) in this work and those from \citetalias{lamassa16}. Values are shown for the soft (\textit{blue}), hard (\textit{orange}), and full (\textit{green}) bands. The \textit{black} solid lines represent the one-to-one relation.}
    \label{fig:fluxcomp}
\end{figure}

In Figure \ref{fig:fluxcomp} we compare fluxes and observed luminosities derived in this work with those obtained by \citetalias{lamassa16}. In the present work, these quantities were derived using best-fit parameters from the spectral analysis (Section \ref{sec:analysis}). \citetalias{lamassa16} derived count rates using the SAS \textsc{emldetect} tool and the Chandra source catalog (\citealt{evans10}) for \xmm and \chandra sources, respectively, and converted to fluxes assuming the same power-law spectrum ($\Gamma=1.7$ for full and hard bands, $\Gamma=2$ for the soft band).
Luminosities were derived with $L = F_{\rm X} \times 4 \pi d_{L}^2$ , where $F_X$ is the flux in the given band and $d_{L}$ is the luminosity distance.
Overall, there is a good correlation for both the flux and luminosity distributions, with few 
outliers (typically sources with large spectral uncertainties due to the low number of counts).

\section{XMM-Newton and Chandra selection functions} \label{app:sens_sim}
\begin{figure}[!tp]
    \center 
    \includegraphics[scale=0.44]{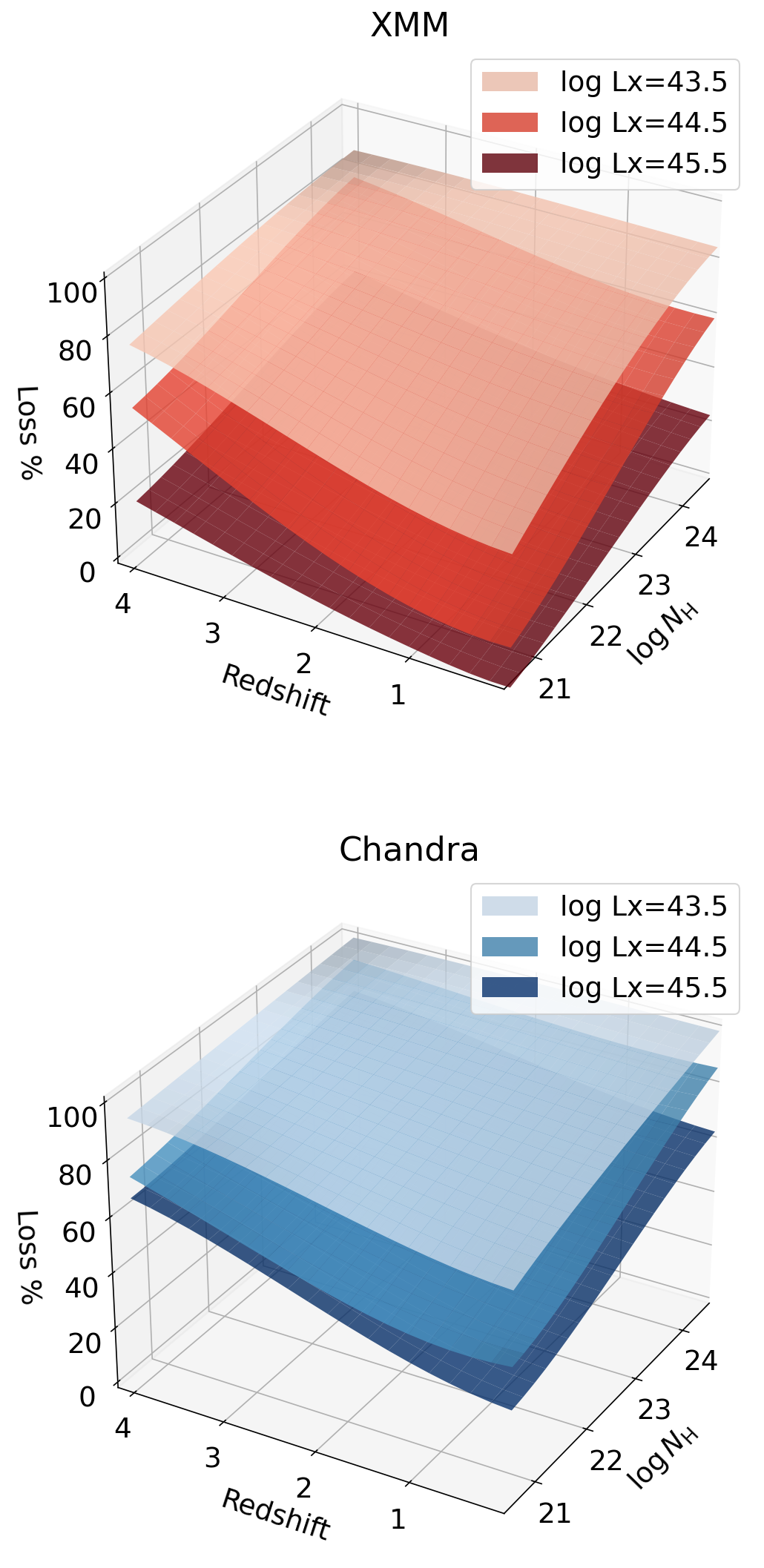}
    \caption{Loss surfaces for \xmm (\textit{top, red}) and \chandra (\textit{bottom, blue}) as a function of $N_{\rm H}$ and redshifts, for different luminosity bins. Representative matrices for both the telescopes were applied.}
    \label{fig:sim_cha_xmm}
\end{figure}

Figure \ref{fig:sim_cha_xmm}\footnote{This figure can be downloaded as a \texttt{numpy} 3D array from GitHub: \\ 
\href{https://github.com/alessandropeca/S82X_sensitivity_correction}{https://github.com/alessandropeca/S82X\_Correction}} shows the simulations described in Section \ref{sec:obs_evol:corr} for \xmm\ (top panel, red scale) and \chandra\ (bottom panel, blue scale) individually. A qualitative comparison between the two panels immediately shows the different efficiencies in finding AGN.
\xmm is more effective than \chandra for almost all the parameter combinations. Only at low luminosity ($\log L_{\rm X}/\mathrm{erg\,\, s}^{-1}=43.5$), high redshift ($z>2$), and high absorption (log $\log N_{\rm H}/\mathrm {cm}^{-2}>23$---i.e. where it is more challenging to detect sources---the two telescopes are comparable, with a difference within $\sim$10-15\%. 
The largest differences appear for high luminosity ($\log L_{\rm X}/\mathrm{erg\,\, s}^{-1}=45.5$) regimes, where \xmm\ has an efficiency a factor up to $\sim$30-40 larger than \chandra.
These differences are due to the combination of different effective areas and exposure times in the S82X, where \xmm\ observations are overall deeper than most of the \chandra exposures.

\newpage\clearpage
\bibliography{sample63}{}
\bibliographystyle{aasjournal}



\end{document}